# Integrative dynamic structural biology unveils conformers essential for the oligomerization of a large GTPase

Thomas-Otavio Peulen[2,9†], Carola S. Hengstenberg[1†], Ralf Biehl[3], Mykola Dimura[2,4], Charlotte Lorenz[3,7], Alessandro Valeri[2,10], Semra Ince[1], Tobias Vöpel[1], Bela Faragó[5], Holger Gohlke[4,8], Johann P. Klare[6*], Andreas M. Stadler[3,7*], Claus A. M. Seidel[2*], Christian Herrmann[1*]

[1]Physical Chemistry I, Faculty of Chemistry and Biochemistry, Ruhr-University Bochum, Universitätsstr. 150, 44780 Bochum, Germany; [2]Chair for Molecular Physical Chemistry, Heinrich-Heine-University Düsseldorf, Universitätsstr. 1, 40225 Düsseldorf, Germany; [3]Jülich Centre for Neutron Science JCNS and Institute for Complex System ICS, Forschungszentrum Jülich GmbH, 52425 Jülich, Germany; [4]Institut für Pharmazeutische und Medizinische Chemie, Heinrich-Heine-Universität Düsseldorf, Universitätsstr. 1, 40225 Düsseldorf, Germany; [5]Institut Laue-Langevin, CS 20156, 38042 Grenoble, France; [6]Macromolecular Structure Group, Department of Physics, University of Osnabrück, Barbarastr. 7, 49076 Osnabrück, Germany, [7]Institute of Physical Chemistry, RWTH Aachen University, Landoltweg 2, 52056 Aachen, Germany; [8]John von Neumann Institute for Computing, Jülich Supercomputing Centre, Institute for Complex Systems-Structural Biochemistry (ICS-6), Forschungszentrum Jülich GmbH, Wilhelm-Johnen-Str., 52425 Jülich, Germany; [9]Present Address: Department of Bioengineering and Therapeutic Sciences, University of California, San Francisco, Mission Bay Byers Hall, 1700 4th Street, San Francisco, CA 94143, USA; [10]Present address: Geomatys, 1000 Avenue Agropolis, 34000 Montpellier, France.

[†] **Contributed equally,** [*] **Corresponding authors**

Guanylate binding proteins (GBPs) are soluble dynamin-like proteins with structured domains that undergo a conformational transition for GTP-controlled oligomerization to exert their function as part of the innate immune system of mammalian cells - attacking intra-cellular parasites by disrupting their membranes. The structural basis and mechanism of this process is unknown. Therefore, we apply neutron spin echo, X-ray scattering, fluorescence, and EPR spectroscopy as techniques for integrative dynamic structural biology to human GBP1 (hGBP1). We mapped hGBP1's essential dynamics from nanoseconds to milliseconds by motional spectra of sub-domains. We find a GTP-independent flexibility of the C-terminal effector domain in the µs-regime and structurally characterize conformers being essential that hGBP1 can open like a pocketknife for oligomerization. This unveils the intrinsic flexibility, a GTP-triggered association of the GTPase-domains and assembly-dependent GTP-hydrolysis as functional design principles of hGBP1 that control its reversible oligomerization in polar assemblies and the subsequent formation of condensates.

**Teaser:** How a pocketknife works on the molecular level

# Introduction

The biological function of proteins is directly linked to their structure, conformational heterogeneity, and their associated conformational dynamics. It is well known that structural flexibilities, heterogeneities, and polymorphisms can enable interactions among biomolecules, promote promiscuity with different binding partners, and are often essential for enzymatic activity.(*1,2*) For a molecular understanding of such biological processes (1) the players of the biological process need to be described by structures, and (2) the associated conformational dynamics need to be characterized in detail. However, if taken out of context the structures of individual macromolecules are often uninformative about function. X-ray crystallography and electron microscopy provide detailed insights on snapshots of conformational states revealing secondary structures of individual domains and domain arrangements. However, to relate structures with their associated function it is imperative to study their conformational dynamics and for a molecular understanding of a biological process all conformational states need to be mapped, ideally watching single molecules move along their transition paths.(*3*)

The relevance of dynamic structural biology is most evident for motor proteins such as myosin or dynamin, where cyclic structural changes are the molecular mechanism for their function. A widespread mechanism exerting such biomolecular function is the binding and cleavage of a suitable substrate to switch between at least two distinct states. Mostly hydrolyzable substrates such as the nucleotides ATP or GTP control structural changes by introducing the substrate hydrolysis as a quasi-irreversible step. Notably, the molecular mechanisms of the functionally relevant dynamics are mostly unknown because structurally flexible intermediates cannot be crystallized. Thus, NMR spectroscopy is often employed to map conformationally excited states and intermediates.(*4*)

For larger proteins the determination of dynamic biomolecular structures is extremely challenging, as there is no single technique that can in parallel observe conformational transitions in biomolecules and determine structures with close to atomistic resolution. To overcome the disadvantages of the individual experimental methods, we employ a new integrative approach that unveils dynamic conformers and domain motions of large multi-domain proteins.(*5,6*) By combining multiple experimental techniques we simultaneously probe protein structures and dynamics and cover time scales from nanoseconds to seconds for dynamic structural biology.

We apply this approach to study molecular mechanisms and design principles of large GTPases, a class of soluble proteins that are important for the innate cell-autonomous immunity in multicellular organisms. These large GTPases, namely guanylate binding proteins (GBPs), belong to the dynamin superfamily and more specifically to the class of interferon-γ induced effector molecules of first cell-autonomous defense.(*7*) GBPs have efficient antimicrobial activity against a wide range of intracellular pathogens such as viruses(*8,9*), bacteria(*10-12*) by assembling of inflammasomes(*13,14*) and by directly attacking the parasites(*15*). In living cells GBP isoforms form polar homo- and hetero-oligomers in different subcellular localizations,(*16,17*) that are involved in the intracellular immune response such as: defense against the vesicular stomatitis virus and the encephalomyocarditis virus,(*8*) suppression of Hepatitis C virus replication,(*9*) promotion of oxidative killing and the delivery of antimicrobial peptides to autophagolysosomes.(*18*)

As a prime example for a GBP we study the human guanylate binding protein 1 (hGBP1). hGBP1 is biochemically well characterized and shows nucleotide-dependent oligomerization.(*19*) In vitro studies demonstrated GTP regulated polymerization of hGBP1 and the formation of polar supramolecular structures.(*20*) Noteworthy, a homolog GBP in mice translocates from the cytosol to endomembranes and attacks the plasma membrane of eukaryotic cellular parasites by the formation of supramolecular complexes during infection.(*15*) An additional feature of GBPs is the GTP induced formation of multimeric complexes in mesoscopic droplet-shaped protein condensates (referred to as vesicle-like structures, VLS) and on parasite membranes. VLS potentially facilitate the controlled formation of productive and supramolecular complexes(*20*) that attack intra-cellular parasites in living cells(*15*).

X-ray crystallography on the full-length hGBP1 revealed a folded and fully structured protein with the typical architecture of a dynamin superfamily member. hGBP1 consists of a large GTPase domain (LG domain), an alpha-helical middle domain, and an elongated, also purely alpha-helical, effector domain comprising the helices α12 and α13, with an overall length of around 120 Å (**Fig. 1A**).(*21*) X-ray crystallography (*19*) and biochemical experiments (*7*) identified the LG domains as interfaces for GTP-analogue induced homo-dimerization. Like for other membrane associated dynamins that form tubular shaped condensates to fuse membranes in cells (*22,23*), cylindrical and tubular structure have been observed for hGBP1 (*20,22*). For hGBP1 neither molecular structures of these tubules nor precursor structures in solution that inform on the assembly mechanism are known. FRET and DEER experiments on the hGBP1-dimer identified two conformers. In the majorly populated hGBP1 dimer, the two C-terminal α13 helices associate.(*24*) This is in line with live-cell experiments that highlight the relevance of helix α13 for the immune response (*12,25,26*). However, an association of the two α13 helices in a hGBP1 dimer requires large-scale structural rearrangements that cannot be explained by the GppNHp bound X-ray crystal structures.(*19*) This highlights the necessity for structural flexibility on the formation pathway of a fully bridged hGBP1 dimer (b-hGBP1:L)$_2$, where nucleotide ligands L (GTP) are bound and the effector domains *and* the LG domains are both associated (**Fig. 1B**). On the formation pathway of the fully bridged hGBP1 dimer, there are at least two intermediates - the ligand complex hGBP1:L and the flexible dimer (f-hGBP1:L)$_2$ (**Fig. 1B**).

We address the question at which step of this pathway hGBP1 becomes flexible (red arrows, **Fig. 1B**). There are three option: either hGBP1's flexibility is substrate independent (*i*), induced by the ligand (*ii*), or induced by the dimerization (*iii*). Consequently, three potential dimerization scenarios could describe the required structural rearrangements (**Fig. 1B**). In the first pathway (**Fig. 1B,** *i*) the structural flexibility is an intrinsic property of the free monomer and already in the absence of substrate; although the flexibility is only needed the dimerization at a later step. In the second pathway (**Fig.1B,** *ii*) the free monomer is predominantly stiff; the binding and/or the hydrolysis of the substrate in the complex hGBP1:L increases the flexibility needed at a later stage for the dimerization. In the third alternative pathway (**Fig.1B,** *iii*) GTP binds to hGBP1 to enable dimerization of the LG domains and the LG domain dimerization triggers an internal rearrangement for effector domains to associate. To sum up, the pathways could be distinguished if one studies the monomeric hGBP1 that has either a single (pathways *ii*, *iii*) or multiple conformations in dynamic exchange (pathway *i*). Otherwise, the different mechanisms are indistinguishable. Hence, to differentiate these pathways, we map the structure and dynamics of the free and the ligand bound hGBP1.

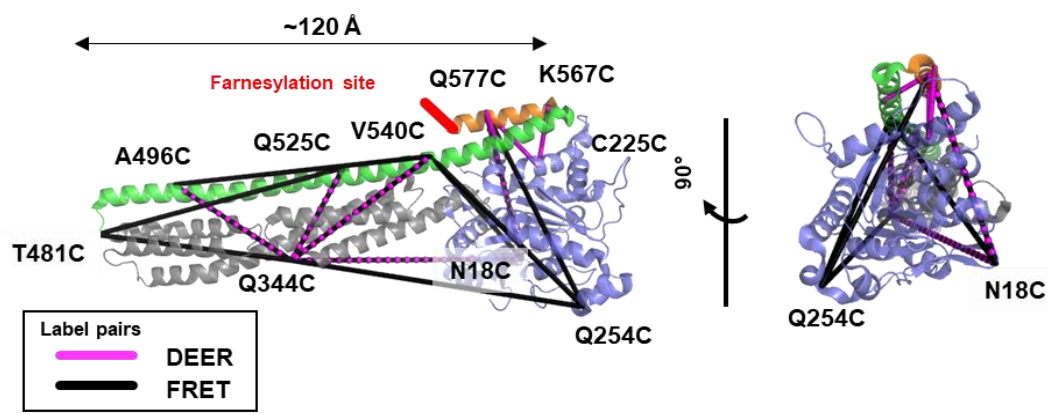

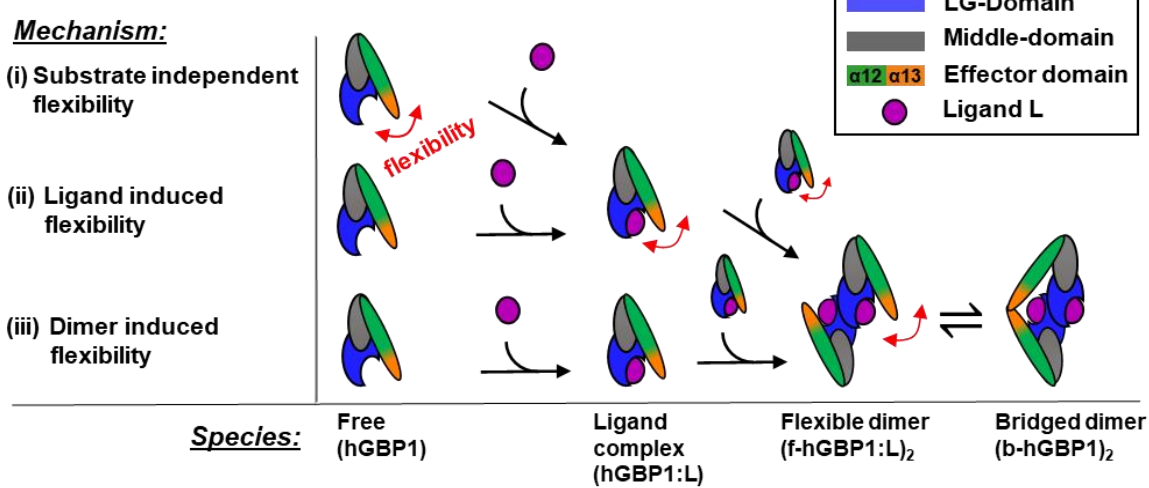

**Figure 1 | Network of pair-wise mutations for DEER and FRET measurements to probe the structural arrangement of the human guanylate binding protein 1 (hGBP1) and potential dimerization pathways.** (**A**) The network is shown on top of the crystal structure (hGBP1, PDB-ID: 1DG3). hGBP1 consists of three domains: the LG domain (blue), a middle domain (gray) and the helices α12/13 (green/orange). The amino acids highlighted by the labels were used to attach spin-labels and fluorophores for DEER-EPR and FRET experiments, respectively. Magenta and black lines connect the DEER pairs and FRET-pairs, respectively. In hGBP1 the C-terminus is post-translationally modified and farnesylated for insertion into parasite membranes (red). (**B**) Potential different pathways for the formation of a functional hGBP1 dimer where the substrate binding LG domains and the helix α13 associate. The association of the helix α13 requires flexibility (highlighted by red arrows). This flexibility could be induced at different stages of a dimerization pathway.

We employ an integrative modeling toolkit for dynamic structural biology to address two objectives: (1) mapping the motions of the monomeric hGBP1 in the absence and the presence of a ligand and (2) resolving the structures of potential hGBP1 conformers in solution. This way, we study the molecular prerequisites for hGBP1 dimerization. We use structural information from small-angle X-ray scattering (SAXS), electron paramagnetic resonance (EPR) spectroscopy by site-directed spin labeling(*27*), ensemble and single-molecule fluorescence spectroscopy(*28*) and dynamic information from neutron spin-echo spectroscopy (NSE) and filtered fluorescence correlation spectroscopy (fFCS)(*5*). We mapped exchange kinetics from nanoseconds to milliseconds and detected at least two new conformational states in hGBP1.

Moreover, interrogating hGBP1's conformational dynamics by a network of 12 FRET pairs (**Fig. 1A**), we generated a temporal spectrum of hGBP1's internal motions. Finally, we discuss potential implications of the detected protein flexibility and conformers controlling the formation of multimers. This allows us to understand the mechanisms excreting the function of this large multi-domain system, i.e., the programmed and controlled oligomerization.

We expect that our findings on the so far unresolved intrinsic flexibility of nucleoside triphosphate processing enzymes will sharpen our view on the importance of conformational dynamics for ligand-controlled allosteric regulation of multi-domain proteins and enzymes far beyond GBPs.

## Results

**Experimental equilibrium distributions**

We combined SAXS, DEER, and FRET experiments to probe distinct structural features of hGBP1 expressed and labeled for DEER and FRET by standard procedures (**Methods 1**). Size exclusion chromatography SAXS (SEC-SAXS) measurements (**Methods 2**) were performed at different protein concentrations (**Fig. S1A**). SEC-SAXS assures the data quality by discriminating aggregates and oligomeric species in the sample immediately before the SAXS data acquisition. A Kratky-plot of the SAXS data (**Fig. 2A**, middle) visualizes that hGBP1's conformation in solution clearly disagrees with the crystal structure of the full-length protein (PDB-ID: 1DG3). *Ab initio* modeling of the SAXS data (**Methods 2**) revealed a shape that suggests an additional kink between the LG and the middle domain (**Fig. 2A**, right).

Orthogonal information to the SAXS data were obtained by DEER (**Methods 3**) and FRET experiments (**Methods 4**), which specifically probe distances between labeling sites (**Fig. 1A**). The results of the DEER and FRET measurements and analyses are exemplified for the dual cysteine variant Q344C/A496C labeled by MTSSL spin probes for DEER experiments (**Fig. 2B**) and by the fluorophores Alexa488 and Alexa647 for ensemble FRET experiments

(**Fig. 2C**).

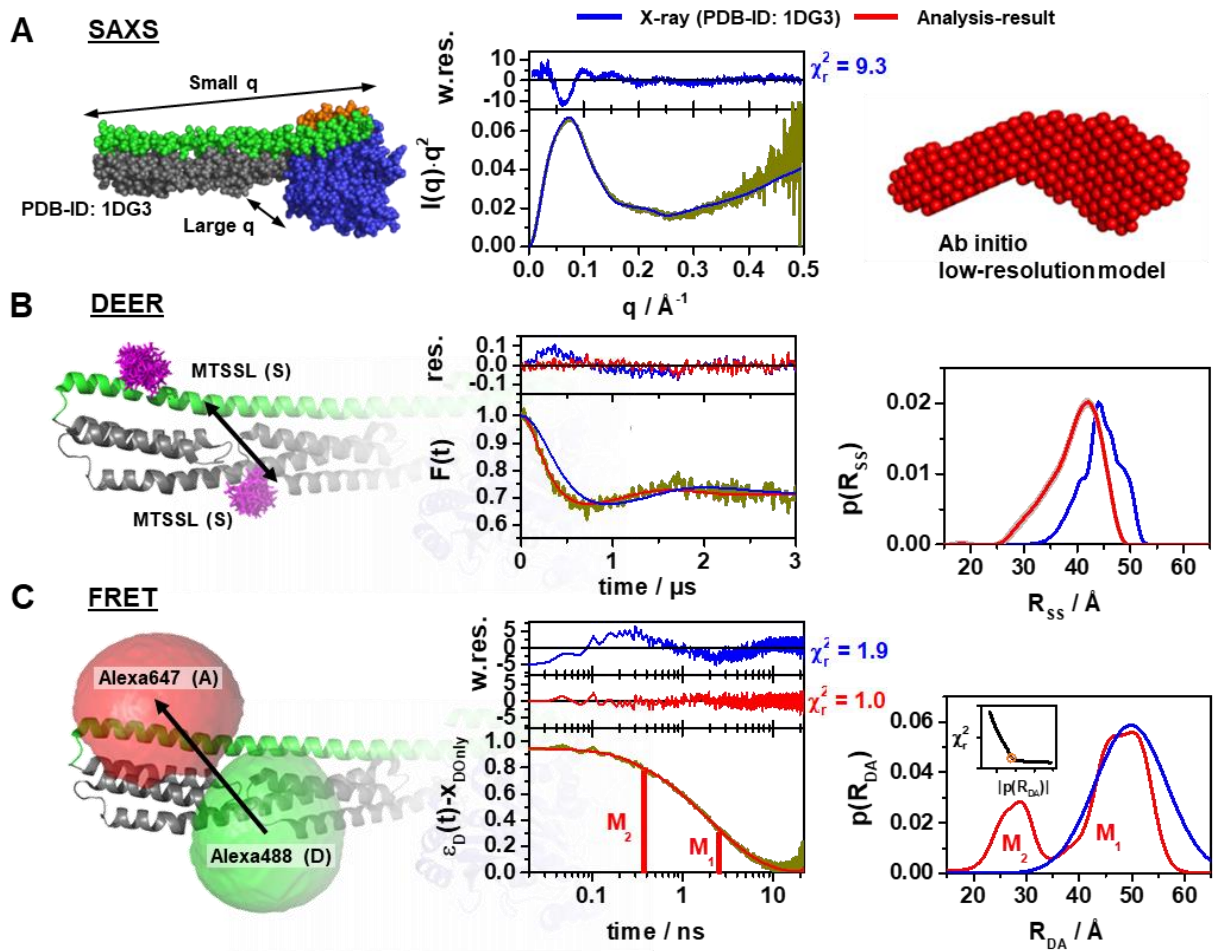

**Figure 2 | Probing the structure of hGBP1 in solution experimentally.** The left panels illustrate the characteristic properties probed by the experimental techniques – (**A**) small angle X-ray scattering (SAXS), (**B**) double electron-electron resonance spectroscopy (DEER), and (**C**) Förster resonance energy transfer spectroscopy (FRET); the middle panels display representations of the experimental ensemble data (dark yellow curves), and the right panels show analysis results thereof. Model-free analyses of the data are presented in red. Predicted experimental data based on a full-length X-ray crystal structure of hGBP1 (PDB-ID: 1DG3) are shown in blue. To the top of the experimental curves, either data noise weighted, w.res., or unweighted residuals, res., are shown (middle panels). In SAXS the scattered intensity $I(q)$ is measured as a function of the scattering vector $q$. For better illustration, $I(q)$ is presented in a Kratky-plot (A: middle). SAXS *ab initio* bead modeling determines an average shape of hGBP1 in solution (A: right). DEER and FRET experiments sense distances between labels that are flexibly coupled to specific labeling sites (exemplified for the double cysteine variant Q344C/A496C). The DEER experiments measured the dipolar coupling between two MTSSL spin-labels (left panel, magenta). FRET experiments measure the energy transfer from a donor fluorophore (D, Alexa488, green) to an acceptor fluorophore (Alexa647, red). In DEER (center) and time-resolved FRET experiments (middle), time-dependent responses of the sample inform on the inter-label distance distributions (right panels). Recovered distance distributions are compared to structural models by simulating the spatial distribution of the labels around their attachment point (left panels). The spatial distributions of the MTSSL-labels (B: left), as well as the donor and acceptor dye (c: left), are shown in magenta, green, and red, respectively. DEER-traces, $F(t)$, analyzed by Tikhonov regularization (red curve) recover inter-spin distance distributions, $p(R_{SS})$. Fluorescence intensity decays of the donor analyzed by the maximum entropy method (MEM) recover donor-acceptor distance distributions, $p(R_{DA})$. The inset displays the L-curve criterion of the MEM reconstruction for the presented data set. The FRET-induced donor decay, $\varepsilon_D(t)$, represents the fluorescence decays.(29) $\varepsilon_D(t)$ is corrected for the fraction of FRET-inactive molecules, $x_{DOnly}$, in the sample. The shape of $\varepsilon_D(t)$ reveals characteristic times (labeled $M_1$ and $M_2$) that correspond to peaks in $p(R_{DA})$. To the right inter-label distance distributions for DEER and FRET are shown.

A model free analysis of the Q344C/A496C DEER data by Tikhonov regularization (**Methods 3**) revealed a clear shift of ~2.5 Å towards shorter distances for the experimental inter-spin distance distribution $p(R_{SS})$ compared to the average distance simulated for an X-ray structure (PDB-ID: 1DG3) using a rotamer library analysis approach (RLA)(*30*) to account for the conformational space of the spin label side chain (**Fig. 2B**, right, **Methods 3**). This indicates that the protein exhibits conformations, where the spin-labels come closer than suggested by the crystal structure. Overall, the $p(R_{SS})$ of all eight DEER measurements were unimodal (**Fig. S2**). Their experimental average distances, $\langle R_{SS,exp} \rangle$, differ from the RLA-predicted distances, $\langle R_{SS,sim} \rangle$, by 1.0 Å to 3.6 Å (**Tab. S1A**). The RLA approach does not account for protein backbone dynamics. Thus, we anticipated finding narrower $p(R_{SS})$ in the simulations than in the experiments. Yet, for the variants Q344C/Q525C and Q344C/V540C the experimental $p(R_{SS})$ are even narrower than the $p(R_{SS})$ predicted by RLA for the crystal structure (**Tab. S1A**). This reduced spread of possible inter-spin distances indicates a reduced conformational freedom of the spin-labeled side chains due to a denser packing of the spin label(s) with the neighboring side chains and/or backbone elements than predicted from the crystal structure.

Our ensemble and single-molecule FRET measurements (**Methods 4**) were paralleled by the following control studies confirming that the probes can accurately report on inter-fluorophore distance distributions, $p(R_{DA})$, of a molecular ensemble representative for the wild-type protein (**Supplementary Note 1**). (*i*) Single-molecule anisotropy measurements (**Fig. S3A, Tab. S1A**) and control samples in the absence of FRET (**Tab. S1B**) validate the model of a mobile dye only weakly quenched by its local environment. (*ii*) Activity assays demonstrate that the dyes and the introduced Cys mutations only weakly affect the protein function (**Supplementary Note 1, Fig. S4**). To recover inter-fluorophore distance distributions, $p(R_{DA})$, we applied ensemble time-correlated single photon counting (eTCSPC) to record high-precision fluorescence intensity decays $f_{DD}^{(DA)}(t)$ and $f_{DD}^{(D0)}(t)$ of donor (D) fluorophores in the presence (DA) and the absence (D0) of acceptor (A) fluorophores, respectively.(*29*) The measured fluorescence decays are available in a public data repository (**Data availability**). We computed the FRET-induced donor decay $\varepsilon_D(t) \equiv f_{D|D}^{(DA)}(t)/f_{D|D}^{(D0)}(t)$ to directly visualize $p(R_{DA})$ in a semi-logarithmic plot of $\varepsilon_D(t)$ where the position (time) and the height (amplitude) of steps recover DA distances and species fractions, respectively.(*29*) $\varepsilon_D(t)$ of the variant Q344C/A496C clearly revealed two distances - a hallmark for hGBP1's conformational heterogeneity. (**Fig. 2C**, center). An analysis of $f_{D|D}^{(DA)}(t)$ by the maximum-entropy method (MEM) resolved a bimodal distance distribution *p(R$_{DA}$)* (**Fig. 2C**, right) with a *major* and *minor* subpopulation. To associated conformational states are referred to as M$_1$, and M$_2$, respectively (**Fig. 2C**, right). For an unambiguous assignment of the experimental distances to the conformational states, all 12 datasets (**Fig. 1**) were analyzed by a joint/global quasi-static homogeneous FRET-model (*29*))for all samples with shared species fractions of M$_1$ = 0.61 and M$_2$ = 0.39(**Tab. S1A**) at room temperature.

To compare theoretical and experimental average DA distances $\langle R_{DA} \rangle$ and distance distributions *p(R$_{DA}$)*, we need (1) a dye model that predicts the spatial distributions of the flexibly linked dyes for given structural models(*31,32*) and (2) reliable uncertainty estimates of the experimental distances (**Supplementary Note 1**). To account for variable interactions of the dyes with the protein surface, we performed accessible contact volume (ACV) simulations of the dye, that consider both, the accessible volume of free dye and the fraction of dye bound at the protein's surface.(*33*) ACV simulations have the attractive feature that the fraction of surface bound dyes needed for calibration is a direct experimental observable registered in the time-resolved anisotropy decays via the residual anisotropies (**Tab. S1A**). Moreover, we validated the ACV approach by confirming that the fluorescence decays of the donor exhibited no significant additional quenching beyond the quenching anticipated for a dye freely diffusing within an ACV (**Supplementary Note 1**).

Overall, in the FRET measurements, M$_1$ agreed better with the X-ray structure than M$_2$ (**Fig. 2C**, right, **Tab. S1A**) - the sum of uncertainty weighted squared deviations, $\chi^2_{FRET}$, for M$_1$ is significantly smaller than for M$_2$ ($\chi^2_{FRET}(M_1, 1DG3) \sim 17$ *vs.* $\chi^2_{FRET}(M_2, 1DG3) \sim 1500$). In an F-test, this corresponds to a p-value>0.999. Thus, considering statistical uncertainties, potential systematic errors, uncertainties of the orientation factor, and uncertainties of the ACVs due to the differences of the donor and acceptor linker length, (**Supplementary Note 1**), we conclude that M$_1$ is more like the X-ray structure than M$_2$. Remarkably, an analysis of the fluorescence decays for the variants A496C/V540C and T481C/Q525C, which were designed to test the stability of helix α12, revealed identical distances for M$_1$ and M$_2$ (**Tab. S1A**). Hence, we corroborate that helix α12 is predominantly extended, like the helix found in the solved crystal structures (e.g., PDB-ID: 1DG3). The variants N18C/Q344C and Q254C/Q344C, probing distances between the middle- to the LG domain, revealed only relatively minor differences between M$_1$ and M$_2$, while for the variants designed to interrogate motions from the middle- to the helices α12/13, M$_1$ and M$_2$ were significantly different.

To sum up, EPR-DEER at cryogenic temperatures detected small deviations to the crystal structure while SAXS and FRET detected clear deviations at room temperature. Presumably, due to the longer linkers used in the FRET experiments leveraging differences, we resolved the experimental DA-distance distributions $p(R_{DA})$ into two states. Temperature dependent measurements revealed that these states found at room temperature are similarly populated at higher intracellular temperatures (**Supplementary Note 1 (section 2), Fig. S4D**).

**Identification and quantification of molecular kinetics**

To probe the conformational dynamics of hGBP1, we performed single-molecule (sm) FRET experiments with Multiparameter Fluorescence Detection (MFD) (**Methods 4**) and Neutron Spin Echo (NSE) experiments (**Methods 5**).(*34,35*) While the NSE experiments are most sensitive up to a correlation time of 200 ns, the filtered fluorescence correlation spectroscopy (fFCS) analyses of our MFD experiments are most sensitive from sub-microseconds to milliseconds. Thus, by combining NSE with MFD-fFCS, we effectively probe for conformational dynamics from nano- to milliseconds.

An analysis result of the NSE data is visualized in **Fig. 3A**, which displays the scattering vector, $q$, dependent effective diffusion coefficient $D_{eff}$ extracted from the initial slope of the NSE spectra measured up to 200 ns (**Fig. S5A**). The translational diffusion coefficient $D_T$ of the protein was obtained from dynamic light scattering (DLS) at the same concentration (**Methods 5**). The diffusion coefficient of a single protein increases from the translational diffusion $D_T$ measured at low $q$ (DLS) due to contributions from rotational diffusion $D_R(q)$ and contributions related to internal protein dynamics $D_{int}(q)$ as the observation length scale $2\pi/q$ covers the protein size. The translational and rotational diffusion coefficients $D_T$ and $D_R(q)$ were calculated and corrected for hydrodynamic interactions and interparticle effects to result in the expected $D_0(q)$ for a rigid body (**Fig. 3A**, black line, **eq. 18**). The measured $D_{eff}(q)$ agrees well with the theoretical calculations accounting for rigid body diffusion alone. A significant additional contribution of internal protein dynamics to the measured effective diffusion coefficients cannot be identified. The same result was obtained by directly optimizing the parameters of an analytical model describing rigid protein-diffusion (**eq. 19**) to the NSE spectra (**Fig. S5B**). Hence, the overall internal protein dynamics may only result in negligible amplitude, i.e., minor overall shape changes, within the observation time up to 200 ns.

To study structure and dynamics from sub-µs to ms we performed MFD smFRET experiments on freely diffusing molecules briefly (milliseconds) diffusing through a confocal detection volume. We performed a burst-wise analysis of the MFD data for each detected molecule to determine the fluorescence weighted average lifetime of the donor, $\langle\tau_{D(A)}\rangle_F$, and the intensity-based FRET efficiency, $E$ (**Methods 4**).(*35*) MFD diagrams, which are multidimensional frequency histograms of parameters determined for single molecules, directly visualize heterogeneities among the molecules. In a MFD-diagram of $E$ and $\langle\tau_{D(A)}\rangle_F$ the "static FRET-line" serves as a reference to detect fast conformational dynamics.(*36*) A shift of a peak from the static FRET line towards longer $\langle\tau_{D(A)}\rangle_F$ indicates fast conformational dynamics within the observation time of the molecules (~ms) (**Methods 4**).(*35,36*) In all 12 FRET variants, only single peaks were visible in the 2D-histograms for the DA labeled molecules (**Fig. 3B**, **Fig. S3A**). In 8 out of 12 variants, we found clear indications of dynamics by a peak shift of the FRET molecules off the static FRET-line towards longer $\langle\tau_{D(A)}\rangle_F$ (**Fig. S3A**). Analogous to relaxation dispersion experiments in NMR, such shifts confirm that $M_1$ and $M_2$ are in an exchange faster than the integration time of the molecules (~milliseconds).(*35,36*) A detailed analysis of the fluorescence decays of the FRET sub-ensembles (**Fig. S3B**) by a two-component model revealed limiting states (**Tab. S1C**) agreeing with the eTCSPC measurements (**Tab. S1A**). Hence, the peak positions in the MFD histograms are consistent with the eTCSPC analysis and are captured by dynamic FRET-lines, which describe the mixing of the two states (**Fig. 3B**, **Fig. S3A**).

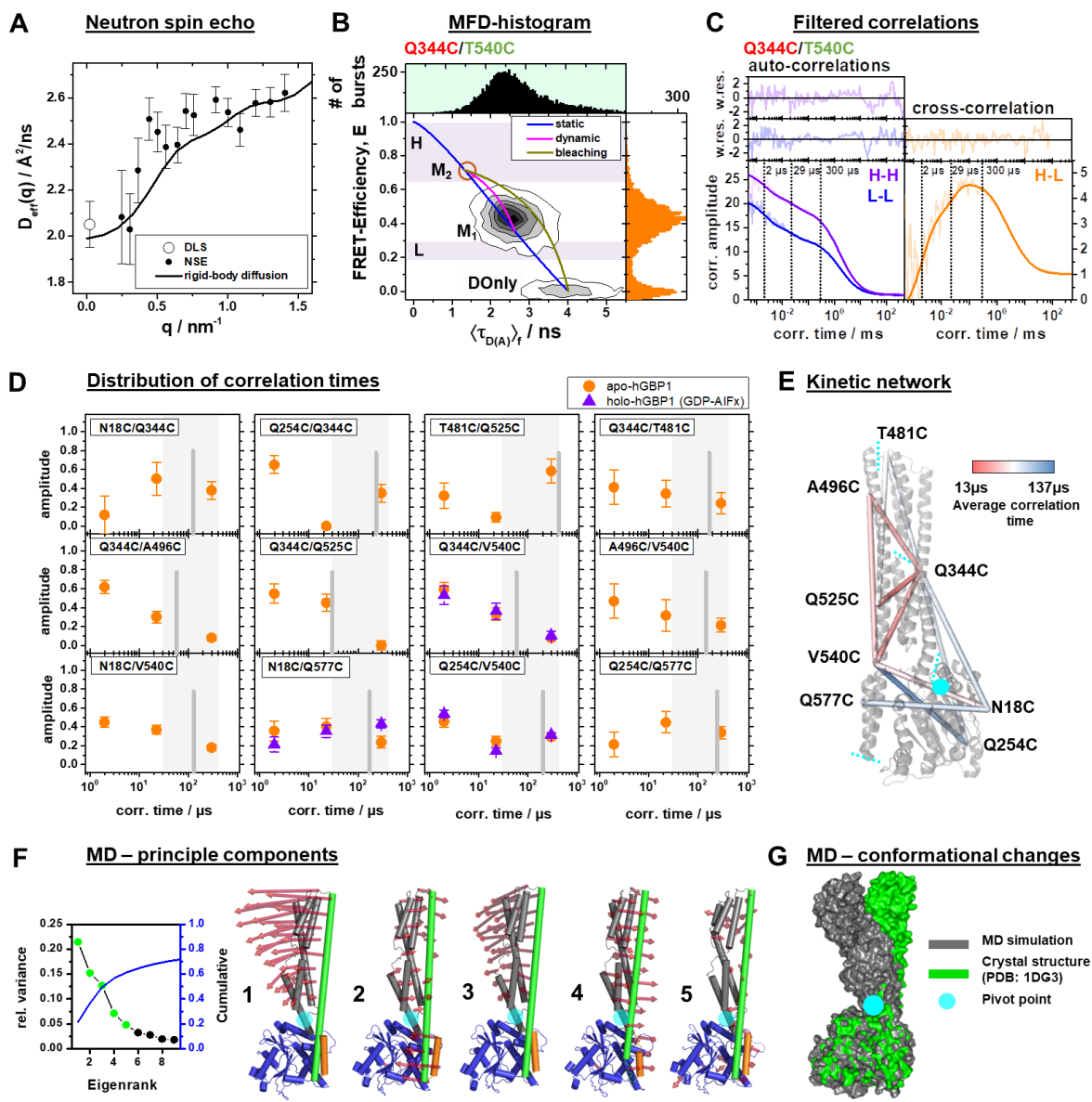

**Figure 3 | Conformational dynamics of hGBP1 studied by NSE, smFRET with multi-parameter fluorescence detection (MFD), and molecular dynamics (MD) simulations.** (**A**) NSE studies of effective diffusion coefficients of hGBP1, $D_{eff}$, determined by NSE and DLS compared to a model describing only the translational and rotational diffusion of the protein by a rigid-body as a function of the scattering vector, $q$. The agreement of the measured effective diffusion coefficients with the calculated values demonstrates insignificant shape changes of hGBP1 as sensed by NSE on fast time scales up to 200 ns. (**B**) A two-dimensional single-molecule histogram displays absolute FRET-efficiencies, $E$, and the fluorescence weighted average lifetimes of the donor in the presence of FRET, $\langle \tau_{D(A)} \rangle$, for the FRET-labeled double cysteine variant Q344C/V540C. The one-dimensional histograms are projections of the 2D histogram. The color of the variant's name indicates the location of the donor (green) and acceptor (red) determined by analysis of time-resolved anisotropy decays. The static-FRET line (blue) relates $E$ and $\langle \tau_{D(A)} \rangle$ for static proteins with a single conformation. The dynamic FRET line (magenta) describes molecules changing their state from $M_1$ to $M_2$ (brown circles) and *vice-versa* while being observed. The states $M_1$ and $M_2$ were identified by eTCSPC (**Tab. S1A**) and sub-ensemble TCSPC (**Fig. S3B**, **Tab. S1C**). Molecules in the state $M_2$ with bleaching acceptors are described by a dark yellow line, which describes the transitions from $M_2$ to the donor only population (DOnly). The highlighted areas refer to the groups of molecules (H for high FRET, L for low FRET) selected for the generation of fluorescence lifetime filters for filtered FCS (fFCS). (**C**) fFCS Q344C/T540C with H and L filters. To the left the species autocorrelation functions (*sACF*) and to the right, a species cross-correlation function (*sCCF*) are shown as semitransparent lines overlaid by model functions shown as solid lines (**eq. 17**). The fFCS model parameters were determined by a joint/global analysis of all 12 FRET-pairs (**Fig. S3C**, **Tab. S2**) revealing at least three correlation times (vertical dotted lines). The weighted residuals for the global analysis are shown to the top. (**D**) The analysis of the fFCS curves assigned to every variant

amplitudes and correlation times shown for the GTP free apo- (orange circles) and GDP-AlF$_x$ bound holo-state (violet triangles). The average correlation times for the variants are shown as gray vertical lines. The gray box highlights the minimum and maximum of the average correlation times for all variants. (**E**) For visualization, the average correlation times of the apo-state were mapped to a crystal structure (PDB-ID: 1DG3, see G) by color-coded bars connecting the Cα-atoms of labeled amino acids. The sections of the five derived rigid bodies are displayed by cyan dashed lines. (**F**) First five principle components of molecular dynamics (MD) and accelerated molecular dynamics (aMD) simulations starting from the crystal structure (PDB-ID: 1DG3). The LG domain, the middle domain, and the helix α12, and α13 are colored in blue, gray, green, and orange, respectively. The red arrows indicate the direction of the motion. The components were scaled by a factor of 1.5 for better visibility. The semi-transparent cyan circle corresponds to a pivot point located at the LG domain. To identify correlated motions in the MD and aMD simulations, we performed principal component analysis (PCA, **Supplementary Note 2**). The first five principal components (PCs) sorted by the magnitude of the eigenvalues, contribute to 60% of the total variance of all simulations. (**G**) Superposition of a MD trajectory frame (gray) deviating the most in RMSD (~8 Å) from the crystal structure (green). Both structural models were aligned to the LG domain.

For two states in dynamic exchange ($M_1 \rightleftharpoons M_2$) under equilibrium conditions, we expected to find a single correlation time. To quantify the precise time-scale(s) of the exchange among the conformers detected by FRET, we analyzed the species cross-correlation functions (*sCCF*) and the species autocorrelation functions (*sACF*) determined by fFCS (**Fig. S3C**) and displayed the results as relaxation time spectra, where the normalized amplitudes are plotted versus the correlation time (**Fig. 3D**). Surprisingly, each individual set of *sCCF* and *sACF* of the 12 FRET pairs required at least two correlation times to be fully described. This is an indication for more complex kinetics or more (kinetic) states, which are unresolved by the analysis of the fluorescence decays. In a global analysis of all 12 variants (**eq. 17**), where we treat all 48 fFCS curves (two *SACF* and *SCCF* per variant) as a single dataset and we recovered three joint correlation times of 2, 23 and 297 µs (**Fig. 3C**). However, the amplitudes of the relaxation times differ significantly so that the average relaxation time varies approximately by one order of magnitude (gray bars in **Fig. 3D**). Intriguingly, this global analysis reveals a variant-specific amplitude distribution of correlation times and highlights significant differences among the variants (**Fig. 3D**, **Fig. S3C**, **Tab. S2**). In most cases, the shortest correlation time has the highest amplitude. This is consistent with the MFD histograms, because we detected shifted/dynamic unimodal peaks. To visualize the dynamics detected by fFCS, we mapped the average correlation times color coded to the FRET network shown on top of a protein X-ray structure (**Fig. 3E**). This visualization highlights that the fast dynamics is mainly associated with the helices α12/13 and the middle domain, while the slow dynamics is predominantly linked to the LG domain.

Referring to the sketch in **Fig. 1B**, we hypothesize that the states $M_1$ and $M_2$ and the transition among them are of functional relevance (pathway *i*). Therefore, we studied the effect on the dynamics exerted by the ligand GDP-AlF$_x$ as a substrate that mimics the holo-state hGBP1:L. The GDP-AlF$_x$ concentration was sufficiently high (100 µM) to fully induce dimerization for hGBP1 at µM concentrations.(*15*) For comparison, the affinity of hGBP1 for mant-GDP is ~3.5 µM and much higher for GDP-AlFx.(*37*) Hence, in the sm-measurements GDP-AlF$_x$ was bound to the LG domain while hGBP1 (20 pM) was still monomeric. We refer to this as the holo-form of the protein and selected a set of variants (N18C/Q577C, Q254C/V540C, Q344C/V540C) for which we found large substrate induced effects at higher hGBP1 concentrations due to oligomerization. Surprisingly, the amplitude distribution of the correlation times is within errors indistinguishable from the measurements of the nucleotide-free apo forms (**Fig. 3D**). Moreover, the FRET observables did not change either.

In conclusion, our integrative study on the structure and dynamics yielded the four major results: (1) We identified heterogeneous conformational ensemble that can be approximated by two majorly populated conformers $M_1$ and $M_2$. $M_1$ is similar to the crystal structure. (2) NSE detects no significant shape changes of hGBP1 on a time-scale up to 200 ns. (3) For the exchange $M_1 \rightleftharpoons M_2$ probed by fFCS, we expected to find a single correlation time but found a complex distribution of correlation times spanning the µs-range. Therefore, we propose for the motion of α12/13 relative to the LG and the middle domain additional intermediate conformational states, resolved by their kinetic fingerprint captured by fFCS. (4) hGBP1's kinetics between the middle domain and α12/13 is unaffected by the presence of a nucleotide analog as a substrate.

**Essential motions determined by molecular dynamics simulations**

We performed molecular dynamics (MD) simulations without experimental restraints to assess the structural dynamics of the full-length crystal structure at the atomistic level and to capture potential motions of hGBP1 (**Methods 6, Supplementary Note 2**). The apo (PDB-ID: 1DG3) and a GTP bound holo-form of hGBP1 were simulated in three replicas by conventional MD simulations for 2 µs each (**Fig. S6A**). Additionally, accelerated molecular dynamics (aMD) simulations, which proved to sample the free-energy landscape of a small protein (58 amino acids) 2000-fold more efficiently(*38*), were performed in two replicas of 200 ns each. Autocorrelation analysis of the RMSD *vs.* the average structure of the MD simulations reveals fast correlation times. The average correlation time in the presence and the absence of GTP were 11 ns and 17 ns (**Fig. S6B**). However, note that the amplitude of the fluctuations is, on average, below an RMSD of 3 Å, which is below the resolution limit of our NSE measurements. In the MD simulations, larger conformational changes (RMSD > 7 Å) with considerable shape changes were very rare events. A principle component analysis revealed kinking motions of the middle domain and helix α12/13 around a pivot point as most dominant motions in the MD simulations **(Fig. 3F)**. A visual inspection of structures deviating most from the mean reveals a kink at the connector of the LG and the middle domain (**Fig. 3G**) consistent with rearrangements required for average shape as recovered by SAXS (**Fig. 2**).

To sum up, the MD simulations cover only time-scales of a few microseconds. Nevertheless, they indicated potential directions of motions and identified a pivot point between the LG and the middle domain. In agreement with NSE on the simulation time-scale, the overall shape is majorly conserved, and large conformational changes are rare events. The helices α12/13 were mobile and exhibited a limited "rolling" motion along the LG and middle domain that could connect the conformers $M_1$ and $M_2$ as suggested by our FRET studies.

**Experimentally guided structural modeling**

We integrate the experimental evidence for alternative conformations beyond the crystal structure into structural models of hGBP1 (**Methods 7**). Considering the specific requirements of label-based methods(*31,32*) we previously demonstrated using synthetic data the reliability of MFD measurements for resolving short-lived conformational states by structural models of a large GTPase.(*33*) Here, we additionally integrate DEER and SAXS data in a joined framework for an unbiased meta-analysis (**Methods 7**) and generate quantitative structural models for hGBP1 in three major steps: (*i*) "*Data acquisition*", (*ii*) "*Model generation*", and (*iii*) "*Model discrimination*" (**Fig. 4A**). In a previous *in silico* benchmark study on the GTPase Alastin, we needed only 29 optimal chosen FRET pairs to achieve an accuracy *vs.* the target structures and a precision below 2 Å.(*33*) For the given set of 12 FRET and 8 DEER pairs of hGBP1 we expect to recover low-resolution models with an average RMSDs in the range of 8-15 Å, and aim to resolve hGBP1's shape, domain arrangement, and topology.

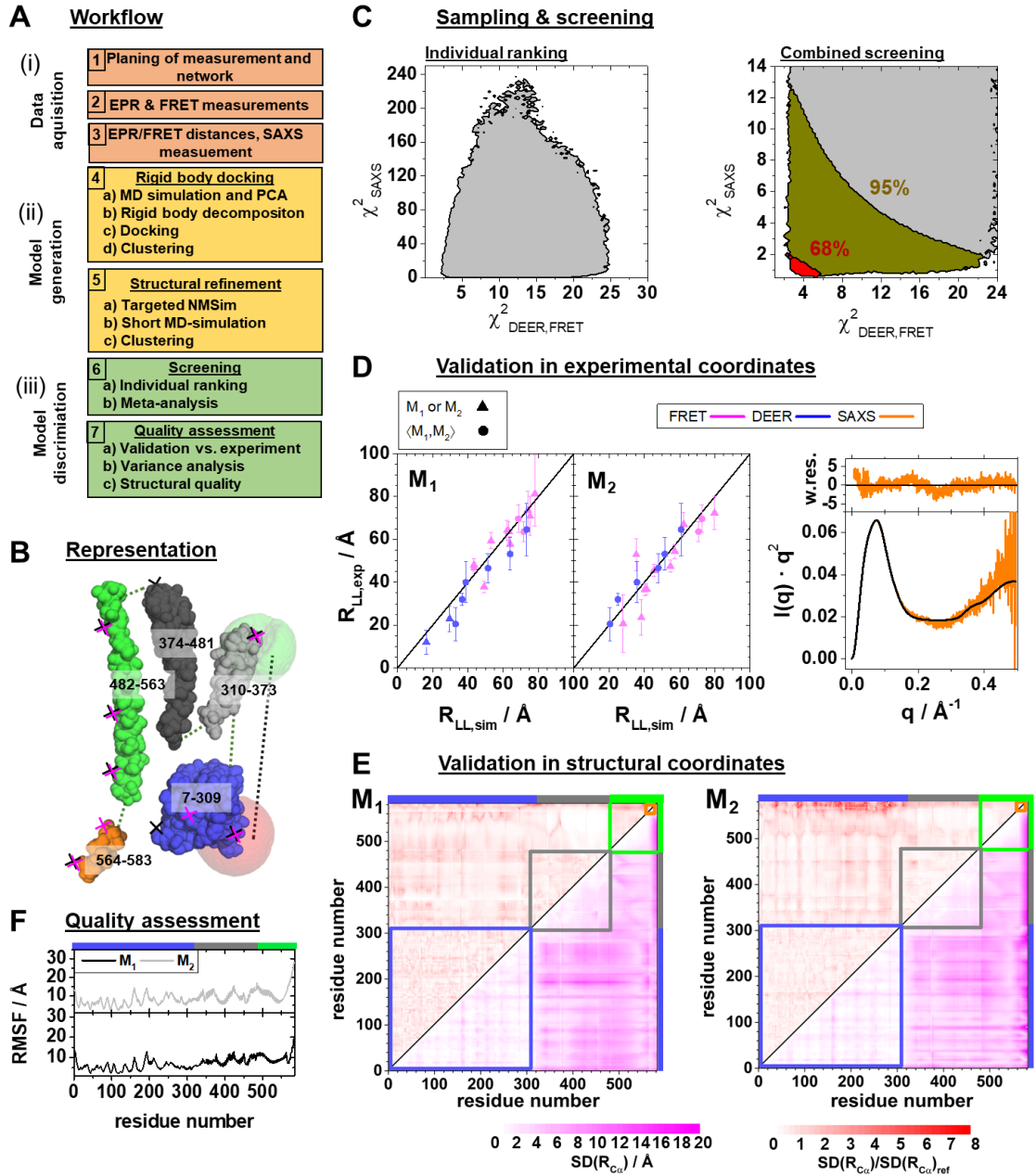

**Figure 4 | Generation and validation of structural models of hGBP1 by integrative modeling using DEER, ensemble FRET, and SAXS data.** (**A**) The applied workflow creates structural models in three steps: "Data acquisition", "Model generation", and a final "Model discrimination" step. The protocol for the generation of structural models combines a coarse-grained rigid body docking (RBD) approach(*32*) with a structural refinement step by NMSim(*39*) and molecular dynamics (MD) simulations. Rigid domains are identified for RBD by MD simulations and principal components analysis (PCA) thereof (**Supplementary Note 2**). Finally, the generated structural models are ranked and discriminated (screening) by the experimental data. (**B**) Representation of hGBP1 by a set of rigid bodies: LG-domain (blue), the middle domain (gray), helix α12 (green), helix α13 (orange). The middle domain was decomposed into two separate bodies. The amino acid ranges of the rigid bodies are given by the shown numbers. The crosses indicate the labeling positions for the FRET (black) and the EPR (magenta) experiments. The average fluorophore positions of an example FRET pair are shown as spheres within the accessible volumes of the dyes which are shown as semi-transparent green (donor) and red (acceptor) surfaces. (**C**) To the left a histogram of $\chi^2_{SAXS}$ (**eq. 22**) and $\chi^2_{DEER,FRET}$ (**eq. 23**) and for all pair of structural models ($M_1$, $M_2$) is shown. $\chi^2_{DEER,FRET}$ and $\chi^2_{SAXS}$ are sums of weighted squared deviations and rank the pairs of structural models. To the right, a meta-analysis (**eq. 25**) fuses the probabilities derived from $\chi^2_{DEER,FRET}$ and $\chi^2_{SAXS}$ to discriminate pairs ($M_1$, $M_2$). The red and the dark yellow areas highlight regions below a p-value of 0.68 and 0.95, respectively. For a given p-value all pairs of structural models with smaller p-values need to be accepted. (**D**) Comparisons of the individual experiments with the pair of structural models best agreeing with all experiments validate the model

by the data. Left diagrams (values see Tab. S3A): Experimental $R_{LL,exp}$ (for DEER $\langle R_{LL,exp}\rangle$ and FRET $\bar{R}_{DA,exp}$) are compared to modeled average inter-label distances $R_{LL,sim}$ (for DEER $\langle R_{LL,sim}\rangle$ and FRET $\bar{R}_{DA,sim}$). We use specific symbols to display inter-label distances $R_{LL,exp}$ for label pairs with distinct (▲) and equal (●) values for $M_1$ and $M_2$, respectively. For EPR $R_{LL,exp}$ represents the average inter-label distance $\langle R_{LL,exp}\rangle$ and for FRET $R_{LL,exp}$ represents the central donor-acceptor distance $\bar{R}_{DA,exp}$. Right diagram: For SAXS, the theoretical scattering curve (black line) is directly compared to the experimental data (orange line) with weighted residuals to the top. (**E**) The standard deviation, SD, of the pair-wise Cα-Cα distance $SD(R_\alpha)$ of the experimentally determined conformational ensemble with a p-value < 0.68 (lower triangles) highlights the variability with possible models for $M_1$ (left) and $M_2$ (right). The $SD(R_\alpha)$ normalized by the smallest possible $SD(R_\alpha)_{ref}$ for the given set of distances and experimental noise (**Methods 7**) validates the selected models in structural coordinates. (**F**) For the structural models with a p-value < 0.68, the root mean square fluctuations of the Cα atoms are displayed for the globally aligned ensemble.

*Data acquisition.* We initially assumed as prior knowledge that the crystal structure of hGBP1 corresponded to the solution structure and designed the above experiments to test this assumption (**Fig. 4A**, steps 1-3). *Model generation.* As we disproved this initial assumption, we employed the experimental data to generate new structural models by modifying our initial model (**Fig. 4A**, steps 4-5). For that, we sampled the experimentally allowed conformational space as vastly as possible by combining simulations of different granularity and computational complexity (**Methods 7**). First, we identify a set of rigid bodies (RBs) (**Fig. 4B, Supplementary Note 4**) using the information on the motions observed in the MD simulations (**Fig. 3F**), an order-parameter based rigidity analysis (**Fig. S6C**), knowledge on the individual domains within the dynamin family (*40,41*), position dependence of FRET and DEER properties (Tab. S1A) and SAXS experiments suggesting a kink in hGBP1's middle domain. To this RB assembly, we applied DEER and ensemble FRET restrains for guided rigid body docking (RBD) (**Methods 7**, **Supplementary Note 5**).(*32*) In this docking step, DEER and FRET restrains were treated by AV and ACV simulations, respectively. Next, all generated RBD structures were corrected for their stereochemistry using NMSim.(*39*) This was achieved by guiding the crystal structure in NMSim towards the RBD structures as templates minimizing the root mean squared deviation (RMSD) taking the uncertainties into account (**Methods 7, Tab. S1A, D**). These refined models were clustered into 343 and 414 groups for the states $M_1$ and $M_2$, respectively (**Methods 7**). Group representatives were used as seeds for short (1-2 ns) MD-simulations of all 343 and 414 group representatives. The MD trajectories were clustered into 3395 and 3357 groups for $M_1$ and $M_2$, respectively, before the model discrimination step by the DEER, FRET, and SAXS data (**Methods 7**).

*Model discrimination.* First, the structural models were ranked by their agreement with the individual techniques, using the quality measures $\chi^2_{SAXS}$ and $\chi^2_{DEER,FRET}$, which capture deviations between the model and the data for SAXS and for the combined DEER and FRET datasets, respectively (**Methods 7**). For maximum parsimony with respect to the modelled conformational states, the DEER, FRET and SAXS measurements were described by two states $M_1$ and $M_2$. Theoretical SAXS curves for all structural models of $M_1$ and $M_2$ were calculated using the computer program CRYSOL. Using the theoretical SAXS curves all possible combinations of structural models for $M_1$ and $M_2$ were ranked by their agreement with the SAXS data in an ensemble analysis (**Fig. 4A**, step 6a; **Fig. 4C, eq. 21**). Like in the model free bead modeling of the SAXS data (**Fig. 2**), for the pair of structural models best agreeing with SAXS the middle domain is kinked towards the LG domain (**Fig. S1C**). The SAXS ensemble analysis reveals species population fractions for $M_1$ in the range of ~0.1-0.7 (**Fig. S1D**, p-value = 0.68). For DEER and FRET measurements, $M_1$ and $M_2$ representatives were ranked by comparing the average simulated inter-label distances $\langle R_{SS,sim} \rangle$ and $\langle R_{DA,sim} \rangle$ with the corresponding experimental distances. The simulated average inter-label distances for $M_1$ and $M_2$ were determined by simulating the distribution of the labels around their attachment point (**Methods 7**).(*30-32,42*) Next, a meta-analysis by Fisher's method fused the experimental data and analysis to rank and discriminate the generated structural models in a statistically meaningful manner (**Fig. 4A**, step 6b). In this step, the meta-analysis considers estimates for the degrees of freedom (dof) of the model and the data (**Methods 7**). This way, we select well-balanced structural models and fully avoid fudge factors equalizing experimental contributions to the model (**Fig. 4C**, Combined screening). A stability test demonstrates that reasonably chosen dofs have only a minor influence on the results (**Fig. S6D**). In the final analysis, a p-value of 0.68 discriminated more than 95% of all structural models (**Fig. 4C**, red area; **Fig. S6E**), leaving 99 and 105 models with average RMSDs of 11.2 Å and 14.5 Å for $M_1$ and $M_2$, respectively. For these structures, uncertainties are largest for α12/13 (**Fig. 4E**).

*Quality assessment*. As the last step, we assessed the quality of the selected structures (**Fig. 4A**, step 7). For DEER and FRET, the consistency of the model with the experiment is demonstrated by comparing simulated distances to the analysis result of the experiment. This is visualized for the pair of structural models best agreeing with DEER, FRET and SAXS combined (**Fig. 4D**, left). For DEER such comparison was used to identify outliers (**Supplementary Note 1, Fig. S4E**). The SAXS measurements were compared to the model by calculated theoretical scattering curve (**Fig. 4D**, right). These representations demonstrate that the models capture the essential features of the experiments within the experimental uncertainties. The experimental uncertainties were propagated to model uncertainties through an exhaustive sampling of the model's conformational space. The model uncertainties are visualized by the experimental standard deviation of pair-wise Cα distances, $SD(R_{C_\alpha})$, (**Fig. 4E**, lower triangles). This representation reveals regions of low and high variability in the structural models. To assess the local quality of the models and check if their variabilities are larger than statistically expected, we compared the experimental precision $SD(R_{C_\alpha})$ to a reference precision $SD(R_{C_\alpha})_{ref}$ by computing the weighted (normalized) precision, $SD(R_{C_\alpha})/SD(R_{C_\alpha})_{ref}$ (**Methods 7**, **Fig. 4E** upper triangles). The weighting reference $SD(R_{C_\alpha})_{ref}$ is the expected precision of "ideal and perfect" model ensembles, determined using the experimental uncertainties under the assumption, that the best experimentally determined model is the ground truth.

For $M_1$, this procedure yields a rather uniform distribution for the weighted precision of the recovered structural models that fluctuates around unity, the theoretical optimum (**Fig. 4E**, left). The distribution of the weighted precision for $M_2$ looks also fine, except close to the C-terminus (end of helix α12 and α13) where the precision of the ensemble is worse than expected (**Fig. 4E**, right), presumably due to granularity of the model or systematic experimental errors.

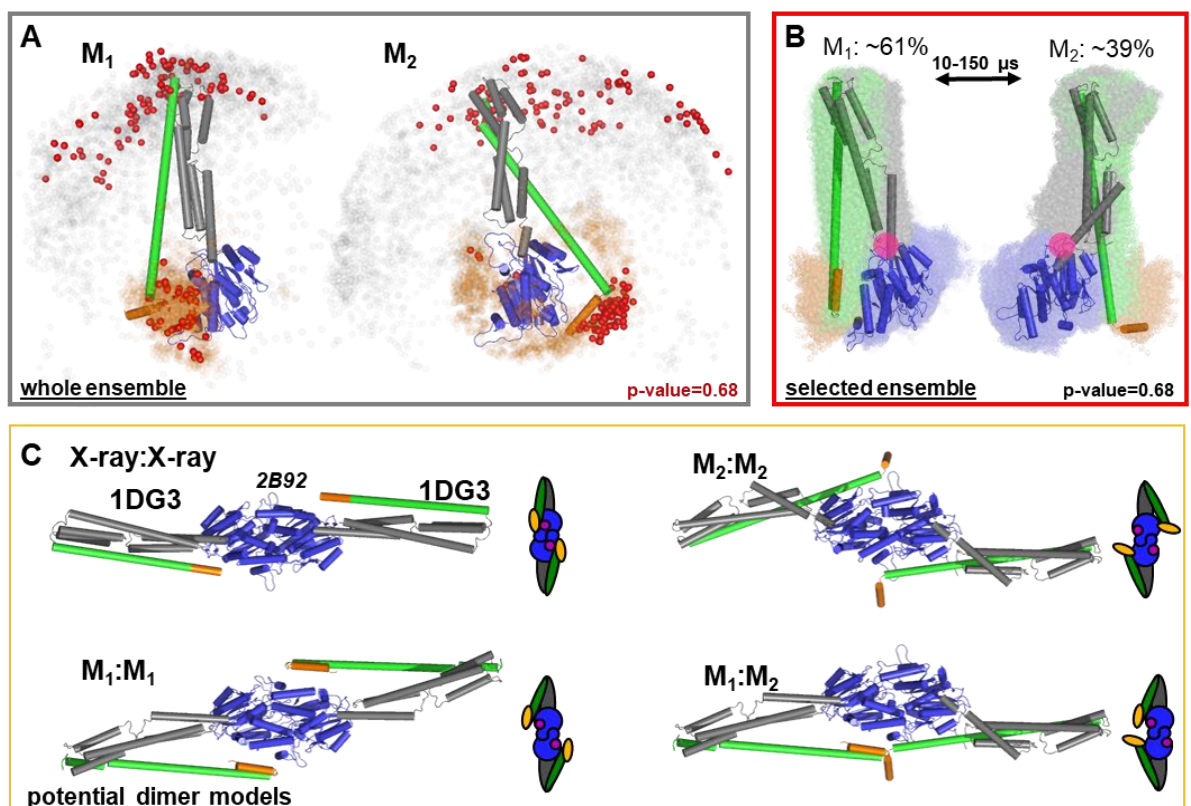

**Figure 5 | Selected conformers and potential dimer models of hGBP1 based on structures generated by integrative modeling of DEER, FRET, and SAXS data.** (**A**) All structural models for $M_1$ and $M_2$ were aligned to the LG domain and are represented by orange and gray dots, indicating the Cα atoms of the amino acids F565 and T481, respectively. The structural model best agreeing with all experiments is shown as cartoon representation. Non-rejected conformations (p-value = 0.68) are shown as red spheres. (**B**) Global alignment of all selected structural models (p-value = 0.68). In the center, the structure best representing the average of the selected ensembles is shown. The conformational transition from $M_1$ to $M_2$ with average correlation times in the range of 10 to 150 μs can be described by a rotation around the connecting region of the LG and the middle domain (pivot point PP, shown as a magenta circle). (**C**) Potential hGBP1:hGBP1 dimer structures constructed by superposing the head-to-head interface of the LG domain (PDB-ID: 2B92) to the full-length crystal structure (1DG3), and both models of the states $M_1$ and $M_2$. The LG and middle domain are colored in blue and gray, respectively. Helices

α12 and α13 are colored in green and orange, respectively. As structural models for $M_1$ and $M_2$, the structures best representing the ensemble of possible conformers are shown.

The heterogeneity of the structural ensembles is judged by their root mean square fluctuations (RMSF) (**Fig. 4F**). The RMSF values of both conformers $M_1$ and $M_2$ depend on the residue number and fluctuate around the expected range of ~ 7 and ~ 9 Å, respectively. To visualize differences among the structural models, we aligned the selected conformers to the LG domain. This demonstrates that in $M_1$ and $M_2$ α12/13 binds at two distinct regions of the LG domain (**Fig. 5A**, red spheres). In $M_1$ α12/13 binds to the same side of the LG domain as in the full-length crystal structure (PDB-ID: 1DG3). In $M_2$ α12/13 is bound to the opposing side of the LG domain. A global alignment of the conformations $M_1$ to $M_2$ and the best representatives of the ensembles visualize the transition between the two states. In our model a rearrangement of residues 306-312 results in a rotation of the middle domain around a pivot point (**Fig. 5B**, magenta circle). Such model describes the experimental data, the relocation of α12/13, and agrees well with global motions identified by PCA of the MD simulations. For the transitions from $M_1$ to $M_2$ α12/13 "rolls" along the LG domain, while the middle domain rotates towards the LG domain. Starting in $M_1$, which is comparable to the crystal structure except for a slight kink of the middle towards the LG domain, the movement of α12/13 stops on the opposite side of the LG domain.

## Discussion

Our experimental findings on the structure of hGBP1 in solution can be approximated by two major conformations $M_1$ and $M_2$, which are in dynamic exchange. We mapped the dynamics of hGBP1 by NSE spectroscopy and fFCS. NSE showed that hGBP1 is a protein without significant detectable shape changes on the ns-timescale up to 200 ns. However, fFCS that analyzed a network of FRET-pairs revealed considerable dynamics on slower time scales (2-300 μs, **Fig. 3**). The distribution of dynamics over such a wide range are indicative for a frustrated potential energy landscape. Structural models for $M_1$ and $M_2$ based on SAXS, DEER and FRET data revealed that the middle domain kinks towards the LG domain and that the helices α12/13 are bound on opposite sides of the LG domain. These findings are self-consistent, as (1) the conformational transition from $M_1$ to $M_2$ and vice versa is complex and may result in a distribution of relaxation times, indicating a rough energy landscape with several intermediates, and (2) the dynamics is mainly associated to α12/13. Analogous to protein folding, where Chung *et al.*(*43*) monitored the transition from the unfolded to the folded state and defined a transition path time, it would be intriguing to define an effective time for the conformational transition from $M_1$ to $M_2$. The conformational transition time would be definitively a convolute between all observed relaxation times (**Fig. 3**, **Tab. S2**) and is expected to be in the sub-millisecond time range. To sum up, the experiments can be described by two conformational states separated by a rugged energy landscape, resulting in slow transition invisible on the NSE timescale. The smFRET measurements demonstrate that this transition is an intrinsic property of hGBP1 that does not depend on the presence of substrate (pathway (i) in **Fig.1B**).

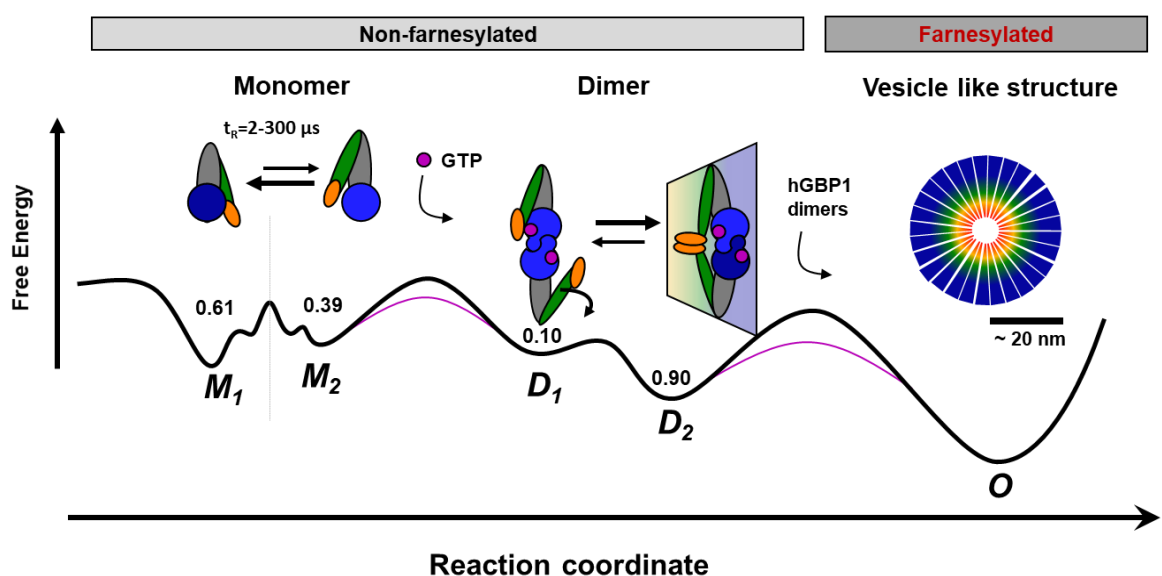

**Figure 6 | Potential oligomerization pathways of the human guanylate binding protein 1 (hGBP1) summarizing current experimental findings (*15,17,20,24*).** In the presence and absence of a nucleotide, hGBP1 is in a conformational exchange between at least two conformational states $M_1$ and $M_2$ with a correlation time of 2-300 µs. Binding of a nucleotide to the LG domain activates hGBP1 for dimerization. After hGBP1 dimerization via the LG domains conformational changes of the middle domains and the helices α12/13 lead to an association of both helices α13. The species fractions for respective populations are given as numbers on top of the wells of a schematic energy landscape (black line). The substrate GTP lowers the activation barrier (red line). Under turn-over of GTP, farnesylated hGBP1 further self-assembles to form highly ordered, micelle-like polymers.

To understand the potential functional relevance of $M_1$ and $M_2$, various observations should be considered. First, the oligomerization of hGBP1 is an important feature that demands flexibility of the structure as deduced from major structural rearrangements described so far.(*20,24,44*) In particular, large movements of the LG, the middle domain and helices α12/13 against each other are required to establish the elongated building blocks of the polymer.(*20*) It is also most conceivable that various dynamically interchanging configurations of the sub-domains need to be sampled to assemble the highly ordered polymer. Dynamins and farnesylated hGBP1 are known to form highly ordered oligomers (*20*) requiring at least two binding sites. We previously showed that non-farnesylated hGBP1 forms dimers via the LG domains (in a head-to-head manner) *and* via helix α13(*24*) in the presence of a GTP analog. This finding is inconsistent with the crystal structures of the full-length protein (PDB-ID: 1DG3 and 1F5N) and for dimers formed by two hGBP1s in the same conformations, as within such dimers the helices α13 are on opposite sides and thus could not associate (**Fig. 5**). However, in a dimer formed of two distinct conformers ($M_1$:$M_2$), the helices α13 are located on the same side of their LG domains. Thus, in line with previous studies, which identified preferred pathways to increase the association yield of protein-protein complexes,(*45*) we suggest that, owing to the conformational flexibility, precursors necessary for oligomerization are already formed spontaneously before binding of the oligomerization-inducing substrate GTP. Remarkably, we detected virtually no substrate induced differences in the amplitude distribution of the correlation times demonstrating that the flexibility is independent of the bound nucleotide. Overall, the findings strongly suggest that the GTP induced dimerization of the GTPase domains and a substrate independent flexibility are needed for a dimerization of the effector domains (pathway *i* in **Fig. 1B**). The substrate solely facilitates hGBP1 association by increasing the affinity of the LG domain as a hub for dimerization.

Structure-wise, we found that the middle domain is kinked with respect to the LG domain as found for other dynamins (*40,41*). Moreover, our data supports two conformations with distinct binding sites of helix a12/13 that can be explained by major rearrangements of the region connecting the middle and the LG domain. Prakash et al. (*21*) described already the interconnecting region of LG and middle domain, which comprise residues 279-310 including a small β-sheet and α-helix 6. The packing of helix α6 (residues 291 to 306) against α1/β1 of the LG domain and against helix α7 of the middle domain was hypothesized to stabilize the relative location of LG and middle domain against each other. Most intriguingly, the Sau group reported on the importance of helix α6 for full catalytic activity of hGBP1 and for oligomer formation. They could also clearly establish the relationship between oligomer formation and defensive activity against hepatitis C virus showing that impairing catalytic activity and oligomer formation by mutations leads also to a decreased antiviral activity.(*46*) These observations support our conclusions as to the importance of the movements around the pivot point located close to α-helix 6. Similar movements have been reported for other dynamin-like proteins, where the GTPase domain rearranges with respect to the middle domain along the catalytic cycle.(*23,47*)

Previous data(*15,17,20,24*) and our new findings in this work lead to a common model which describes the reaction pathway of hGBP1 from a monomer to the formation of mesoscale droplets *in vitro* and living cells (**Fig. 6**). All structural requirements for this multi-step conformational rearrangement for positioning the two interaction sites and defining the molecular polarity are already predefined in the monomeric hGBP1 molecule. In the absence of substrate and other GBP molecules, hGBP1 adopts at least two distinct conformational states.

Upon addition of GTP the LG domain is capable of binding to another protomer, whilst the conformational dynamics appear to remain unchanged. When two GTP bound hGBP1s associate, a head-to-head dimer either in a $M_1:M_1$, $M_2:M_2$ or a $M_1:M_2$ configuration is formed. As the $M_1:M_2$ dimer has a higher stability, in $M_1:M_2$ the α13 helices of the two subunits associate, the equilibrium is shifted towards the $M_1:M_2$ dimers.(*24*)

Notably, *in vivo* helix α13 is farnesylated at the end of a "CaaX" motif(*7*). Thus, hGBP1 provides a membrane anchor and now, hGBP1 dimers supposedly act as amphiphilic particles in the build-up of supramolecular structures. (*20*) This suggests, in line with the common knowledge that amphiphilic Janus particles can form liquid phases(*48*), that hGPB1 forms protein condensates and droplets, also referred to as vesicle like structures (VLS) in living cells. (*15*) *In vitro* studies (*20*) of farnesylated hGBP1 showed that the polymerization is driven by the GTP-dependent high-affinity association of the N-termini and the association of the lipophilic C-terminus. For hGBP1, the amphiphilicity depends on the presence of GTP. Beyond pure affinity driven phase separations in living cells(*49*), the time-dependent hydrolysis of GTP affects the lifetime of the GPB multimers and is responsible for their dynamic formation and dissolution. This allows for a fine controlled and orchestrated attack on a parasite by GPBs in response to the signaling cascade in activated cells (*15*).

It is also interesting to compare these features of hGBP1 with properties of proteins that exhibit phase separation and form condensates. The recent literature for phase separation (*50*) especially discusses proteins that form assemblies composed of modular interaction domains or large intrinsically disordered regions (IDR), which introduce flexibility to the scaffold of the condensate. An additional important factor for phase separation are multivalent interactions. If one compares these findings with the behavior of the well-folded hGBP1, it becomes obvious that hGBP1's intrinsic flexibility, multivalency and amphiphilicity after opening might be analogously essential for the formation of condensates. To conclude: (1) hGBP1 is multivalent with interaction sites of distinct affinities that define the polarity of the formed molecular assembly. The high affinities of LG domains ensure formation of a dimeric encounter complex already at low concentrations in the first step. The conformational flexibility of hGBP1's effector domain promotes the second key step for multimerization - the association of helices α13 that makes the dimer amphiphilic. (2) The affinity of hGBP1 for membranes might be increased by a polybasic sequence directly adjacent to the CaaX box ($^{582}$KMRRRK$^{587}$).(*16,51*) Thus, an association of the α13 helices does not only further stabilize the hGBP1 dimer, but also increase amphiphilicity and promote membrane association. (3) The GTP hydrolysis by the LG domain controls the formation of the reversible formation of multimeric complexes by GBPs.

In a more general view, our results on hGBP1 demonstrate that the exchange between distinct protein conformations is usually encoded in its design (pathway *i*, **Fig. 1A**). Thus, the conformational flexibility of a protein can already be a characteristic of the apo form although this property is only relevant for a later stage of the protein's functional cycle, for example in a complex with its ligand, substrates and other proteins, respectively. Considering, for example, the movement of the substrate-dependent conformational transitions in the finger subdomain of a DNA polymerase,(*52*) it is obvious that these opening and closing movements are essential for catalyzing polymerization under ambient conditions. The rule that functionally relevant conformational equilibria may be predefined by protein design also applies to other steps in protein function. In future, when considering additional quantitative live-cell studies such integrative approaches may provide a molecular picture on complex biological processes like intracellular immune response.

## Materials and Methods

### 1   Protein expression and labeling

*Expression and purification*. All cysteine variants are based on cysteine-free hGBP1 (C12A/ C82A/ C225S/ C235A/ C270A/ C311S/ C396A/ C407S/ C589S) and were constructed in a pQE80L vector (Qiagen, Germany) following the instructions of the QuikChange site-directed mutagenesis kit (Stratagene, USA) according to (*24,53*). Previously, these mutations were shown to only weakly affect hGBP1's function.(*24,53*) New cysteines were introduced at various positions of interest (N18C, Q254C, Q344C, T481C, A496C, Q525C, 540C, Q577C). The mutagenesis was verified by DNA sequencing with a 3130xl sequencer (Applied Biosystems, USA). hGBP1 was expressed in *E. coli* BL21(DE3) and purified following the protocol described previously.(*37*) A Cobalt-NTA-Superflow was used for affinity chromatography. No glycerol was added to any buffer as it did not make any detectable differences. To not interfere with the following labeling reactions, the storage buffer did not contain DTT or DTE. Protein concentrations were determined by absorption at 280 nm according Gill and Hippel using an extinction coefficient of 45,400 $M^{-1}$ $cm^{-1}$. Tests of enzyme activity and function demonstrate that the effect of mutations and labeling on hGBP1's function is small (**Supplementary Note 1**, section 2).

*Protein labeling.* FRET labeling was performed in two steps. First, the protein was incubated with a deficit amount of Alexa647N maleimide C2 (Alexa647) (Invitrogen, Germany). To start the first labeling reaction, a solution with a hGBP1 concentration 100-300 µM in labeling buffer containing 50 mM Tris-HCl (pH 7.4), 5 mM $MgCl_2$, 250 mM NaCl was gently mixed with a 1.5-fold molar excess of Alexa647. After 1 hour incubation on ice, the unbound dye was removed using a HiPrep 26/20 S25 desalting column (GE Healthcare, Germany) with a flow rate of 0.5 ml/min. After this first labeling step, double, single and unlabeled proteins were separated based on the charge difference introduced by the coupled dyes using anion exchange chromatography on a ResourceQ column (GE Healthcare, Germany) and a salt gradient running from 0-500 mM NaCl over 120 ml at a pH of 7.4 and flow rate of 2.0 ml/min. The peaks in the elugram were analyzed for their degree of labeling (dol) by measuring their absorption by UV/Vis spectroscopy at a wavelength of 280 nm and 651 nm. The fraction with the highest, single-acceptor labeled protein amount was labeled with a 4-fold molar excess of Alexa488 C5 maleimide (Alexa488). The unreacted dye was separated as described for the first labeling step. Finally, the dol for both dyes was determined (usually 70-100% for each dye). The dols were determined by absorption using 71,000 $M^{-1}$ $cm^{-1}$ and 265,000 $M^{-1}$ $cm^{-1}$ as extinction coefficients for Alexa488 and Alexa647, respectively. The labeled proteins were aliquoted into buffer containing 50 mM Tris-HCl (pH 7.9), 5 mM $MgCl_2$, 2 mM DTT, shock-frozen in liquid nitrogen and stored at -80 °C.

The spin labeling reactions were conducted at 4 °C for 3 hours using an 8-fold excess of (1-Oxyl-2,2,5,5-tetramethylpyrroline-3-methyl) methanethiosulfonate (MTSSL) as a spin label (Enzo Life Sciences GmbH, Germany). The reaction was performed in 50 mM Tris, 5 mM $MgCl_2$ dissolved in $D_2O$ at pH 7.4. Unbound spin labels were removed with Zeba Spin Desalting Columns (Thermo Fisher Scientific GmbH, Germany) equilibrated with 50 mM Tris, 5 mM $MgCl_2$ dissolved in $D_2O$ at pH 7.4. Concentrations were determined as described before. Labeling efficiencies were determined by double integration of CW room temperature (RT) EPR spectra by comparison of the EPR samples to samples of known concentration. In all cases, the labeling efficiencies were ~90-100%.

## 2  Small angle X-ray scattering

*Experimental methods.* Small-angle X-ray scattering (SAXS) was measured on the beamlines X33 at the Doris III storage ring, DESY and on BM29 at the ESRF(*54*) using X-ray wavelengths of 1.5 Å and 1 Å, respectively. On BM29 a size exclusion column (Superdex 200 10/300 GL, GE Healthcare) coupled to the SAXS beamline was used (SEC-SAXS). The scattering vector $q$ is defined as $q = 4\pi/\lambda \cdot \sin(\theta/2)$ with the incident wavelength λ and the scattering angle θ.

The measurements cover an effective $q$ range from 0.015 to 0.40 Å$^{-1}$ for X33 data and 0.006 to 0.49 Å$^{-1}$ for BM29 data.

SAXS allows determining the shape and low-resolution structure of proteins in solution by the measured scattering intensity $I(q)$, which is proportional to the form factor $F(q)$ multiplied by the structure factor $S(q)$.(*55*) $F(q)$ informs about the electron distribution in the protein, while $S(q)$ contains $q$-dependent modulations due to protein-protein interactions occurring at higher protein concentration. At sufficiently low protein concentrations (in the limit of $c \to 0$) the structure factor converges towards unity. A concentration series (hGBP1 concentrations of 1.1, 2.1, 5.0, 11.5 and 29.9 mg/mL) was recorded on X33, whereas on BM29 two SEC-SAXS runs using protein concentrations of 2 mg/mL and 16 mg/mL that were loaded on the SEC column (used buffer for both SAXS and SEC-SAXS experiments: 50 mM TRIS, 5 mM MgCl$_2$, 150 mM NaCl at pH 7.9) have been performed. The SEC-SAXS data were averaged over the elution peak. The obtained SEC-SAXS data of the used high and low protein solutions were overlapping validating the infinite dilution limit. Therefore, the SEC-SAXS data recorded at the high protein concentration were used for further data analysis. An automated sample changer was used for sample loading and cleaning of the sample cell on X33. The storage temperature of the sample changer and the temperature during X-ray exposure in the sample cell were 10°C. The buffer was measured before and after each protein sample as a check of consistency. For each sample, eight frames with an exposure time of 15 sec each were recorded to avoid radiation damage. The absence of radiation damage was verified by comparing the measured individual frames. The frames without radiation damage were merged. On BM29 X-ray frames with exposure time of 1 sec were continuously recorded. The scattering contribution of the buffer and the sample cell was subtracted from the measured protein solutions. Measured background corrected SAXS intensities $I(q,c)$ of the hGBP1 solutions are shown in **Fig. S1A**. $I(q,c)$ were scaled by the protein concentration $c$ and extrapolated ($c \to 0$) to determine the form factor $I(q,0)$ of the protein at infinite dilution. At larger $q$-values, where the structure factor equals unity, the extrapolated form factor overlapped with the SAXS data of the highest protein concentration within the error bars. Therefore, for better statistics the extrapolated form factor at small $q$-values and the data of the 29.9 mg/mL solution at larger scattering vectors were merged. The structure factor $S(q,c)$ (**Fig. S1B**) was extracted by $S(q,c) = I(q,c) / (c \cdot I(q,0))$ and fitted by a Percus-Yevik structure factor including the correction of Kotlarchyk *et al.* for asymmetric particles resulting in an effective hard sphere radius of 2.2 nm.(*56*)

*Analysis methods.* Data was analyzed using the ATSAS software package.(*57*) Theoretical scattering curves of the crystallographic and the simulated structures of the monomer were calculated and fitted to the experimental SAXS curves using the computer program CRYSOL. The distance distribution function *P*(*r*) was determined using the program DATGNOM. *Ab initio* models were generated using the program DAMMIF. In total 20 *ab initio* models were generated, averaged and the filtered model was used. Normalized spatial discrepancy (NSD) values of the different DAMMIF models were between 0.8 and 0.9 indicative of good agreement between generated *ab initio* models. The resolution of the obtained *ab initio* model is 29±2 Å as evaluated by the resolution assessment algorithm.

## 3 EPR spectroscopy

*Experimental methods.* Pulse EPR (DEER) experiments were performed at X-band frequencies (~9.4 GHz) with a Bruker Elexsys 580 spectrometer equipped with a split-ring resonator (Bruker Flexline ER 4118X-MS3) in a continuous flow helium cryostat (CF935; Oxford Instruments) controlled by an Oxford Intelligent Temperature Controller ITC 503S adjusted to stabilize a sample temperature of 50 K. Sample conditions for the EPR experiments were 100 µM protein in 100 mM NaCl, 50 mM Tris-HCl, 5 mM $MgCl_2$, pH 7.4 dissolved in $D_2O$ with 12.5 % (v/v) glycerol-$d_8$. Further details are described by Vöpel at el. (*24*).

DEER inter spin-distance measurements were performed using the four-pulse DEER sequence (*58,59*):

$$\frac{\pi}{2}(\nu_{obs}) - \tau_1 - \pi(\nu_{obs}) - t' - \pi(\nu_{pump}) - (\tau_1 + \tau_2 - t') - \pi(\nu_{obs}) - echo \quad (1)$$

with observer pulse ($\nu_{obs}$) lengths of 16 ns for $\pi/2$ and 32 ns for $\pi$ pulses and a pump pulse ($\nu_{pump}$) length of 12 ns. A two-step phase cycling (+ ‹x›, - ‹x›) was performed on $\pi/2(\nu_{obs})$. Time *t'* was varied with fixed values for $\tau_1$ and $\tau_2$. The dipolar evolution time is given by $t = t' - \tau_1$. Data were analyzed only for $t > 0$. The resonator was overcoupled to $Q \sim 100$. The pump frequency $\nu_{pump}$ was set to the center of the resonator dip (coinciding with the maximum of the EPR absorption spectrum. The observer frequency $\nu_{obs}$ was set ~65 MHz higher, at the low field local maximum of the EPR spectrum. Deuterium modulation was averaged by adding traces recorded with eight different $\tau_1$ values, starting at $\tau_{1,0} = 400$ ns and incrementing by $\Delta\tau_1 = 56$ ns. Data points were collected in 8 ns time steps or, if the absence of fractions in the distance distribution below an appropriate threshold was checked experimentally, in 16 ns time steps. The total measurement time for each sample was 4 - 24 h.

*Analysis methods.* The DEER data was analyzed using the software DeerAnalysis which implements a Tikhonov regularization.(*60*) Background correction of the DEER signal dipolar evolution function *V*(*t*) (normalized to unity at the time *t* = 0)

$$V(t) = F(t) \cdot V_{background}(t), \tag{2}$$

was performed assuming an isotropic distribution of the spin-labeled hGBP1 molecules in frozen solution that is described by

$$V_{background}(t) = \exp(-k \cdot t). \tag{3}$$

Briefly, the resulting form factor $F(t)$ is modulated with the dipolar frequency

$$\omega_{DD}(R_{SS}, \theta) = \frac{1}{4\pi} \cdot \frac{g^2 \mu_B^2 \mu_0}{\hbar} \cdot \frac{1}{R_{SS}^3} \cdot (3\cos^2\theta - 1), \tag{4}$$

that is proportional to the cube of the inverse of the inter-spin distance $R_{SS}$ ($\mu_B$: Bohr magneton; $\mu_0$: magnetic field constant; $\theta$: angle between the external magnetic field and the vector connecting the two spins, for nitroxide spin labels the $g$ values of both spins can be approximated with the isotropic value $g \approx 2.006$). Analysis of the form factor $F(t)$ in terms of a distance distribution $p(R_{SS})$ was performed by a Tikhonov regularization. A simulated time domain signal

$$S(t) = K(t, R_{SS}) \cdot p(R_{SS}) \tag{5}$$

from a given distance distribution $p(R_{SS})$ was calculated by means of a kernel function

$$K(t, R_{SS}) = \int_0^1 \cos[(3x^2 - 1) \cdot \omega_{DD} \cdot t] \, dx \tag{6}$$

with $\omega_{DD}(R_{SS}) = \frac{2\pi \cdot 52.04 \, MHz \, nm^{-3}}{R_{SS}^3}$ for nitroxide spin labels.

The optimum $p(R_{SS})$ was found by minimizing the objective function

$$G_\alpha(P) = \|S(t) - V_{local}(t)\|^2 + \alpha \cdot \left\|\frac{d^2}{dr^2} p(R_{SS})\right\|^2. \tag{7}$$

The regularization parameter $\alpha$ was varied to find the best compromise between smoothness, i.e., the suppression of artifacts introduced by noise, and resolution of $p(R_{SS})$. The optimum regularization parameter was determined by the L-curve criterion, where the logarithm of the smoothness $\left\|\frac{d^2}{dr^2} p(R_{SS})\right\|^2$ of $p(R_{SS})$ is plotted against the logarithm of the mean square deviation $\|S(t) - V_{local}(t)\|^2$, allowing to choose the distance distribution with maximum smoothness representing a good fit to the experimental data.

Theoretical inter spin label distance distributions for MTS spin labels attached to structural models have been calculated using the rotamer library analysis (RLA) implemented in the freely available software MMM (*30*).

## 4 Fluorescence spectroscopy

*Experimental methods - eTCSPC.* Ensemble time-correlated single-photon-counting (eTCSPC) measurements of the donor fluorescence decay histograms were either performed on an IBH-5000U (HORIBA Jobin Yvon IBH Ltd., UK) equipped with a 470 nm diode laser LDH-P-C 470 (Picoquant GmbH, Germany) operated at 8 MHz or on a EasyTau300 (PicoQuant, Germany) equipped with an R3809U-50 MCP-PMT detector (Hamamatsu) and a BDL-SMN 465 nm diode laser (Becker & Hickl, Germany) operated at 20 MHz. The donor fluorescence was detected at an emission wavelength of 520 nm using a slit-width that resulted in a spectral resolution of 16 nm in the emission path of the machines. A cut-off filter (495 nm) in the detection path additionally reduced the contribution of the scattered light. All measurements were conducted at room temperature under magic-angle conditions. Typically, $14 \cdot 10^6$ to $20 \cdot 10^6$ photons were recorded at TAC channel-width of 14.1 ps (IBH-5000U) or 8 ps (EasyTau300). When needed, the analysis considers differential non-linearities of the instruments by multiplying the model function with a smoothed and normalized instrument response of uncorrelated room light. The fits cover the full instrument response function (IRF) and 99.9% of the total fluorescence. The IRFs had typically FWHM of 254 ps (IBH-5000U) or 85 ps (PicoQuant EasyTau300).

*Experimental methods - Single-molecule (sm) spectroscopy.* A beam of linearly polarized pulsed argon-ion laser (Sabre®, Coherent) was used to excite freely diffusing molecules through a corrected Olympus objective (UPLAPO 60X, 1.2 NA collar (0.17)). The laser was operated at 496 nm and 73.5 MHz. An excitation power of 120 µW at the objective has been used during experiments. The fluorescence light was collected through the same objective and spatially filtered by a 100 µm pinhole which defines an effective confocal detection volume of ~3 fl. A polarizing beam-splitter divided the collected fluorescence light into its parallel and perpendicular components. Next, the fluorescence light passed a dichroic beam splitter that defines a "green" and "red" wavelength range (below and above 595 nm, respectively). After passing through band pass filters (AHF, HQ 520/35 and HQ 720/150) single photons were detected by two "green" (either τ-SPADs, PicoQuant, Germany or MPD-SPADs, Micro Photon Devices, Italy) and two "red" detectors (APD SPCM-AQR-14, Perkin Elmer, Germany). Two SPC 132 single photon counting boards (Becker & Hickel, Berlin) have recorded the detected photons stream. Thus, for each detected photon the arrival time after the laser pulse, the time since the last photon and detection channel number (so, polarization and color) were recorded.

*Conditions.* All ensemble and single-molecule FRET experiments were performed at room temperature in 50 mM Tris-HCl buffer (pH 7.4) containing 5 mM $MgCl_2$ and 150 mM NaCl. All ensemble measurements were performed at concentrations of labeled protein of approximately 200 nM. The single-molecule (sm) measurements were performed at concentrations of labeled protein of approximately 20 pM to assure that only single-molecules were detected. All sm MFD-measurements probing the hGBP1 apo state were performed under two conditions: (*i*) without unlabeled protein, and (*ii*) with 7.5 µM unlabeled protein to minimize the loss of labeled molecules due to adsorption in the measurement chamber. Both conditions gave comparable results. Due to the higher counting statistics, all results of the apo state reported in this work have been obtained for condition *ii*. To study also the ligand bound holo state hGBP1:L (Fig. 1B) by fFCS in Fig. 4D, we used the ligand GDP-$AlF_x$ as a non-hydrolyzable substrate. The ligand GDP-$AlF_x$ is formed *in situ* by diluting a stock solution with 30 mM $AlCl_3$ and 1 M NaF by 1:100 in the standard buffer containing 100 µM GDP and 20 pM labeled protein without unlabeled protein (condition *i*). All fFCS measurements of the hGBP1 apo and holo state, respectively. were performed under condition *i*.

*Fluorescence decay analysis.* Fluorescence intensity decays of the donor in the presence, $f_{D|D}^{(DA)}(t)$, and the absence of FRET, $f_{D|D}^{(D0)}(t)$, inform on DA distance distributions, $p(R_{DA})$. However, the local environment of the dyes may result in complex fluorescence decays of the donor $f_{D|D}^{(D0)}(t)$ and the acceptor $f_{A|A}^{(AD)}(t)$ even in the absence of FRET. Such sample-specific fluorescence properties were accounted for by donor and acceptor reference samples using single cysteine variants. $f_{D|D}^{(D0)}(t)$ and $f_{A|A}^{(A0)}(t)$ were formally described by multi-exponential model functions:

$$f_{D|D}^{(D0)}(t) = \sum_i x_D^{(i)} \exp\left(-\frac{t}{\tau_D^{(i)}}\right) \tag{8}$$

$$f_{A|A}^{(DA)}(t) = \sum_i x_A^{(i)} \exp\left(-\frac{t}{\tau_A^{(i)}}\right)$$

Here *D*/*D* refers to the donor fluorescence under the condition of donor excitation and *A*/*A* refers to the acceptor fluorescence under acceptor excitation. The individual species fractions $x_D^{(i)}$ and $x_A^{(i)}$ and lifetimes of the donor $\tau_D^{(i)}$ and the acceptor $\tau_A^{(i)}$ are summarized in **Tab. S1**.

We assume that the same distribution of FRET-rate constants quenches all fluorescent states of the donor (quasi-static homogeneous model (29)). Thus, $f_{D|D}^{(DA)}(t)$ can be expressed by:

$$f_{D|D}^{(DA)}(t) = f_{D|D}^{(D0)}(t) \cdot \sum_i x_{RET}^{(i)} \exp\left(-t \cdot k_{RET}^{(i)}\right) = f_{D|D}^{(D0)}(t) \cdot \epsilon_D(t). \tag{9}$$

Where $\varepsilon_D(t)$ is the FRET-induced donor decay. The MFD measurements demonstrate that the major fraction of the dyes is mobile (**Supplementary Note 1**). Therefore, we approximate the $\kappa^2$ by 2/3 and relate $\varepsilon_D(t)$ to the $p(R_{DA})$ by:

$$\epsilon_D(t) = \int_{R_{DA}} p(R_{DA}) \cdot \exp\left(-t \cdot k_0 \cdot \left(\frac{R_0}{R_{DA}}\right)^6\right) dR_{DA} + x_{DOnly}. \tag{10}$$

Here, $R_0$ is the Förster-radius ($R_0 = 52$ Å) and $k_0 = 1/\tau_0$ is the radiative rate constant of the unquenched dye ($\tau_0 = 4$ ns). In $\varepsilon_D(t)$ incomplete labeled molecules lacking an acceptor and molecules with bleached acceptors are considered by the fraction of FRET-inactive, $x_{DOnly}$.

For rigorous uncertainty estimates $p(R_{DA})$ was modeled by a linear combination of normal distributions. Overall, a superposition of two normal distributions with a central distance $\bar{R}_{DA}^{(1,2)}$ and a width $w_{DA}$ was sufficient to describe the data:

$$p(R_{DA}) = \frac{1}{\sqrt{\frac{\pi}{2}} \cdot w_{DA}} \left[ x_1 e^{-\left(\frac{2\left(R_{DA} - \bar{R}_{DA}^{(1)}\right)}{w_{DA}}\right)^2} + (1 - x_1) e^{-\left(\frac{2\left(R_{DA} - \bar{R}_{DA}^{(2)}\right)}{w_{DA}}\right)^2} \right] \tag{11}$$

In the analysis of the seTCSPC data, the FRET-sensitized emission of the acceptor, $f_{A|D}^{DA}(t)$, was considered to reduce the overall photon noise and a typical width of 12 Å was consistent with the data. $f_{A|D}^{(DA)}(t)$ was described by the convolution of $f_{A|A}^{(DA)}(t)$, and $f_{D|D}^{(DA)}(t)$:

$$f_{A|D}^{(DA)}(t) = f_{D|D}^{(D0)}(t) \cdot \epsilon_D(t) \otimes f_{A|A}^{(DA)}(t) \tag{12}$$

All $f(t)$s were fitted by model functions using the iterative re-convolution approach.(*61*) Here, the parameters of a model function $g(t)$ were optimized to the data by using the modified Levenberg–Marquardt algorithm. The model function $g(t)$ considers experimental nuisances as scattered light and a constant background:

$$g(t) = N_F \cdot f(t) \otimes IRF(t) + N_{BG} \cdot IRF(t) + bg \tag{13}$$

$N_F$ is the number of fluorescence photons, $N_{BG}$ is the number of background photons due to Rayleigh or Raman scattering and $bg$ is a constant offset attributed to detector dark counts and afterpulsing. In seTCSPC, the fraction of scattered light and the constant background was calculated by the experimental integration time and the buffer reference measurements. In eTCSPC, the fraction of scattered light and the constant offset were free fitting parameters. Finally, $g(t)$ was scaled to the data by the experimental number of photons and fitted to the experimental data.

All fluorescence decays were fitted by ChiSurf, a custom software package tailored for the global analysis of multiple fluorescence experiments (https://github.com/Fluorescence-Tools/ChiSurf).(*29*) Statistical errors were estimated by sampling the parameter space (*62*) and applying an F-test at a confidence level of 95%.

*Burst-wise analysis.* Briefly, as the first step in the burst-wise analysis, fluorescence bursts were discriminated from the background signal of 1–2 kHz of the single-molecule measurements by applying an intensity threshold criterion. Next, the anisotropy and the fluorescence averaged lifetime, $\langle \tau_{D(A)} \rangle_F$, were determined for each burst. Moreover, the background, the detection efficiency-ratio of the "green" and "red" detectors, and the spectral cross-talk were considered to determine the FRET efficiency, $E$, of every burst.(*35*) The species averaged fluorescence lifetime of the donor in the absence of an acceptor $\langle \tau_{D(0)} \rangle_x$, $\langle \tau_{D(A)} \rangle_F$, and the FRET efficiency estimate the mean $\langle \tau_{D(A)} \rangle_x = (1 - E) \cdot \langle \tau_{D(0)} \rangle_x$ and variance $var(\tau_{D(A)}) = \langle \tau_{D(A)} \rangle_F \cdot \langle \tau_{D(A)} \rangle_x - \langle \tau_{D(A)} \rangle_x^2$ of the burst averaged fluorescence lifetimes distribution. This highlights conformational dynamics by a non-zero variance (**Fig. S3A**). For a detailed analysis of the sub-ensemble the fluorescence photons of multiple burst were integrated into joint fluorescence decay histograms (seTCSPC, **Fig. S3B**). The seTCSPC fluorescence decays were analyzed as described above.

*FRET-lines.* By relating fluorescence parameters, FRET lines serve as a visual guide to interpret histograms of MFD parameters determined for individual molecules. The fluorescence weighted lifetime of the donor, $\langle \tau_{D(A)} \rangle_F$, and the FRET efficiency $E$ were related by FRET-lines by a methodology similar as previously described.(*36*) First, FRET-rate constant distributions, $p(k_{RET})$, were calculated for a given set of model parameters. Next, $p(k_{RET})$ was converted to the averages $\langle \tau_{D(A)} \rangle_F$ and $E$. This results in a parametric relation between $\langle \tau_{D(A)} \rangle_F$ and $E$, called a FRET-line. We use two types of FRET-lines: dynamic and static FRET-lines. Dynamic FRET-lines describe the mixing of typically two states. A static FRET-line relates $\langle \tau_{D(A)} \rangle_F$ to $E$ for all molecules that are static within their observation time (the burst duration). Static molecules are identified by populations in a MFD histogram located on the static FRET-line. The FRET-lines were calculated using the scripting capability of ChiSurf assuming states with normal distributed distance and are calibrated for sample-specific fluorescence properties, i.e., donor and acceptor fluorescence quantum yields, the fraction of acceptor in power dependent dark states (cis-state in Alexa647), and complex fluorescence decays of the donor in the absence of FRET.

*Filtered species cross-correlations.* Filtered FCS increases the contrast by a set of state-specific filters applied to the recorded photon stream. For every FRET pair a specific set of filters, $w_j^{(i)}$, was generated using experimental fluorescence bursts for high (H) and low (L) FRET states as previously described and listed in Tab. S2.(*5*) Using these filters species cross correlation functions $G^{(n,m)}(t_c)$ were calculated by weighted signal intensities $S_j(t)$:

$$G^{(n,m)}(t_c) = \frac{\langle F^{(n)}(t) \cdot F^{(m)}(t+t_c) \rangle}{\langle F^{(n)}(t) \rangle \cdot \langle F^{(m)}(t+t_c) \rangle} \text{ with } F^{(n)}(t) = \left( \sum_{j=1}^{d \cdot L} w_j^{(n)} \cdot S_j(t) \right) \tag{14}$$

Herein *n* and *m* are the two species (either H or L), *d* is the number of detectors, *L* is the number of TAC channels, and $S_j(t)$ is the signal recorded in the TAC-channel *j*. The choice of *n* and *m* defines the type of the correlation function. If *n* equals *m*, $G^{(n,n)}(t_c)$ is a species autocorrelation function (*sACF*), otherwise $G^{(n,m)}(t_c)$ is a species cross-correlation function (*sCCF*).(*5*) Overall four correlation curves were generated per sample: two species auto - $sACF^{H,H}(t_c)$, $sACF^{L,L}(t_c)$ and two species cross - $sCCF^{H,L}(t_c)$, $sCCF^{L,H}(t_c)$ correlation curves. All curves were fitted by a model which factorizes $G^{(n,m)}(t_c)$ into a diffusion-, $G_{diff}^{(n,m)}(t_c)$, and a kinetic- term $G_{kin}^{(n,m)}(t_c)$:

$$G^{(n,m)}(t_c) = 1 + \frac{1}{N_{eff}^{(n,m)}} \cdot G_{\text{Diff}}^{(n,m)}(t_c) \cdot G_{\text{kin}}^{(n,m)}(t_c). \tag{15}$$

Here, $N^{(n,m)}$ is the effective number of molecules. The *sACF*s were fitted by individual effective numbers of molecules. The two *sCCF*s shared a single effective number of molecules.

We assume that the same diffusion term can describe all correlation curves of a sample and that the molecules diffuse in a 3D Gaussian illumination/detection profile. Under these assumptions $G_{diff}^{(n,m)}(t_c)$ is

$$G_{\text{Diff}} = \left(1 + \frac{t_c}{t_{\text{Diff}}}\right)^{-1} \left(1 + \left(\frac{\omega_0}{z_0}\right)^2 \left(\frac{t_c}{t_{\text{diff}}}\right)\right)^{-1/2}, \tag{16}$$

where $t_{diff}$ the characteristic diffusion time and $\omega_0$ and $z_0$ are the radii of the focal and the axial plane, respectively, where the intensity decayed to $1/e^2$ of the maximum's intensity.

The kinetic terms of the *sACF* and the *sCCF* were formally described by:

$$G_{kin}^{L,H}(t_c) = \left(1 - A_0^{LH} \cdot \left(A_1 \cdot e^{-t_c/t_{c,1}} + A_2 \cdot e^{-t_c/t_{c,2}} + A_3 \cdot e^{-t_c/t_{c,3}}\right)\right) \tag{17}$$

$$G_{kin}^{H,L}(t_c) = \left(1 - A_0^{HL} \cdot \left(A_1 \cdot e^{-t_c/t_{c,1}} + A_2 \cdot e^{-t_c/t_{c,2}} + A_3 \cdot e^{-t_c/t_{c,3}}\right)\right) \cdot \left(1 - A_b^{HL} \cdot e^{-t_c/t_b}\right)$$

$$G_{kin}^{L,L}(t_c) = \left(1 + A_1^{LL}\left(e^{-t_c/t_{c,1}} - 1\right) + A_2^{LL}\left(e^{-t_c/t_{c,2}} - 1\right) + A_3^{LL}\left(e^{-t_c/t_{c,3}} - 1\right)\right)$$

$$G_{kin}^{H,H}(t_c) = \left(1 + A_1^{HH}\left(e^{-t_c/t_{c,1}} - 1\right) + A_2^{HH}\left(e^{-t_c/t_{c,2}} - 1\right) + A_3^{HH}\left(e^{-t_c/t_{c,3}} - 1\right)\right) \cdot \left(1 + A_b^{HH}\left(e^{-t_c/t_b} - 1\right)\right)$$

.

Here, $A_0$ defines the amplitude of the anti-correlation; $A_b$ accounts for acceptors bleaching in the high-FRET state; $t_b$ is the characteristic bleaching time of the acceptor (under the given conditions typically 5-10 ms); $A_1$, $A_2$ and $A_3$ together with $t_{c,1}$, $t_{c,2}$ and $t_{c,3}$ define the anti-correlation time spectrum of the H to L and L to H transitions. The sum of $A_1$, $A_2$ and $A_3$ was constrained to unity. The correlation times $t_{c,1}$, $t_{c,2}$ and $t_{c,3}$ were global parameters shared among all samples. $A_1$, $A_2$ to $A_3$ were sample specific. The amplitudes $A_1^{HH}, A_2^{HH}, A_3^{HH}$ and $A_1^{LL}, A_2^{LL}, A_3^{LL}$ of the sACFs were non-global parameters optimized for every curve individually. Overall 48 correlation curves of 12 samples were analyzed as a joint dataset. The uncertainties of the amplitudes and correlation times were determined by support plane analysis that considers the mean and the standard deviation of the individual correlation channels. Estimates for the mean and the standard deviation of the correlation channels were determined by splitting individual measurements. The global data analysis of the FCS curves was performed using ChiSurf.

## 5 Neutron spin-echo spectroscopy

NSE was measured on IN15 at the Institut Laue-Langevin, Grenoble, France. Four incident neutron wavelengths with 8, 10, and 12.2, and 17.5 Å were used. The buffer composition for NSE experiments was 50 mM TRIS, 5 mM $MgCl_2$, 150 mM NaCl at pD 7.9 in heavy water (99.9 atom % D). The protein concentration was 30 mg/mL. The measured NSE spectra are shown in **Fig. S5A**. Effective diffusion coefficients $D_{eff}$ were determined from the initial slope of the NSE spectra by using a cumulant analysis $\frac{I(q,t)}{I(q,0)} = \exp\left(K_1 t + \frac{1}{2} K_2 t^2\right)$ with $D_{eff} = -\frac{K_1}{q^2}$.

The rigid body diffusion $D_0(q)$ of a structural model at infinite dilution was calculated according to (*34*):

$$D_0(q) = \frac{1}{q^2 F(q)} \sum_{j,k} \left\langle b_j \exp(-i\vec{q}\vec{r}_j) \begin{pmatrix} \vec{q} \\ \vec{q} \times \vec{r}_j \end{pmatrix} \widehat{D} \begin{pmatrix} \vec{q} \\ \vec{q} \times \vec{r}_k \end{pmatrix} b_k \exp(i\vec{q}\vec{r}_k) \right\rangle \tag{18}$$

where $\widehat{D}$ is the 6x6 diffusion tensor, which was calculated using the HYDROPRO program.(*63*) $D_0(q)$ was calculated for the hGBP1 crystal structure (PDB-ID: 1DG3) and the best representing $M_2$ structure. The population values have been determined from fits to the SAXS data with 69% best representing $M_2$ structure and 31% crystal structure at the temperature of 10°C.

The full NSE spectra were described by rigid body diffusion and internal protein dynamics according to (*64*):

$$I(q,t)/I(q,0) = [(1 - A(q)) + A(q)\,exp(-\Gamma t)] \cdot \qquad (19)$$

$$exp\left(-q^2 D_t \frac{H_t}{S(q)} t\right) \left(\sum_{l=0}^{15} S_l(q)\, exp(-l(l+1)D_r H_r t)\right) / \sum_{l=0}^{15} S_l(q) \text{ with } S_l(q) =$$

$$\sum_m \left|\sum_i b_i j_l(q r_i) Y_{l,m}(\Omega_i)\right|^2$$

where $D_t$ and $D_r$ are the calculated scalar translational and rotational diffusion coefficients found in the trace of $\widehat{D}$ of the rigid protein at infinite dilution from the structural models. Rotational diffusion of the rigid protein were expressed in spherical harmonics with spherical Bessel functions $j_l(qr_i)$, spherical harmonics $Y_{l,m}$ and scattering length densities $b_i$ of atoms at positions $r_i$. Here the crystal structure was used as a base. $D_t$ and $D_r$ were chosen according to the mixture of crystal structure and best representing $M_2$ structure. Direct interaction and hydrodynamic interactions were accounted for by the corrections $D_{t,eff}(q) = D_t H_t/S(q)$ and $D_{r,eff} = H_r D_r$. Interparticle interactions were considered by the structure factor $S(q)$ as measured by SANS. $H_t$ and $H_r$, reduce the effective translational and rotational diffusion coefficients. $H_t$ is related to the intrinsic viscosity $[\eta]$ by $H_t=1-c[\eta]$ and $H_r$ can be approximated by $1-H_r=(1-H_t)/3$ for spherical particles,(*65*) which might underestimate $H_r$ for large asymmetric particles. Internal protein dynamics was described by an exponential decay with a *q*-independent rate $\Gamma$, and a *q*-dependent contribution $A(q)$ of internal dynamics to the NSE spectra.

The parameters $H_t$, $H_r$ the relaxation time $\lambda$ and the amplitudes $A(q)$ (**eq. 19**) were simultaneously optimized to all NSE spectra (**Fig. S5**). The fits show a small contribution of internal dynamics with amplitudes close to the error bars and seemingly long relaxation times, but not strong enough to be determined unambiguously. Fitting the spectra without additional internal dynamics shows an excellent description of the data (**Fig. S5**) with $H_t = 0.61 \pm 0.01$ and $H_r = 0.72 \pm 0.03$ as the only fitting parameters.

Dynamic light scattering was measured on a Zetasizer Nano ZS instrument (Malvern Instruments, Malvern, United Kingdom) in $D_2O$ buffer identical to that used in the NSE experiment. Autocorrelation functions were analyzed by the CONTIN like algorithm (*66*).

## 6 Simulations

*MD simulations & principal component analysis.* We performed molecular dynamics (MD) and accelerated MD (aMD) simulations to identify collective degrees of freedom, essential movements, and correlated domain motions of hGBP1 by Principal Component Analysis (PCA) (all references for the methodology are given in **Supplementary Note 2**). The simulations were started from a known crystal structure of the full-length protein (PDB code: 1DG3) protonated with the program PROPKA at a pH of 7.4, neutralized by adding counter ions and solvated in an octahedral box of TIP3P water with a water shell of 12 Å around the solute. The obtained system was used to perform unbiased MD simulations and aMD simulations. The Amber14 package of molecular simulation software(*67*) and the ff14SB force field were used to perform five unrestrained all-atom MD simulations. Three of the five simulations were conventional MD (2 µs each) and two aMD simulations (200 ns each). The "Particle Mesh Ewald" method was utilized to treat long-range electrostatic interactions; the SHAKE algorithm was applied to bonds involving hydrogen atoms. For all MD simulations, the mass of solute hydrogen atoms was increased to 3.024 Da and the mass of heavy atoms was decreased respectively according to the hydrogen mass repartitioning method.(*68*) The time step in all MD simulations was 4 fs with a direct-space, non-bonded cutoff of 8 Å. For initial minimization, 17500 steps of steepest descent and conjugate gradient minimization were performed; harmonic restraints with force constants of 25 kcal·mol$^{-1}$ Å$^{-2}$, 5 kcal·mol$^{-1}$·Å$^{-2}$, and zero during 2500, 10000, and 5000 steps, respectively, were applied to the solute atoms. Afterwards, 50 ps of NVT simulations (MD simulations with a constant number of particles, volume, and temperature) were conducted to heat up the system to 100 K, followed by 300 ps of NPT simulations (MD simulations with a constant number of particles, barostat and temperature) to adjust the density of the simulation box to a pressure of 1 atm and to heat the system to 300 K. A harmonic potential with a force constant of 10 kcal·mol$^{-1}$ Å$^{-2}$ was applied to the solute atoms at this initial stage. In the following 100 ps NVT simulations the restraints on the solute atoms were gradually reduced from 10 kcal·mol$^{-1}$ Å$^{-2}$ to zero. As final equilibration step 200 ps of unrestrained NVT simulations were performed. Boost parameters for aMD were chosen by the method as previously suggested.(*38*)

## 7 Integrative modeling

The generation of structural models follows the workflow (**Fig. 4A**) presented in the main text. A key prerequisite for integrative modeling is the simulation of experimental observables for a given set of structural models. To generate an integrative structural model, the degrees of freedom, i.e., the model needs to be defined, structural models need to be generated, the structural models need to be ranked, i.e., evaluated against the experimental data, and experiments need to be combined in a meta-analysis.

*Simulation of experimental parameters*

Theoretical SAXS scattering curves for the structural models were calculated using the established software CRYSOL.(*55*) DEER and the FRET inter-label distance distributions $p_{sim}(R_{LL}, M)$ were simulated by accessible volume (AV) simulations. The experimental inter-label distances were compared to the simulated average distances (**Fig. 4D**). For a given protein conformation M the average simulated distance for all label linker conformations $\langle R_{LL,\text{sim}} \rangle$ is

$$\langle R_{LL,\text{sim}} \rangle (M) = \int R_{LL} \cdot p_{sim}(R_{LL}, M) dR_{LL}. \tag{20}$$

Because of the different meaning of the experimental DEER and FRET inter-label distances, the modeled average inter-spin distances $\langle R_{LL,sim} \rangle$ and the center to center inter-dye distances $\bar{R}_{DA,sim}$ are denoted in Fig. 4D with the general symbol $R_{LL,sim}$.

The DEER AV simulations were calibrated against established rotamer library approaches (**Fig. S2B**). The AV simulations of the fluorophores were refined using experimental anisotropies to account for dyes bound to the molecular surface in accessible contact volume (ACV) simulations. The model for the fluorescent dyes and the transferability of the results were validated by reference measurements and protein activity measurements (**Supplementary Note 1**).

In detail, the spatial distribution of the labels was modeled by accessible volume (AV) simulations weighted by the fraction of dyes in contact with the protein – accessible contact volume (ACV).(*31,69,70*) The used ACV simulations determine all sterically allowed positions of a label, which is approximated by ellipsoids, using a geometric search algorithm, and weight the fraction of dyes in contact with the protein by experimental anisotropies (**Supplementary Note 1**).(*33*) The center of the ellipsoid was connected by a linker to the $C_\beta$-atoms of the reactive amino-acid. The linker extends from the reactive group to the center of the dipole of the labels. The fluorophores were simulated with previously published parameters determined by the spatial dimensions of the dyes.(*31,32*) The donor (Alexa Fluor 488 C5 maleimide, Alexa488) and the acceptor fluorophore (Alexa Fluor 647 C2 maleimide, Alexa647) were modeled using a linker width $L_{width}$ of 4.5 Å and linker-length $L_{link}$ of 20.5 Å and 22 Å for Alexa488 and Alexa647, respectively. The radii of the ellipsoids ($R_{dye1}$, $R_{dye2}$ and $R_{dye3}$) for Alexa488 were 5.0 Å, 4.5 Å and 1.5 Å and for Alexa647 11.0 Å, 4.7 Å and 1.5 Å, respectively. The residual anisotropy was used as an estimate for the fraction of dyes bound to the surface of the protein for screening by accessible surface volume simulations.(*33*) The parameters for the methanethiosulfonate (MTSSL) spin labels were determined by comparing ACV simulated $p(R_{SS})$ to established simulation approaches(*30,42*) resulting in linker-length of 8.5 Å, a linker-width of 4.5 Å, and an ellipsoid radius of 4.0 Å (**Fig. S2B**).

The analysis of the fluorescence data provided per variant two central distances $\bar{R}$ that were assigned based on their relative population to the identified conformations $M_1$ and $M_2$ (**eq. 11**) while the model free DEER analysis yields distance distributions (**eqs. 5, 6**) that were considered by their average distance $\langle R_{LL,exp} \rangle$. Note, contrary to the simulated average distance, the experimental average is a linear combination of the distances of the two co-existing conformations.

*Definition of an integrative structural model*

In the model definition, the experimental constraints and the constraints imposed by the model need to be defined. To describe our experimental observables, we use in the first step a decomposition of the protein into rigid bodies (RB). The used RBD-framework represents proteins as an assembly of flexible linked rigid bodies interacting via a very soft repulsion (clash) potential which tolerates atomic overlaps to a certain degree.(*32*) Essential steps for the generation of structural models by rigid body docking (RBD) is the segmentation of the protein into rigid bodies (RB). Consistent with MD simulations and the biochemical pre-knowledge on the existence of different domain, hGBP1 was decomposed into its individual domains: the LG domain (aa 1-309), the middle domain (aa 310-481) and the helices α12 (aa 482-563) and α13 (aa 564-583) for RBD (**Supplementary Note 3, Fig. S6C, Fig. 4B**). To allow for internal reorganization the middle domain is represented by two rigid bodies (aa 310-373, aa 374-481). The N- to the C-terminal parts of the rigid bodies were connected via bonds with a weak quadratic potential. Such reduced model does not allow for bending of the individual domains. Therefore, we used a very soft clash-potential (**Supplementary Note 4**).

*Generation of structural models*

We use in a first step coarse-grained rigid-body (RB) models and experimental constraints from DEER and FRET, to sample the experimentally allowed conformational space as vast as possible. As first step to generate structural models we use RB docking (RBD) with DEER and FRET restrains. Here, average distances between the labels were determined by modeling their spatial distribution of the labels around their attachment point by accessible volume (AV) simulations.(*32*) Deviations between the modeled and the experimental FRET and DEER distances were minimized by driving initial random configurations the rigid-body assembly towards an optimal conformation (**Supplementary Note 4**). The restraints are compiled in the supplement (**Tab. S3**). This docking procedure was repeated 20,000 times for $M_1$ and $M_2$ to generate structural models refined by subsequent NMSim and MD simulations (**Fig. 4A**).

Next, the structural models generated by RBD were refined by the computationally more demanding normal mode based all-atom multiscale NMSim. NMSim generates representations with stereochemical accurate conformations by a three-step protocol and incorporates information about preferred directions of protein motions into a geometric simulation algorithm.(*39*) We used the RBD structures as a target for NMSim to optimize the stereochemistry. In targeted NMSim the conformational change vector is formulated as a linear combination of the modes calculated for the starting structure (the crystal structure) weighted by the proximity to the target structure (the RBD structure). This way, the normal modes that overlap best with the direction of conformational change contribute more to the direction of motion in NMSim.

Next, the structural models refined by NMSim were clustered into 343 and 414 groups by their $C_\alpha$ RMSD for the states $M_1$ and $M_2$, respectively, using hierarchical agglomerative clustering with complete linkage and distance threshold of 5 Å. As final step, conventional MD simulations on the group representatives were performed for 2 ns (**Methods 6**). The MD trajectories were clustered using hierarchical agglomerative clustering with complete linkage and distance threshold of 2 Å into 3395 and 3357 groups for $M_1$ and $M_2$, respectively.

*Individual ranking of structural models*

To filter (screen) structural models, the calculation of probabilities, the (dis)agreement of the model with the data needs to be measured. Here, the disagreement of the simulated and experimental data was measured by weighted sums of squared deviations, $\chi^2$. The structural models were compared to the SAXS and to the combined DEER and FRET dataset by $\chi^2_{SAXS}$ and $\chi^2_{DEER,FRET}$, respectively.

For a consistent description of the FRET, DEER, and the SAXS experiments, the experimental scattering curve was described by a mixture of the conformations. Here, as a first step, theoretical scattering curves for all proposed conformations were calculated using the program CRYSOL. Next, model functions $I_{model}(q, M_1, M_2)$ for all possible combinations of structural models for the states $M_1$ and $M_2$ were calculated. The model functions were linear combinations of $F_{M1}(q)$ and $F_{M2}(q)$, the theoretical scattering curves for $M_1$ and $M_2$, respectively.

$$I_{model}(q, M_1, M_2) = x_{M1} \cdot F_{M1}(q) + (1 - x_{M1}) \cdot F_{M2}(q) \qquad (21)$$

To determine the initially unknown fraction of molecules in the $M_1$ state, $x_{M1}$, the sum of weighted squared deviations between the experiment and the data $\chi^2_{SAXS}$ to the measured data, $I_{\exp(q)}$ was minimized.

$$\chi^2_{SAXS}(M_1, M_2) = \frac{1}{N} \sum_{i=1}^{N} \left( \frac{I_{\exp(q_i)} - I_{model}(q_i, M_1, M_2)}{\sigma(q_i)} \right)^2 \tag{22}$$

Above, $\sigma(q_i)$ is the noise of the experimental scattering curve and $N$ is the number of detection channels.

For the combined DEER and FRET dataset, $\chi^2_{DEER,FRET}$ measures the disagreement between simulated distances and experimental distances considering the asymmetric (deviation dependent) uncertainty of the distances. For a pair of structural models ($M_1$, $M_2$) we approximate $\chi^2_{DEER,FRET}$ by:

$$\chi^2_{DEER}(M_1, M_2)$$
$$\approx \sum_i \left( \frac{\langle R^{(i)}_{LL,exp} \rangle (M_1) - \langle R^{(i)}_{LL,sim} \rangle (M_1)}{w^{(i)}(M_1)/2} \right)^2$$
$$+ \sum_i \left( \frac{\langle R^{(i)}_{LL,exp} \rangle (M_2) - \langle R^{(i)}_{LL,sim} \rangle (M_2)}{w^{(i)}(M_2)/2} \right)^2$$

$$\chi^2_{FRET}(M_1, M_2) \approx \sum_i \left( \frac{\bar{R}^{(i)}_{DA,exp}(M_1) - \langle R^{(i)}_{DA,sim} \rangle (M_1)}{\Delta^{(i)}(M_1)} \right)^2 + \sum_i \left( \frac{\bar{R}^{(i)}_{DA,exp}(M_2) - \langle R^{(i)}_{DA,sim} \rangle (M_2)}{\Delta^{(i)}(M_2)} \right)^2$$

$$\chi^2_{DEER,FRET}(M_1, M_2) = \chi^2_{DEER}(M_1, M_2) + \chi^2_{FRET}(M_1, M_2). \tag{23}$$

Here, $\bar{R}^{(i)}_{DA,exp}(M_1)$ and $\bar{R}^{(i)}_{DA,exp}(M_2)$ are the central experimental donor-acceptor FRET distances assigned to $M_1$ and $M_2$. $\langle R^{(i)}_{LL,exp} \rangle (M_{\{1,2\}})$ is the average label-label distance in DEER experiments. Modeled average inter-label distances $\langle R^{(i)}_{LL,sim} \rangle (M_{\{1,2\}})$ correspond to the average simulated label-label distance $\langle R_{LL,sim} \rangle$ for DEER and average simulated donor-acceptor distance and $\langle R^{(i)}_{DA,sim} \rangle (M_{\{1,2\}})$ for FRET, which is a good approximation for the central donor-acceptor distance $\bar{R}_{DA,exp}$ of a symmetric distance distribution being used for analysis (**eq. 11**). Uncertainties that depend on the sign of the deviation between the model and the data were considered by the half width of the distance distribution $w^{(i)}(M_{\{1,2\}})/2$ for DEER and estimate for the uncertainty of the central distance $\Delta^{(i)}(M_{\{1,2\}})$ for FRET.

*Model discrimination & quality assessment*

The experimental technique assesses different structural aspects with uncertainties thereof, e.g. inter-label distance distributions in DEER, FRET *vs.* average shapes in SAXS. Thus, the balance of the techniques for modeling and screening, captured by relative weights, will affect the final model. A well-balanced model weights the different experiments by estimates of their relative information content. This way balanced absolute probabilities, which depend on accurate estimates of the degrees of freedom for the model and the data, for a structural model can be calculated. By combining these probabilities in a meta-analysis, a well-balanced structural model combining diverse techniques can be recovered. Here, we combined the label-based FRET, DEER measurements with SAXS measurements to a well-balanced integrative structural model.

The values for $\chi^2_{DEER,FRET}$ and $\chi^2_{SAXS}$ assess the quality in a pair of structural models with respect to the experiment. We use these $\chi^2$ values to identify/filter models that are significantly worse than the best possible pair of structures for the respective methods. For that, we compare pairs of $\chi^2$ values for structural models by an F-test (The ratio $x \coloneqq \chi^2_1/\chi^2_2$ is F-distributed). For two $\chi^2$-values with corresponding degrees of freedom $d_1$ and $d_2$ the cumulative F distribution is:

$$F(x, d_1, d_2) = I_{\frac{d_1 x}{d_1 x + d_2}}\left(\frac{d_1}{2}, \frac{d_2}{2}\right). \tag{24}$$

Here, $I$ is the regularized incomplete beta function. To relate the F-value $x$ to a probability $\alpha$, for given $\chi^2_1$ and $\chi^2_2$ and significantly different $d_1$, and $d_2$, we must compute the inverse of the cumulative F distribution.

Here, we compare the $\chi^2$ value of all possible combinations of structural models ($M_1$, $M_2$) and experimental techniques DEER/FRET and SAXS F-values to the $\chi^2$ value of best pair of structures ($x = \chi^2/\min(\chi^2)$). These models have the same dofs ($d_1 = d_2 = \text{dof}$). Hence, we first identify the best model and compute $x$ for all pairs of models. Next, we determine the degrees of freedom, dof, that are calculated by the degrees of freedom of the data, dof$_d$, and the degrees of freedom of the model, dof$_m$, i.e. dof=dof$_m$-dof$_d$. With $x$ and dof we compute the probability $\alpha$ that a model is significantly worse than the best model.

The dof$_m$ was estimated by a PCA applied to all structural models. PCA revealed that 10 principal components explain more than 90% of the total variance. Hence, we conclude that dof$_m$ ~ 10. For DEER/FRET, dof$_{d,DEER/FRET}$ was estimated by correcting the total number of inter-label distances (see **Tab. S3A**: 22 FRET, 8 DEER) for duplicates and for redundant mutual information content. This was accomplished by determining the number of informative distances via a greedy backward elimination feature selection algorithm for our total ensemble ((*33*), see Fig. 5) so that the precision of the obtained corresponded to our experimental one. In this way, we obtained a dof$_{d,DEER/FRET}$ = 22 (**Fig. S6D**) - a value that is close to the number of independent label-pair positions of 23. For SAXS measurements the number of Shannon channels is typically in the range of 10 to 23. For our measurements, the number of Shannon channels approximately 18-22 (*71,72*). We used the number of Shannon channels as an initial estimate for the dof of the SAXS measurements, dof$_{d,SAXS}$, and we varied dof$_{d,SAXS}$ in the range of 10 to 24. We found only minor effects of dof$_{d,SAXS}$ on $\alpha_{SAXS}$, the SAXS discimination power of the models, and used dof$_{d,SXAS}$ =17 to discriminate structural models (for details see **Fig. S6D**). Using these estimates of dof$_m$ and dof$_d$, $\alpha$ for DEER/FRET,$\alpha_{DEER,FRET}$, and SAXS, $\alpha_{SAXS}$, were calculated for all pairs (M$_1$, M$_2$). Next, $\alpha_{DEER,FRET}$ and $\alpha_{SAXS}$ were combined in a meta-analysis to a joint probability of discriminating a pair.

$\alpha_{DEER,FRET}$ and $\alpha_{SAXS}$ measure how likely a pair (M$_1$, M$_2$) is dissimilar from the best pair for DEER/FRET and SAXS, respectively. To combine DEER/FRET and SAXS we used the probability *p* that a pair is similar, *p = 1-α*. Note, *p* for DEER/FRET, *p$_{DEER,FRET}$*, and for SAXS, *p$_{SAXS}$*, considers the degrees of freedom for the system and data. Moreover, *p$_{DEER,FRET}$* and *p$_{SAXS}$* are independent. Thus, Fisher's method was applied to fuse datasets in a meta-analysis. Fisher's method combines probabilities of *k* independent tests (here *k* = 2) into a combined $\chi^2_{2k}$ with 2*k* degrees of freedom. For *p$_{DEER,FRET}$* and *p$_{SAXS}$*, the combined probability is

$$\chi^2_{2k} \sim -2 \cdot \sum_{i=1}^{k} \ln(p_i) = -2 \cdot \ln\left(\prod_{i=1}^{k} p_i\right) \quad (25)$$

$$= -2 \cdot \ln(p_{DEER,FRET} \cdot p_{SAXS}).$$

Thus, $\chi^2_{2k}$ is chi-squared distributed with 4 combined degrees of freedom. In this way a $\chi^2_{2k}$ value was determined for every (M$_1$, M$_2$), and pairs (M$_1$, M$_2$) were discriminated by a chi-squared test with 4 degrees of freedom.

*Assessment of model precision & quality in Fig 4E*

To assess the local quality of the models, the inter-residue distances between all $C_\alpha$ atoms, $R_{C_\alpha}$, and the standard deviation, $SD(R_{C_\alpha})$, of the distribution of $R_{C_\alpha}$ were calculated for all models as a measure for the experimental model precision (**Fig. 4E**, lower triangles). Next, we checked if these variabilities are larger than statistically expected. For this, we compared the experimental precision $SD(R_{C_\alpha})$ to the precision $SD(R_{C_\alpha})_{ref}$ of a ground truth model ensemble as an "ideal and perfect" reference by computing the weighted (normalized) precision, $SD(R_{C_\alpha})/SD(R_{C_\alpha})_{ref}$. Due to (*i*) the incomplete experimental information on the model, (*ii*) the uncertainties of the experiments, and (*iii*) imprecisions of the model, we anticipate a limited resolution of the model even for ideal experiments. We calculated the reference precision of the ground truth ensembles in two steps. At first, we use the models for $M_1$ and $M_2$ of the experimental ensemble that describe our FRET and EPR data best. Next, we use the distances corresponding to the best models and our experimental errors in **Tab. S3A** to generate the ideal reference ensemble by our structural modeling pipeline (**Fig. 4A**) so that we could compute the theoretical inter-residue distance distributions and precisions. The finally computed distributions of the weighted precisions, $SD(R_{C_\alpha})/SD(R_{C_\alpha})_{ref}$ allow us to test whether the modeled conformational ensemble approaches the theoretical optimum ratios around unity or whether systematic deviations indicate problems in the modelling.

Please note that the above procedure provides only an estimate for the reference model precision and corresponding variability of $R_{C_\alpha}$. For a correctly estimated model precision with the corresponding $SD(R_{C_\alpha})_{ref}$, the weighted precision $SD(R_{C_\alpha})/SD(R_{C_\alpha})_{ref}$ theoretically has the meaning of an F-value. Such an F-value for pair-wise estimates of the model precision as $SD(R_{C_\alpha})$ could be used for estimating the probability that the model insufficiently describes the data within their experimental noise. This procedure could yield such estimates for residue pairs that facilitate the detection of the model defects and limitations.

**Data availability**

The following material is available at Zenodo (doi 10.5281/zenodo.1490101): (i) fluorescence decays recorded by eTCSPC used to compute the distance restraints in Tab. S3A, (ii) single-molecule multiparameter fluorescence data: raw data, fluorescence decays of FRET sub-ensembles (seTCSPC), fFCS curves, (iii) scripts for structural modeling of conformational ensembles through integrative/hybrid (I/H) methods using FRET, DEER and SAXS. The experimental SAXS data and the *ab initio* analysis thereof are available in the SASBDB at https://www.sasbdb.org/data/SASDDD6/e1d68arhp4/. The generated conformational ensembles will be uploaded to Zenodo and later deposited at the PDB-Dev. Further datasets generated during and/or analyzed during the current study are available from the corresponding author on reasonable request.

**Code availability**

Most general custom-made software is directly available from http://www.mpc.hhu.de/en/software. General algorithms and source code is published under

https://github.com/Fluorescence-Tools. Additional computer code custom-made for this publication is available upon request from the corresponding authors.

**Competing Interests**

All authors declare that they have no competing interests.

## Acknowledgments

This work was supported by DFG grants RESOLV (EXC 1069) and HE 2679/6-1 to CH, SE 1195/17-1 to CAMS, KL2077/1-2 to JPK and STA 1325/2-1 to AS. TOP and CL wish to acknowledge the support of the International Helmholtz Research School of Biophysics and Soft Matter (BioSoft). A part of this research was supported by the European Research Council through the Advanced Grant 2014 hybridFRET (671208) to CAMS. We are grateful for computational support and infrastructure provided by the "Zentrum für Informations- und Medientechnologie" (ZIM) at the Heinrich Heine University Düsseldorf and the computing time provided by the John von Neumann Institute for Computing (NIC) to HG on the supercomputer JURECA at Jülich Supercomputing Centre (JSC) (user ID: HKF7). This work is based upon experiments performed on the instruments BM29 at the European Synchrotron Radiation Facility (ESRF), X33 at the Doris III storage ring, DESY, and IN15 at the Institut Laue-Langevin (ILL). We acknowledge the ESRF, the EMBL and the ILL for provision of synchrotron and neutron radiation facilities and we would like to thank Drs. Martha Brennich and Clement Blanchet for assistance in using BM29 and X33.

## Author contributions

TOP, CSH, RB, MD, CL, HG, JPK, AS, CAMS, and CH wrote the manuscript. MD performed the molecular simulations under the supervision of HG. CSH prepared samples for smFRET and performed protein activity assays. SI and CL prepared samples for SAXS measurements. TV prepared sampled for EPR measurements. CL and AS performed and analyzed SAXS measurements. RB, AS and BF performed NSE measurements and analysis. TOP, CSH, and AV performed the smFRET measurements under the supervision of CAMS. TOP analyzed the smFRET measurements. JPK performed and analyzed the EPR measurements. TOP combined the FRET, EPR, and SAXS measurements in a meta-analysis for integrative modeling. CH, CAMS, JPK and AS planned and supervised the research project.

## Additional information

Supplementary Information accompanies this paper.

# Supplementary Information

## Integrative dynamic structural biology unveils conformers essential for the oligomerization of a large GTPase

**Thomas-Otavio Peulen, Carola S. Hengstenberg et al.**

| | |
|---|---|
| Supplementary Figure 1 | Small-angle X-ray scattering measurements on the nucleotide free hGBP1. (**A**) Measured SAXS data of hGBP1 at different protein concentrations. (**B**) Structure factor extracted from the SAXS data. (**C**) Pair of structural models selected corresponding best to the SAXS scattering data. (**D**) Fitted species fractions. |
| Supplementary Figure 2 | DEER-spectroscopy on a network of MTSSL spin-labeled pairs resolves pairwise inter-label distance distributions. |
| Supplementary Figure 3 | Single-molecule fluorescence measurements and analysis (**A**) Multi-parameter fluorescence detection histograms. (**B**) Sub-ensemble fluorescence decays. (**C**) Filtered fluorescence correlation. spectroscopy. |
| Supplementary Figure 4 | Quality controls for labeling based methods. (**A**) Protein activity measurements (**B**) dye model consistency (**C**) temperature dependent state population (**D**) dimerization activity of labeled species (**E**) state assignment consistency. |
| Supplementary Figure 5 | Internal dynamics on the nanosecond time-scale by Neutron spin echo spectroscopy (NSE). |
| Supplementary Figure 6 | (**A**, **B**) Analysis of molecular dynamics simulations. (**C**) Identification of flexible regions. (**D**, **E**) Stability and significance analysis of integrative models. |
| Supplementary Figure 7 | Assessment of the conformers within the conformational space covered by MD simulations using FRET- and EPR-data. |
| Supplementary Table 1 | Inter-label distance analysis of DEER and ensemble fluorescence decay measurements (eTCSPC). (**A**) Analysis results of the DEER, FRET ensemble fluorescence decays measurements, and residual anisotropies. (**B**) Reference fluorescence lifetimes of Alexa647 and Alexa488 maleimide coupled to different single cysteine hGBP1 variants. (**C**) Complementary inter-dye distance analysis of donor and sensitized acceptor fluorescence decays of sub-ensemble (seTCSPC) obtained from of single-molecule FRET experiments. (**D**) Uncertainties of the average inter-dye distances determined by eTCSPC measurements. |
| Supplementary Table 2 | Filtered fluorescence correlation spectroscopy analysis results. |
| Supplementary Table 3 | (**A**) Experimental restraints for rigid body docking. (**B**) Additional restraints used for rigid body docking. |
| Supplementary Note 1 | Quality assessment of labeled samples for fluorescence spectroscopy, uncertainty estimation, and consistency analysis. |
| Supplementary Note 2 | MD simulations and PC Analysis |
| Supplementary Note 3 | Identification of rigid domain and rigid body decomposition |
| Supplementary Note 4 | Rigid body docking |

# Supplementary Figures

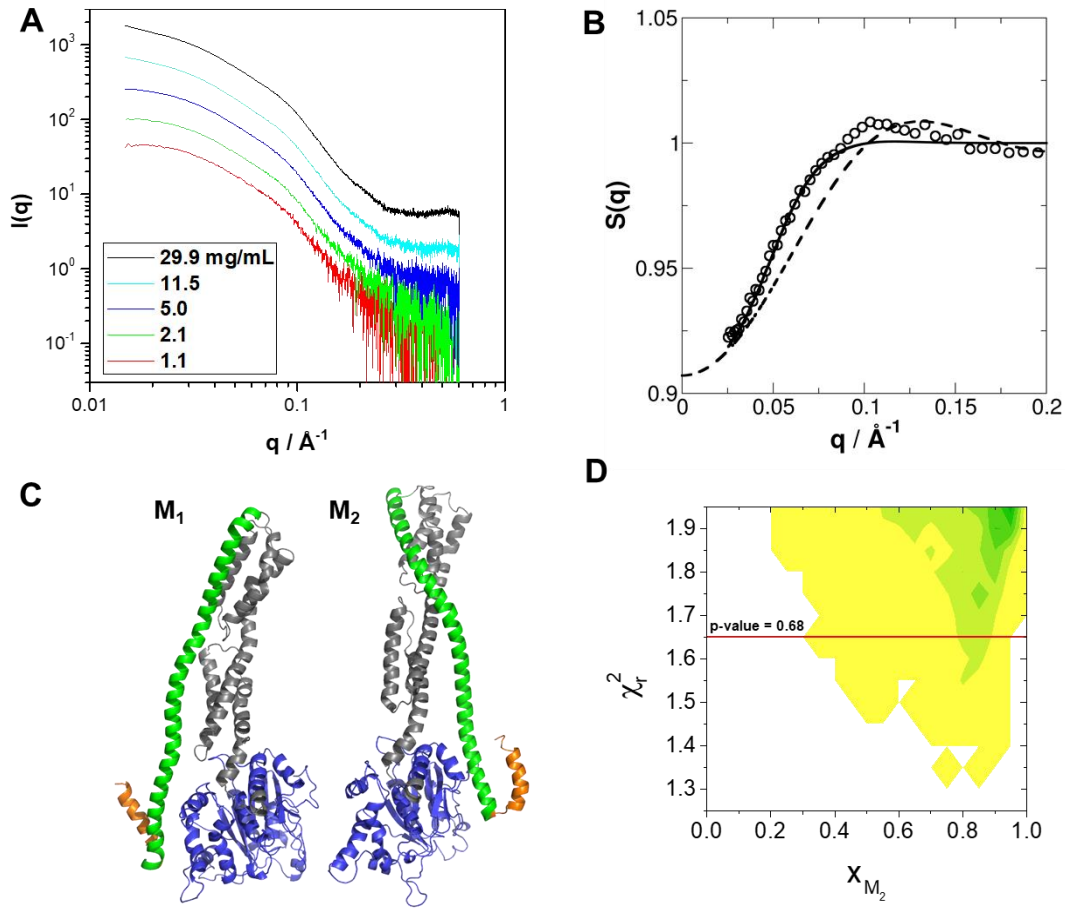

**Supplementary Figure 1 |** Small-angle X-ray scattering measurements on the nucleotide-free hGBP1. (**A**) Measured SAXS data of hGBP1 at different protein concentrations. The scattering curves are not normalized by the protein concentration. (**B**) Structure factor of the 29.9 mg/mL solution extracted from the SAXS data. The structure factor is obtained by the background corrected SAXS curves at highest concentration scaled through division by the form factor (empty circles). The fitted structure factors according to the Percus-Yevik structure factor include the correction for the protein asymmetry factor *beta* (full line) (*58,75*). For comparison the uncorrected structure factor without asymmetry factor is given (stitched line). Data are averaged at larger wave vectors to reduce noise. (**C**) The pair of structural models ($M_1$, $M_2$) selected from the structural ensemble generated by DEER- and FRET-measurements best corresponding to the SAXS scattering data is shown in a cartoon representation. Both structural models are aligned to the LG domain. (**D**) The fitted fractions of the conformer $M_2$, $x_{M_2}$, for all combinations ($M_1$, $M_2$) generated by rigid body docking (RBD) using the DEER and FRET constraints are shown in dependence of the reduced sum of the squared deviation, $\chi_r^2$, in a 2D histogram. The red line corresponds to a p-value of 0.68. Pairs of structural models above the red line are discriminated.

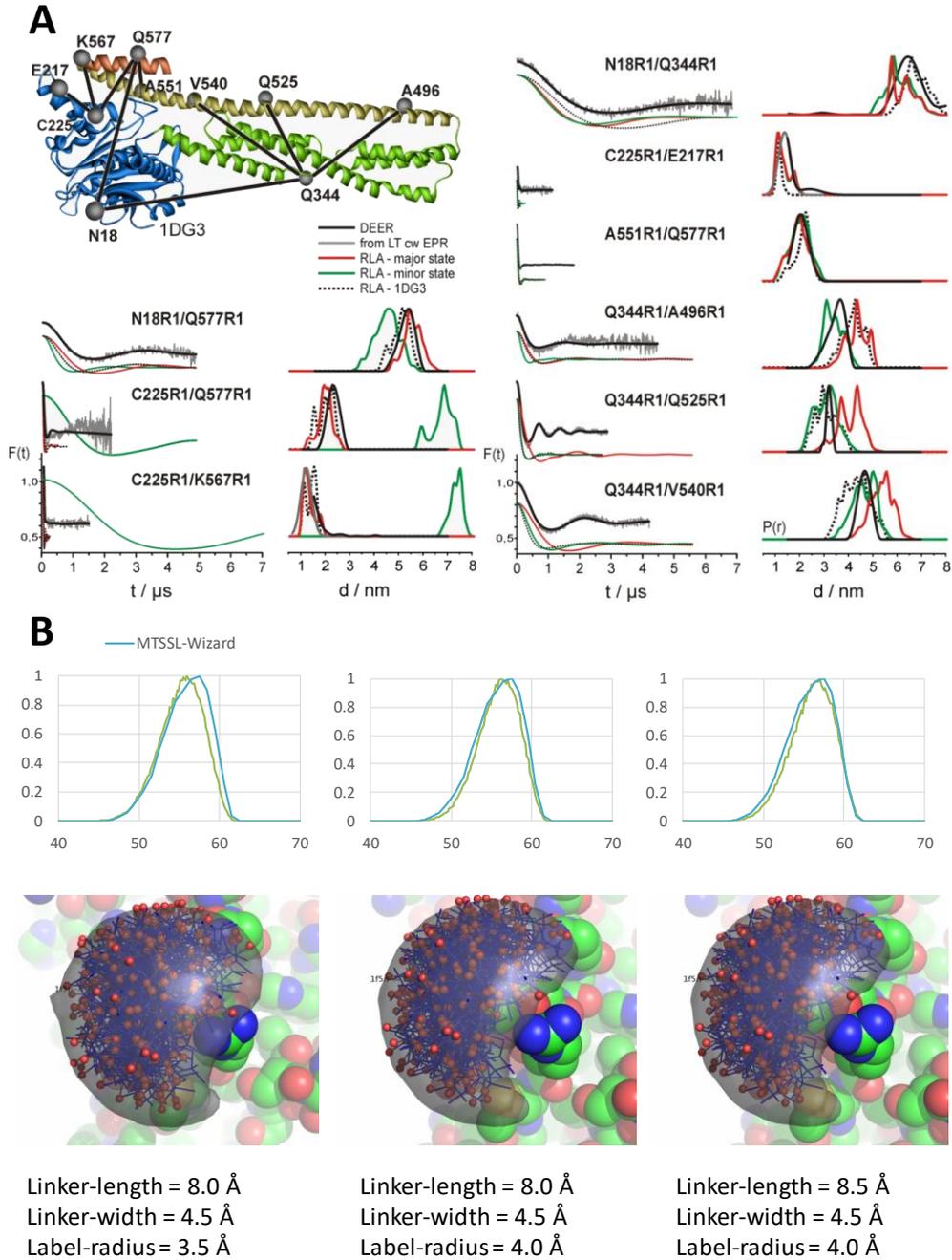

**Supplementary Figure 2 |** DEER-spectroscopy on a network of MTSSL spin-labeled pairs of the hGBP1 resolves pairwise inter-label distance distributions. (**A**) At the top, the network of spin-labeled hGBP1 is shown superposed to a crystal structure of hGBP1. A rotamer library analysis (RLA) simulates for the crystal structure (PDB-ID: 1DG3), the FRET major state ($M_1$), the minor state ($M_2$) inter-spin distance distributions. To the left, experimental background corrected DEER-traces and simulated DEER-traces based on a RLA of different structural models; to the right, inter-spin distance distribution as determined by Tikhonov regularization of the experimental DEER-trace. (**B**) Parametrization of the EPR-MTSSL label for accessible volume calculations. Top the distance distributions for the spin-pair N18C/Q577C of the hGBP1 crystal structure (PDB-ID: 1DG3) as calculated by the MTSSL-Wizard (*43*) is overlaid by the distance distribution as calculated by accessible volume calculations with the parameter set as provided below. For visual comparison, the rotamers are overlaid with the accessible volume calculated for the labeling position N18C. To parameterize the MTSSL-label we used the variant N18C/Q577C as reference and optimized the simulated linker-length, the label-radius and the linker-width until the distance distribution as determined by the AV-calculations agrees best with the distance distributions as determined by the MTSSL-Wizard(*43*) and MMM(*30*). The best agreement was found using a linker-length of 8.5 Å, a linker-width of 4.5 Å and a label-radius of 4.0 Å. All rigid body dockings were performed using this parameter set.

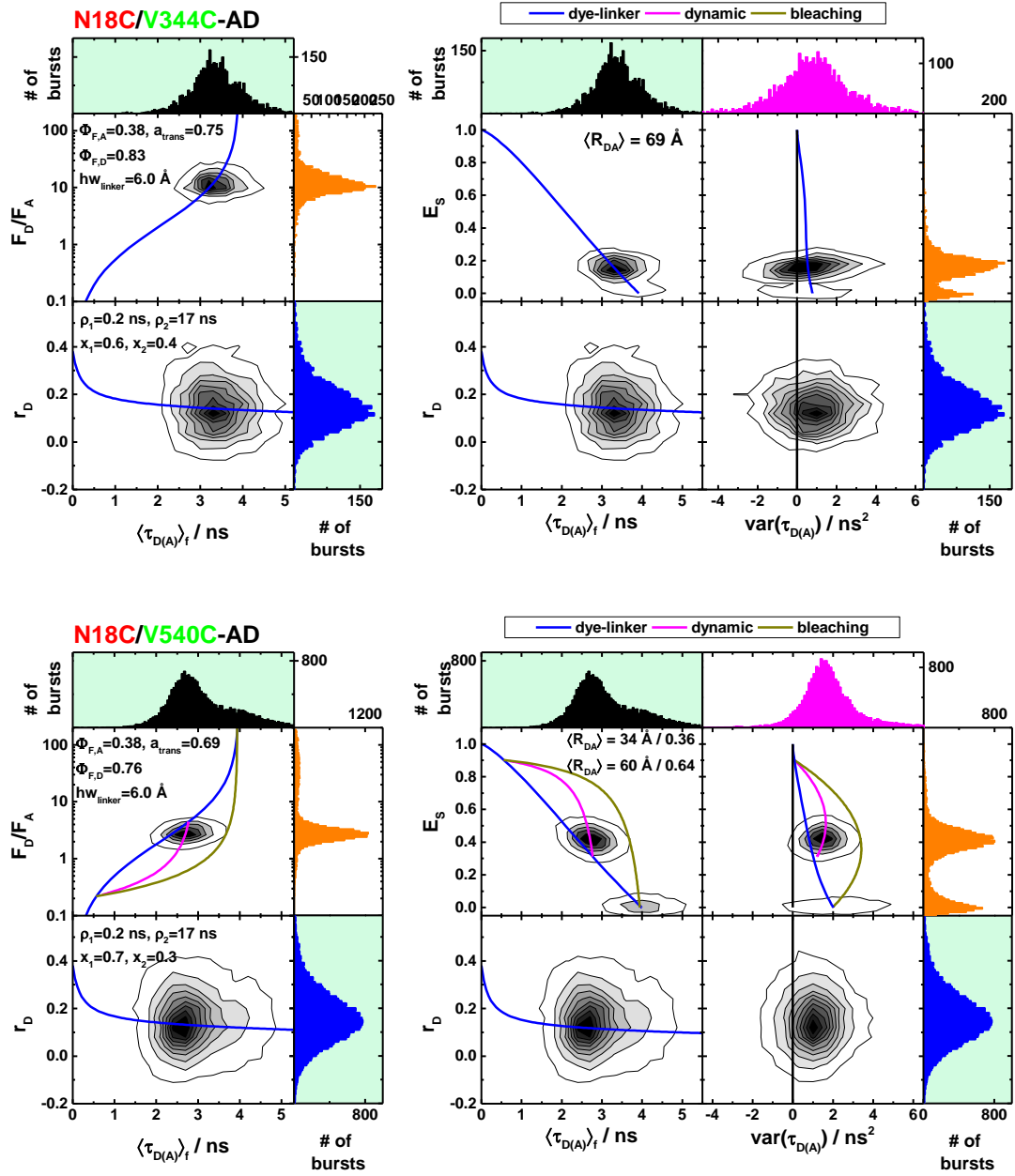

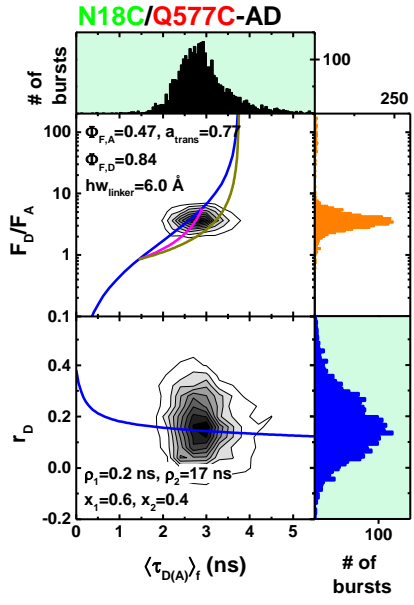
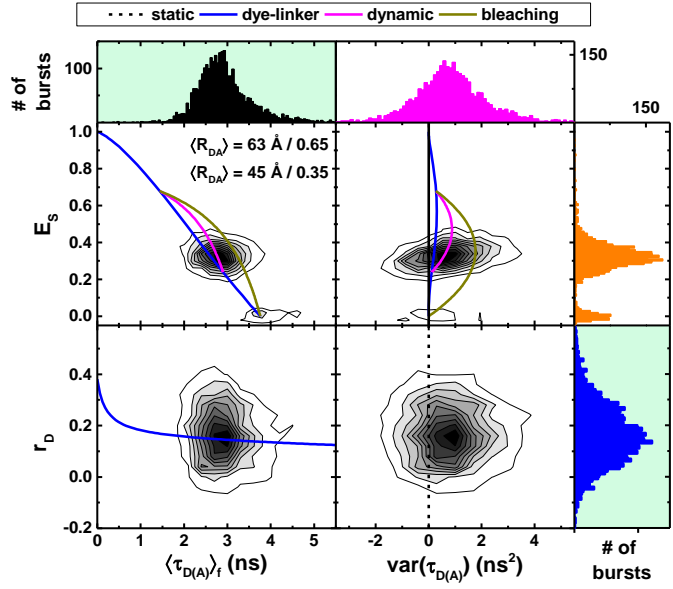
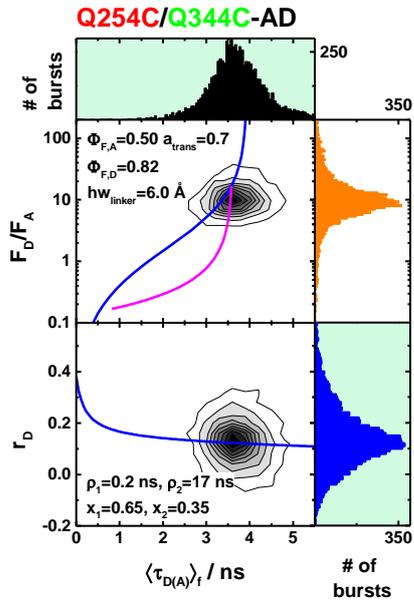
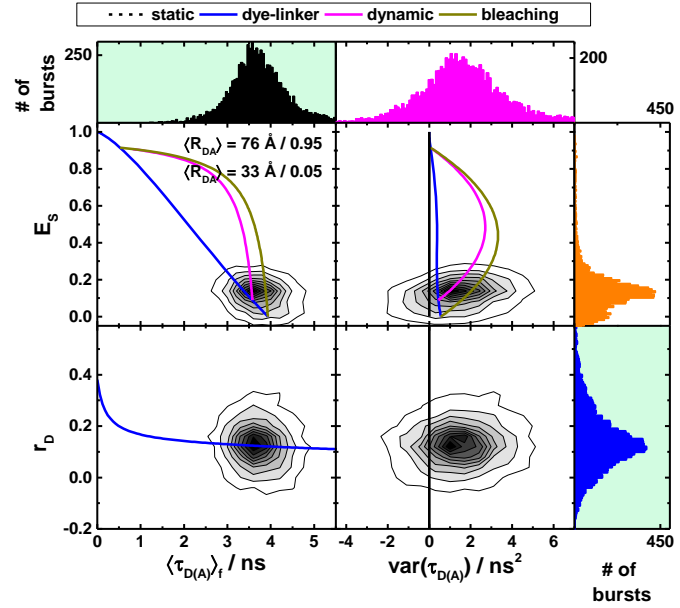

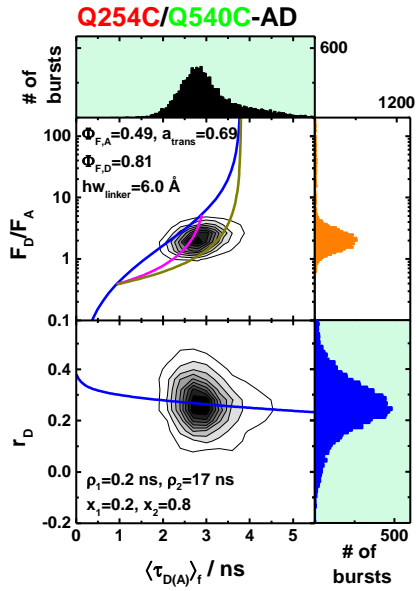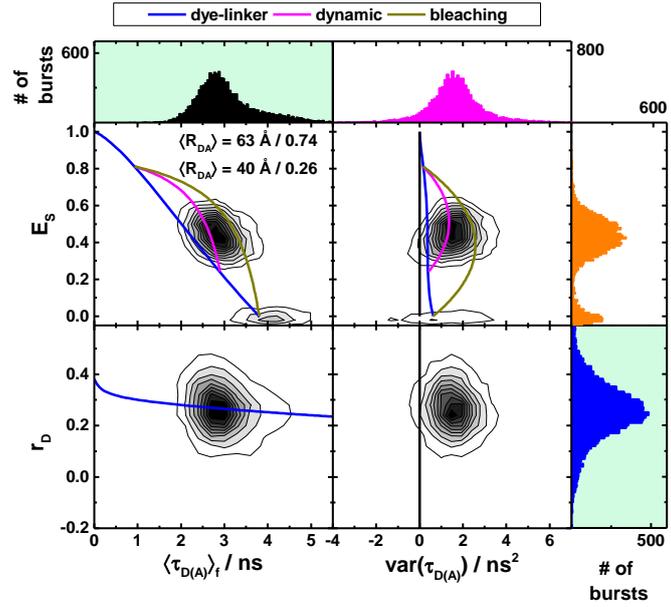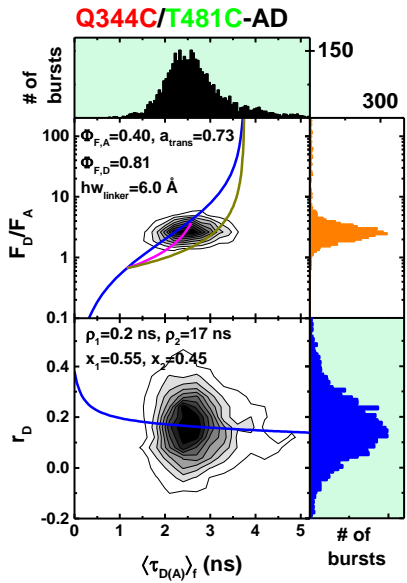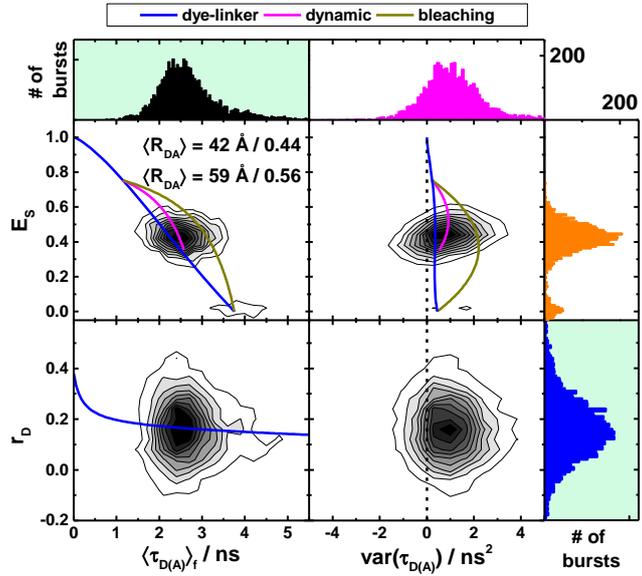

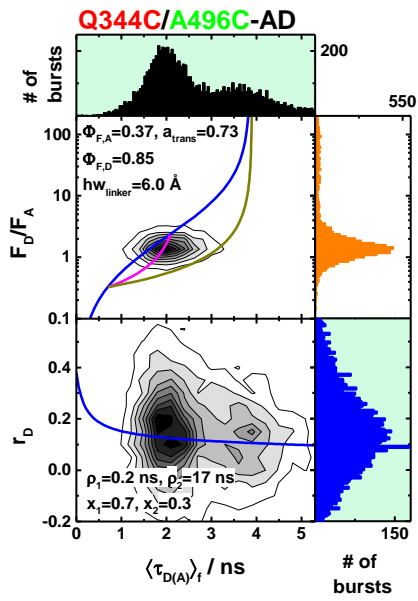
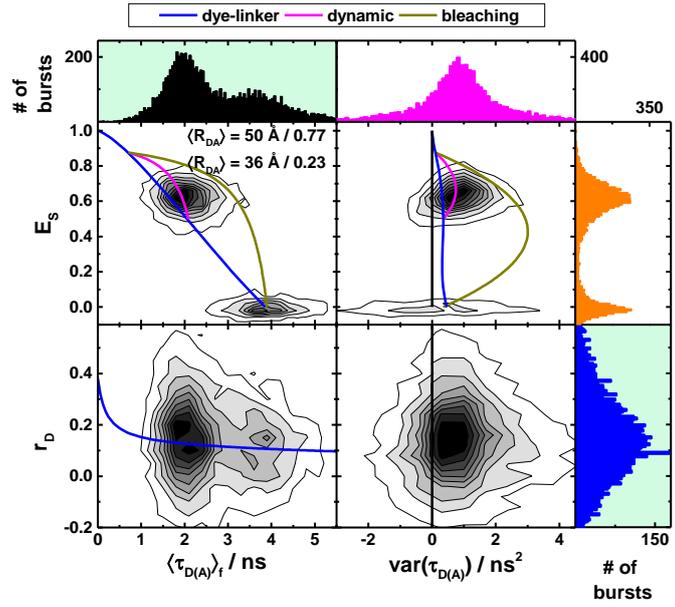
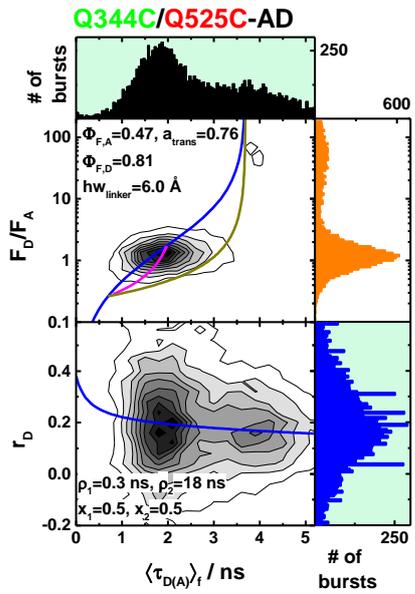
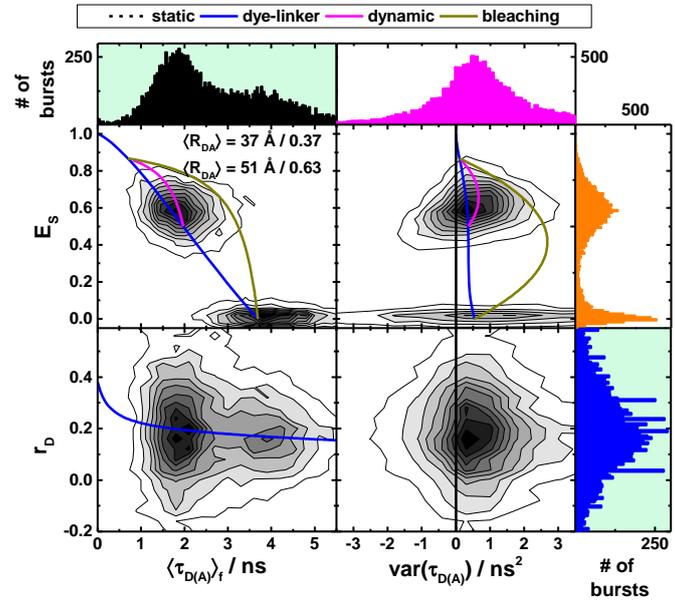

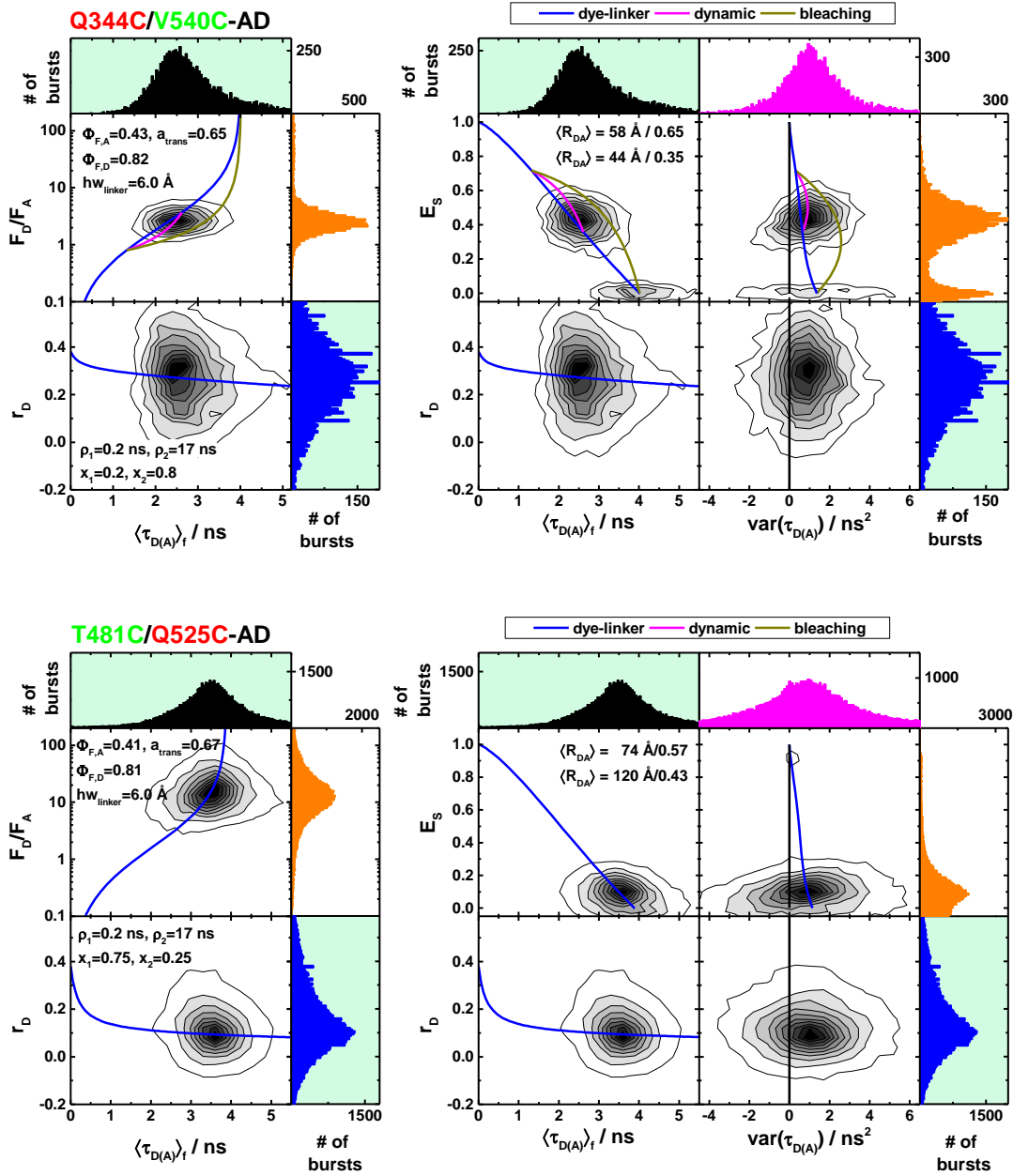

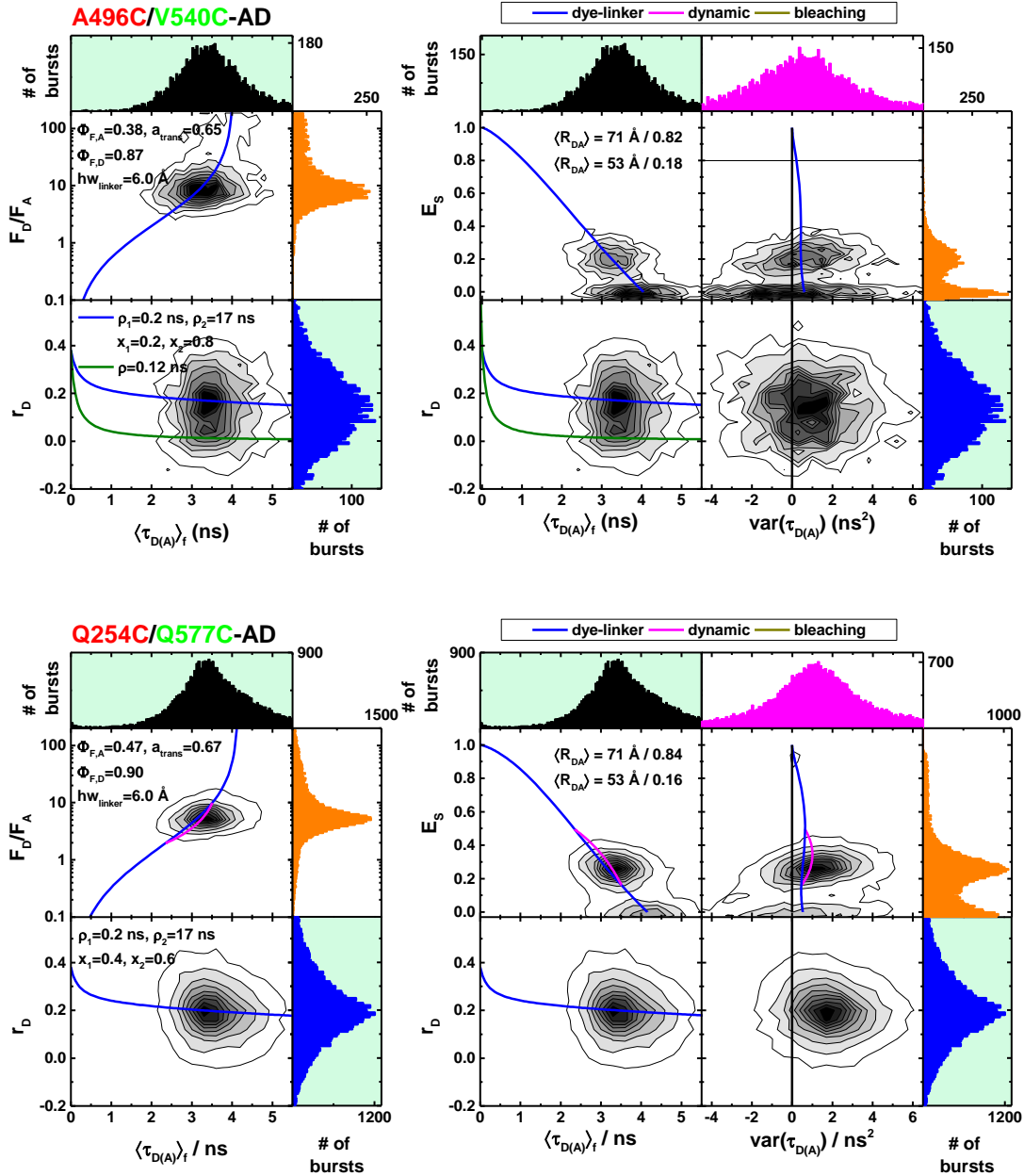

**Supplementary Figure 3A | Single-molecule fluorescence measurements.** Multi-parameter fluorescence detection histograms of different variants of Alexa488 and Alexa647 labeled human guanylate binding protein 1. The dashed blue lines are either static FRET-lines considering linker broadening (top panels) or Perrin-equations for a dye with two rotational correlation times (bottom panels, **Supplementary Note 1, eq. 1**). $x_1$ and $x_2$ refer to the fraction of fast and slow rotating dyes, respectively. For variants with states of different FRET efficiencies, dynamic FRET-lines connecting these states are shown as magenta solid line. The dark-yellow lines describe the acceptor bleaching from high-FRET states. The data are displayed in histograms of the donor-acceptor fluorescence intensity-ratio, $F_D/F_A$, the steady-state transfer-efficiency, $E_S$, and the mean fluorescence averaged lifetime of the donor in the presence of the acceptor $\langle \tau_{D(A)} \rangle_F$. The variance of the donor-acceptor lifetime $var(\tau_{DA})$ was calculated for every detected fluorescence burst. The color of the FRET-pair name indicated the most probable position of the donor (green) and acceptor dye (red); as inset the fluorescence quantum yield of the acceptor $\Phi_{F,A}$ and the donor $\Phi_{F,D}$ are shown. The fraction of the acceptor Alexa647 in trans-conformation, $a_{trans}$, was determined by FCS and is shown as inset.

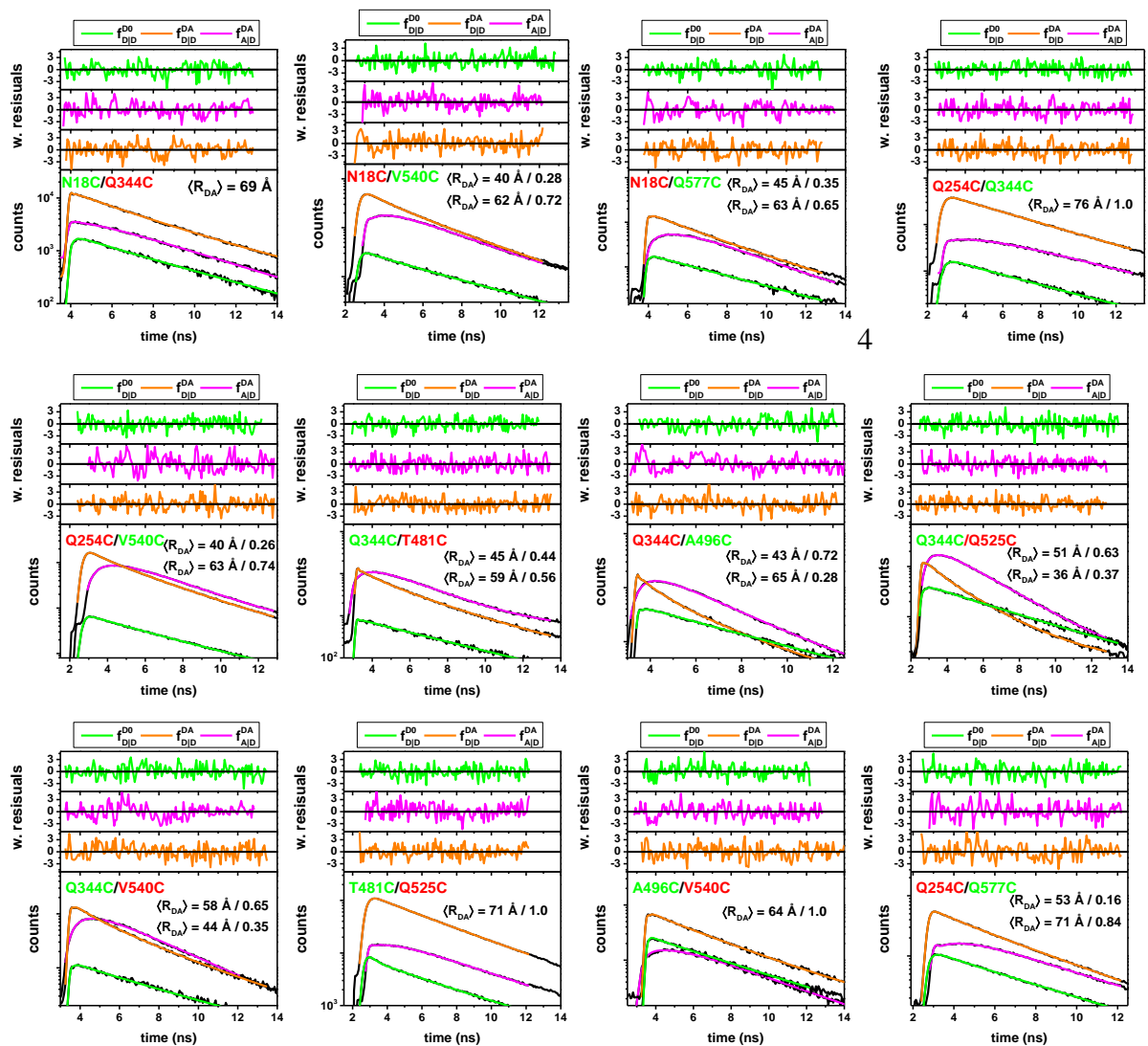

**Supplementary Figure 3B |** Sub-ensemble fluorescence intensity decays of single-molecule FRET measurements on different FRET-labeled (Alexa488, Alexa647) variant of the human guanylate binding protein 1. In green the fluorescence intensity decay of the donor-only fraction of the respective sample are shown. The donor-only fraction was selected by the acceptor intensity. In orange the time-resolved fluorescence intensity of the donor in the presence of the acceptor is shown. In magenta the FRET-sensitized acceptor emission is displayed. The fluorescence decays were jointly analyzed by a weighted combination of normal distributed donor-acceptor distances. The mean and the fractions of the normal distributions are reported by the insets, e.g., the variant Q254C/Q577C was described by two normal distributions, with average distances of 53 Å and 71 Å with fractions of 0.16 and 0.84, respectively.

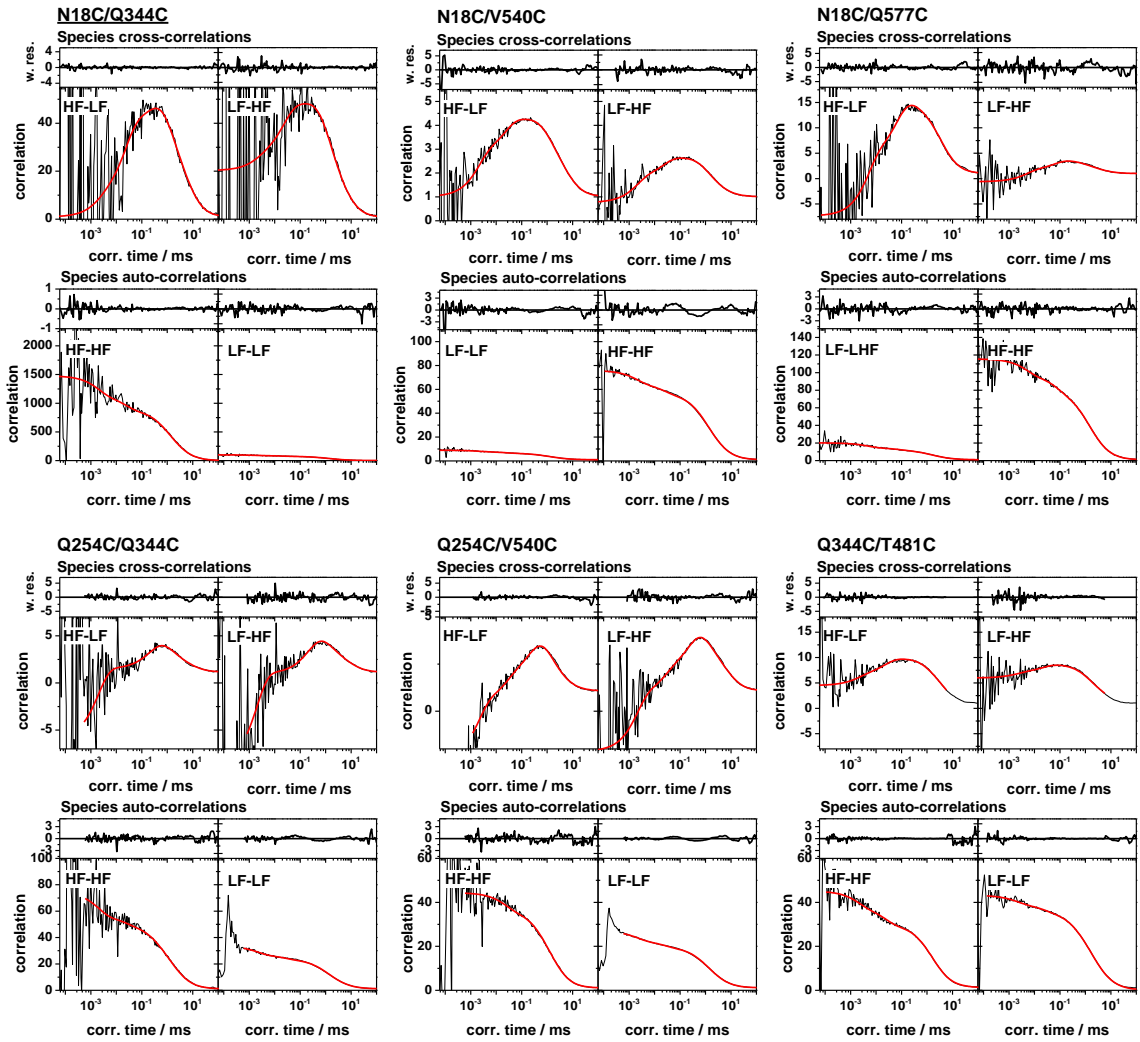

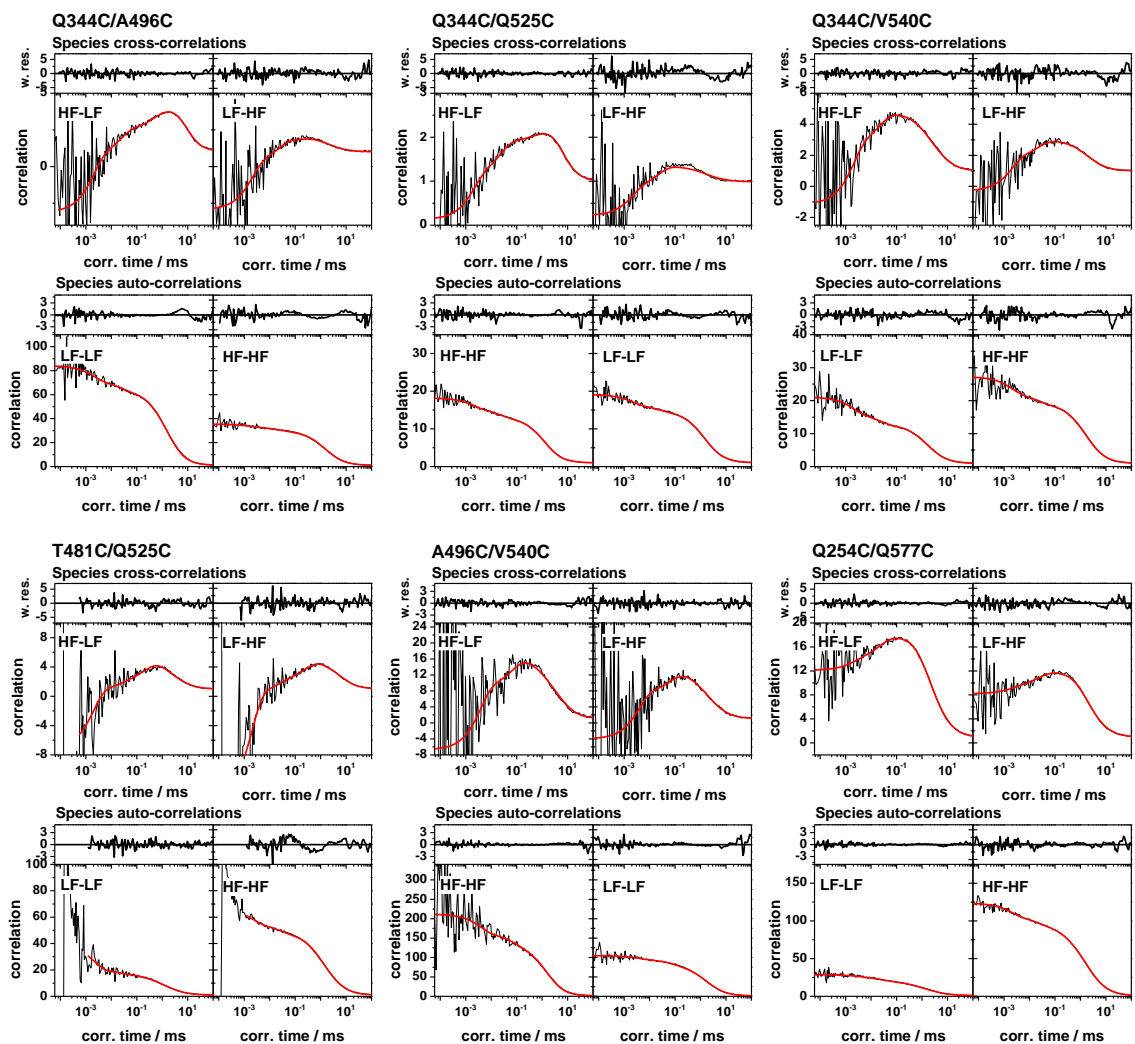

**Supplementary Figure 3C |** Filtered fluorescence correlation spectroscopy of FRET-labeled variants for the human guanylate binding protein 1 probeing hGBP1's internal dynamics from µs to ms. In the MFD histograms high FRET, H, and low FRET, L, species were identified to generate a variant specific set of filters. These filters were used to calculate two species cross-correlation functions, *sCCF*s, and two species autocorrelation functions, *sACF*s. For every variant the *sCCF*s and the *sACF*s are shown to the top and bottom, respectively. The sACFs and the sCCFs of all variants were analyzed by a global model (red lines, **Methods 4**, **eq. 19**) with three correlation times. The weighted residuals of the model and the data are shown to the top of the *sACF*s and the *sCCF*s. The displayed correlation curves correspond to the fluorescence intensity weighted average correlation curves obtained by subsetting the measurement. The fluorescence intensity weighted averages correspond to the displayed mean correlation amplitudes of the individual correlation channels.

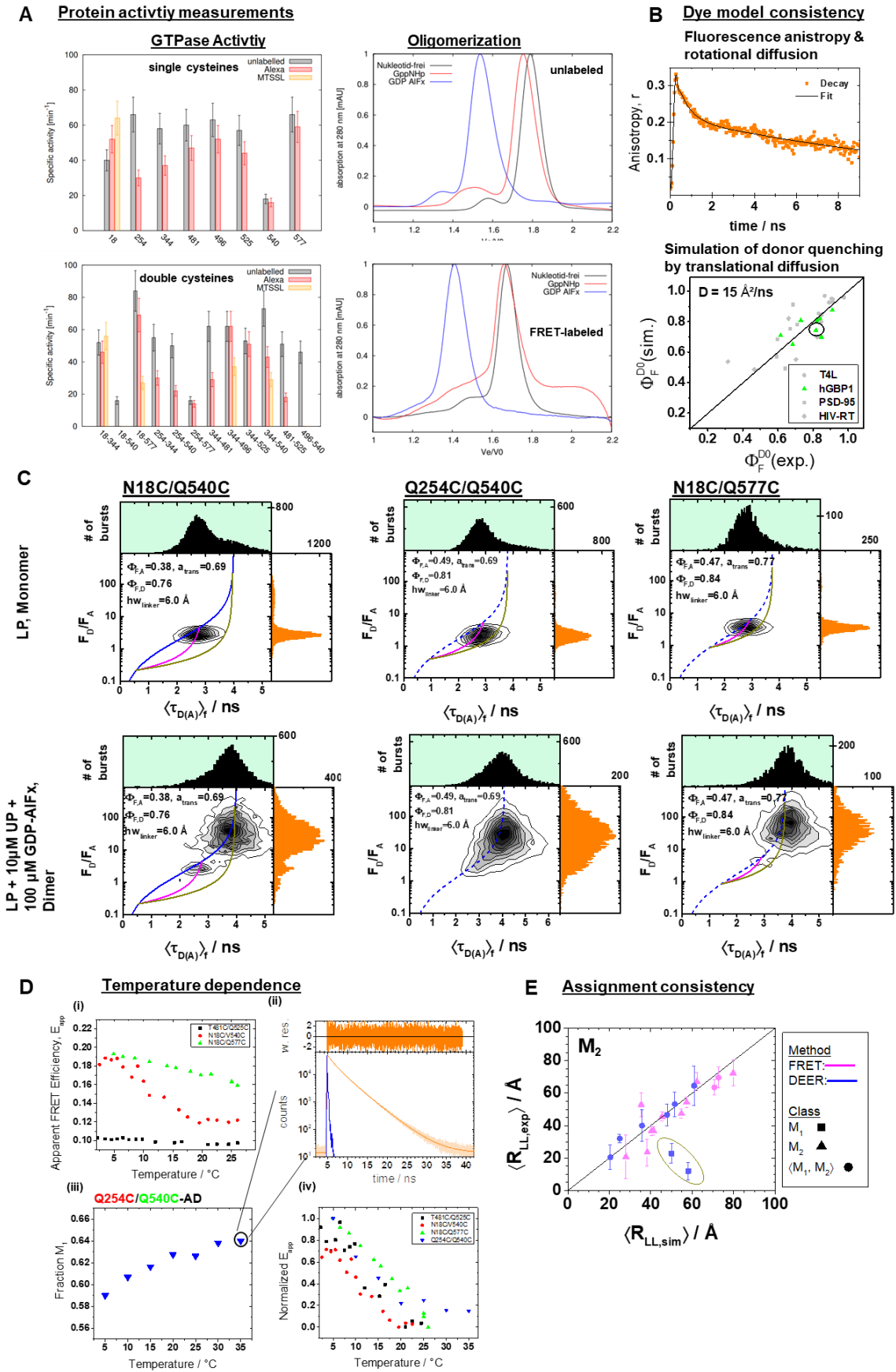

**Supplementary Figure 4 |** Quality controls for labeling based methods. (**A**) Assessment of the labeling on the protein activity by comparison of the GTPase activity (left) and the self-oligomerization of hGBP1. The effect of labeling on GTPase activity of hGBP1 as measured by the specific activities of 1μM single cysteine hGBP1

mutants at 25°C, either unlabeled or modified by Alexa488 or MTSSL at their free cysteines. Specific activities of the hGBP1 variants labeled by Alexa488 and Alexa647. The effect of the labels on the oligomerization of hGBP1 was assessed by size exclusion chromatography of 20 μM of unlabeled (top, right) and double labelled hGBP1 Cys9 (bottom, right) in the presence of 150 to 200 μM GppNHp or GDP AlFx or in the absence of any nucleotide. (**B**) The fluorescence properties of the dye were studied by time-resolved anisotropies (left). The donor Alexa488 was predominantly freely rotating. This is highlighted by the fast-initial decay of the time-resolved anisotropy, *r*(*t*). (**C**) Temperature dependent FRET measurements: (i) apparent FRET efficiency $\boldsymbol{E_{app}} = \boldsymbol{1}/(\boldsymbol{1} + \boldsymbol{S_G}/\boldsymbol{S_R})$ ($\boldsymbol{S_G}$ and $\boldsymbol{S_R}$ are the measured (uncorrected) green and red fluorescence intensities) as measured on a steady-state fluorometer. (ii) Time-resolved fluorescence decay of the hGBP1 variant Q254C/Q540C for a temperature of 35 °C. (iii) Temperature dependence of the population of the state M1 for the variant Q254C/Q540C as determined by an analysis of the associated time-resolved fluorescence decays. (iv) Normalized changes of the fluorescence observables in dependence of the temperature. (**D**) Multiparameter single-molecule fluorescence measurements of a set of comparable hGBP1 variants that is weakly affected in their GTP hydrolysis to different extent by the introduced mutations and labels. The Labeling positions N18C and Q254C are on opposing sites of the molecule. LP and UP refer to labeled protein and unlabeled protein, respectively. In the presence of UP and GDP-AlFx hGBP1 forms a dimer and undergoes significant conformational changes. These conformational changes were detected for the variants with weakly (N18C/Q577C) and variants stronger affected in their GTP hydrolysis (Q254C/Q540C) & (N18C/Q577C). The mutation Q577C has for the labeled and the unlabeled hGBP1 no effect on the specific activity. The mutation Q540C affects GTP hydrolysis activity of the labeled and the unlabeled hGBP1 equally strong. The mutation Q254C affects the GTP hydrolysis activity only the presence of a dye. (**E**) A consistency analysis reveals that two DEER datasets (encircled in yellow) resolve $M_1$ instead of an averaged state of the two states, $\langle \boldsymbol{M_1}, \boldsymbol{M_2} \rangle$. The deviation between the simulated and the experimental observables beyond the noise of the other measurements identify two distances assigned to $M_2$ as a mis-assignment.

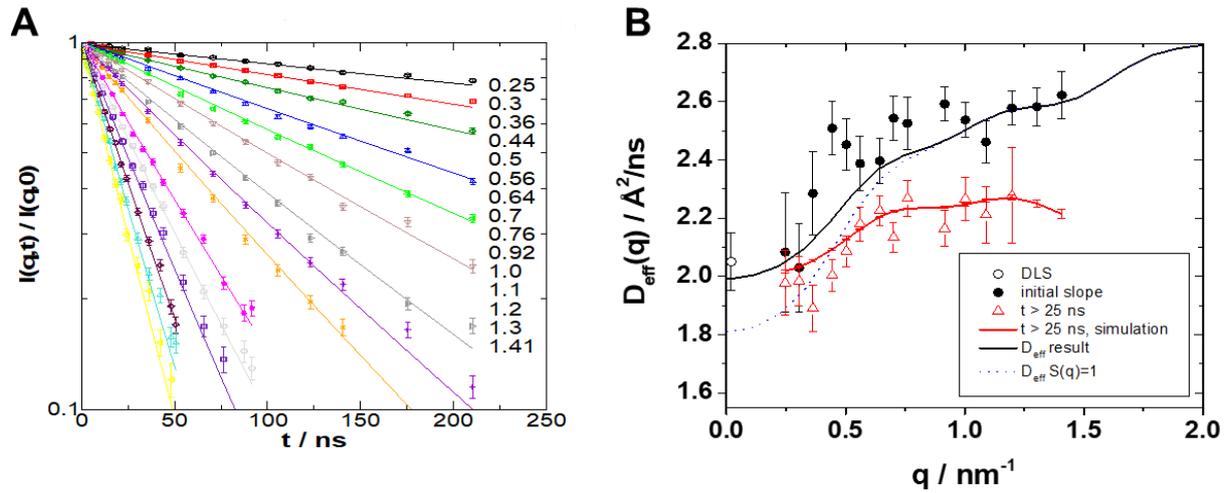

**Supplementary Figure 5 |** Neutron spin echo spectroscopy (NSE) on the hGBP1 resolves internal dynamics on the nanosecond time-scale. **(A)** Intermediate scattering function as measured by NSE with fits according to the rigid body models (**Methods 5**). The numbers to the right show the respective wave-vectors from top down with $q$ in nm$^{-1}$. **(B)** Effective diffusion coefficients determined by NSE together with rigid body diffusion calculated from the protein structure. Circles and black line correspond to values derived from the initial slope of the NSE spectra and for rigid body diffusion at infinite dilution. The strong $q$-dependent increase is entirely due to the elongated shape of the protein. Triangles and red line correspond values obtained from exponential fits to the NSE spectra and to theoretical curves (**Methods 5**, **eq. 21**) without internal dynamics for $t > 25$ ns. It is evident that rigid body diffusion includes a fast component that is visible at short times below 25 ns.

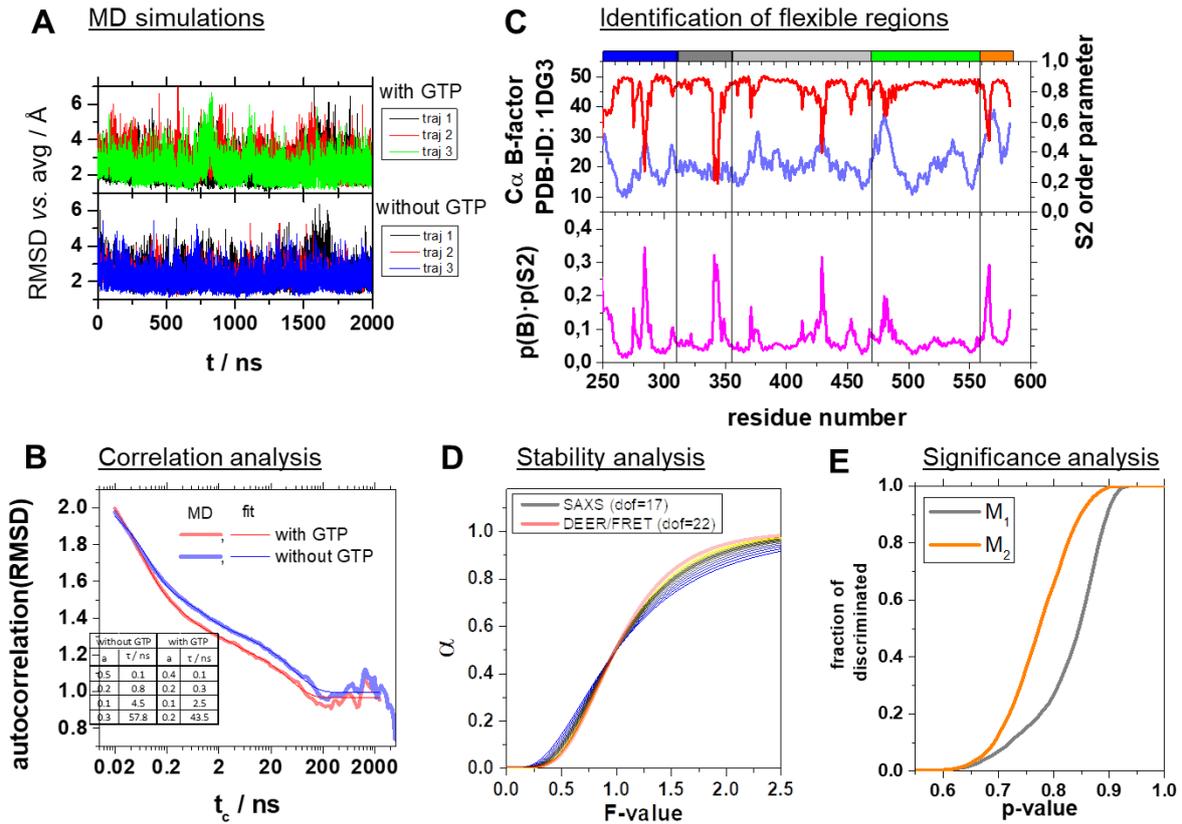

**Supplementary Figure 6 | Analysis of molecular dynamics simulations, identification of flexible regions, and structure generation and discrimination**. (**A**) Root mean squared deviation (RMSD) *vs.* the average structure of three repeats of 2 μs MD simulations in the presence and absence of GTP bound to the binding pocket of the LG domain. (**B**) Autocorrelation analysis of the RMSDs. The autocorrelation function of the RMSD vs. the average structure of the MD simulations in the presence (red) and absence (blue) of GTP are shown as light-colored thick lines. $t_c$ is the correlation time. The darker colored red and blue lines correspond to a multiexponential model ($\sum_{i=1}^{4} a_i \exp\left(-\frac{t_c}{\tau_i}\right) + b$). The amplitudes $a_i$ and the characteristic times $\tau_i$ are given in the table shown as an inset. (**C**) To the top the crystallographic B-factors of the Cα atoms (PDB-ID: 1DG3) and the NH $S^2$ order parameters calculated from the MD simulations are shown. The bottom graph illustrated the product of the B-factor normalized to the range of (0,1] and the NH $S^2$ order parameter. (**D**) Cumulative probability α that for a given F-value a proposed structural model is significantly worse than the best-found structural model. The experimental degrees of freedom ($dof_{d,SAXS}$) for SAXS was varied from 11 to 24 taking the values 11, 12, 13, 14, 15, 16, 17, 18, 20, 22, 24 with colors varying from blue to yellow. The cumulative probabilities were calculated using the best model as a reference ($x = \chi^2/\min(\chi^2)$) and an estimate of $dof_m \sim 10$ for the degrees of freedom of the model (eq. 24). The $dof_{d,SAXS}$ was varied to asses the influence of the relative weights of DEER, FRET and SAXS in eq. 25. (**E**) Fraction of discriminated structures *vs.* p-value of discriminating a pair of structure from the best structure.

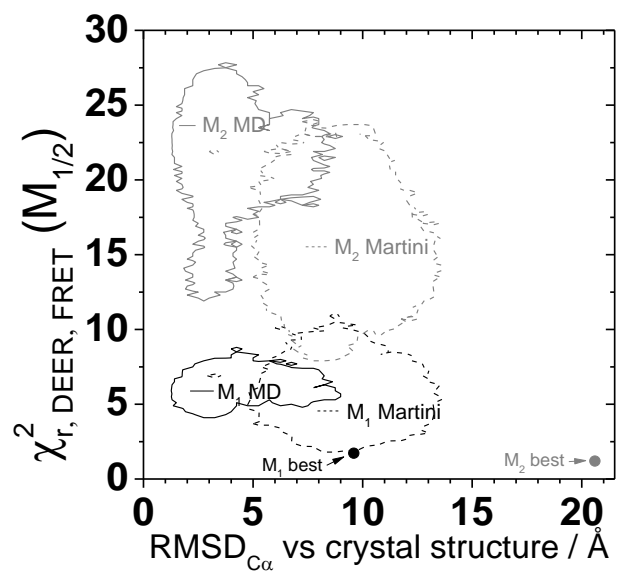

**Supplementary Figure 7 | Assessment of the conformers within the conformational space covered by MD simulations using FRET- and EPR-data (Supplementary Table 3A).** We evaluated the conformers obtained by multi-resolution MD simulations (all-atom MD and coarse-grained MD with the Martini force field of Barz et al (*39*) using our FRET positioning and screening (FPS) toolkit (*32*) to compute the quality parameter $\chi^2_{DEER,FRET}$ (eq. 23). The structural features of the conformers are described by the measure, $RMSD_{C\alpha}$ versus the hGBP1 crystal structure (PDB-ID: 1DG3). For comparison, we added the parameters of the best representative conformers for $M_1$ and $M_2$ obtained by our integrative structural modeling (**Fig. 5**).

# Supplementary Tables

**Supplementary Table 1A** | Inter-label distance analysis of DEER measurements, ensemble fluorescence decays (eTCSPC), and residual donor fluorescence anisotropies. Average distances between the spin-labels are referred to as $\langle R_{LL,exp} \rangle$. The width of the inter-spin distance distribution is $w$. The center values of the donor-acceptor distance distribution correspond to $\bar{R}_{DA,exp}(M_1)$ and $\bar{R}_{DA,exp}(M_2)$ for the states, $M_1$ and $M_2$, respectively. The average donor-acceptor distance and the inter-spin distance simulated for the full-length crystal structure of hGBP1 (PDB-ID: 1DG3) are $\bar{R}_{DA,sim}$ and $\langle R_{LL,sim} \rangle$, respectively, with corresponding distribution widths $w$. The uncertainty estimates of central distance of a state determined by FRET is $\Delta(M_{\{1,2\}})$.

| Category/ | DEER[a] | | | | ensemble FRET[b] | | | | Fluorescence Anisotropy[e] |
|---|---|---|---|---|---|---|---|---|---|
| Type | Experiment | | Simulation[c] | | Experiment | | Simulation[d] | | Experiment |
| State | Average over all states | | Crystal (PDB-ID: 1DG3) | | $M_1$ | $M_2$ | Crystal (PDB-ID: 1DG3) | | Donor Alexa488 |
| Joint species fractions $x_1$, $x_2$ | | | | | 0.61 | 0.39 | | | |
| Variant | $\langle R_{LL,exp} \rangle$ /Å | $w$ /Å | $\langle R_{LL,sim} \rangle$ /Å | $w$ /Å | $\bar{R}_{DA,exp}(M_1)$ $\pm\Delta(M_1)$/Å | $\bar{R}_{DA,exp}(M_2)$ $\pm\Delta(M_2)$/Å | $\bar{R}_{DA,sim}$ /Å | $w$ /Å | $r_\infty$ |
| N18C/Q344C | 64.6 | 12.2 | 66.5 | 9.5 | 73.6±8.6 | 67.0±5.5 | 72.4 | 12.1 | 0.15 |
| N18C/V540C | | | | | 57.8±3.6 | 36.6±2.7 | 63.6 | 8.8 | 0.11 |
| N18C/Q577C | 53.2 | 7.6 | 50.5 | 7.5 | 64.2±4.7 | 47.4±2.8 | 60.1 | 10.5 | 0.15 |
| C225/K567C | 12.0 | 5.5 | 13.0 | 6.5 | | | | | |
| C225/Q577C | 22.9 | 6.0 | 19.5 | 10.0 | | | | | |
| Q254C/Q344C | | | | | 81.3±16.5 | 72.3±7.8 | 73.9 | 12.2 | 0.13 |
| Q254C/V540C | | | | | 63.6±4.6 | 36.8±2.7 | 60.9 | 12.4 | 0.30 |
| Q254C/V577C | | | | | 70.8±9.1 | 52.9±7.4 | 73.1 | 8.9 | 0.17 |
| Q344C/T481C | | | | | 37.9±2.6 | 54.5±3.3 | 57.6 | 11.2 | 0.11 |
| Q344C/A496C | 40.0 | 9.6 | 42.5 | 10.0 | 48.0±2.9 | 23.5±8.2 | 48.4 | 10.6 | 0.19 |
| Q344C/Q525C | 32.0 | 5.4 | 29.5 | 10.2 | 46.7±2.8 | 20.7±13.3 | 41.5 | 10.7 | 0.30 |
| Q344C/V540C | 46.6 | 6.8 | 43.0 | 14.5 | 59.3±3.8 | 45.5±2.8 | 48.7 | 11.9 | 0.10 |
| T481C/Q525C | | | | | 69.6±6.4 | 69.5±6.4 | 71.2 | 11.0 | 0.30 |
| A496C/V540C | | | | | 63.6±4.6 | 63.6±4.6 | 66.8 | 11.6 | 0.23 |
| A551C/Q577C | 20.5 | 7.7 | 22.5 | 5.0 | | | | | |

[a] DEER distance distributions for calculation of average inter-spin distances and width were determined by Tikhonov regularization of the experimental DEER-traces (**eq. 7**). [b] The ensemble fluorescence decays were jointly analyzed by a quasi-static homogeneous model (*29*) with two FRET species with the species fractions $x_1$ and $x_2$ as well as a D-only species (**eqs. 10, 13**) using the donor properties in **Tab. S1B** and a Förster Radius $R_0 = 52$ Å. Moreover, the model accounted for the distance distribution with a typical width of 12 Å caused by the flexible dye-linkers (**eq. 11**). The reported uncertainty estimates, indicated by ±, include statistical uncertainties, potential systematic errors of the references, uncertainties of the orientation factor determined by the anisotropy of donor samples, and uncertainties of the AVs due to the differences of the donor and acceptor linker length (**Note S1 section 5**). The individual components are listed in **Tab. S1D**. Reference measurements of single D and A labeled variants are summarized in **Tab. S1B**, respectively. [c] For EPR-DEER the inter-spin distance distribution was calculated by a rotamer library analysis (see **Methods 7**). [d] The inter-fluorophores distance distribution and the corresponding average distance and width were calculated by accessible volume simulations.(*32*) [e] Residual anisotropies of Alexa488 in FRET labeled (Alexa488, Alexa647) variants of the human guanylate binding protein 1 determined by an analysis of the MFD histograms using a Perrin equation for a bio-exponential anisotropy decay (**Fig. S3**)

**Supplementary Table 1B |** Fluorescence lifetimes, $\tau$, with corresponding species fractions, $x$, of Alexa647 and Alexa488 maleimide coupled to different single cysteine variants of the hGBP1 determined by an analysis of the fluorescence intensity decays measured by ensemble TCSPC.

| | Alexa647[a] | | | | | | | | Alexa488[a] | | | | | |
|---|---|---|---|---|---|---|---|---|---|---|---|---|---|---|
| Variant | $x_1$ | $\tau_1$ / ns | $x_2$ | $\tau_2$ / ns | $x_3$ | $\tau_3$ / ns | $\langle\tau\rangle_x$ / ns [b] | $\Phi_{F,A}$ [c] | $x_1$ [a] | $\tau_1$ / ns | $x_2$ | $\tau_2$ / ns | $\langle\tau\rangle_x$ / ns [b] | $\Phi_{F,D}$ [c] |
| N18C  | 0.39 | 1.85 | 0.49 | 1.22 | 0.12 | 0.10 | 1.33 | 0.40 | 0.82 | 4.15 | 0.18 | 1.35 | 3.65 | 0.82 |
| Q254C | 0.58 | 2.23 | 0.42 | 1.42 |      |      | 1.89 | 0.57 | 0.69 | 3.60 | 0.31 | 0.53 | 2.65 | 0.59 |
| Q344C | 0.58 | 2.06 | 0.42 | 1.09 |      |      | 1.75 | 0.52 | 0.94 | 3.80 | 0.06 | 1.00 | 3.63 | 0.81 |
| T481C | 0.43 | 1.89 | 0.57 | 1.32 |      |      | 1.57 | 0.43 | 0.93 | 3.78 | 0.07 | 1.07 | 3.59 | 0.81 |
| A496C | 0.43 | 1.21 | 0.57 | 1.88 |      |      | 1.59 | 0.48 | 0.84 | 3.89 | 0.16 | 1.14 | 3.44 | 0.77 |
| Q525C | 0.65 | 1.93 | 0.35 | 1.08 |      |      | 1.63 | 0.49 | 0.80 | 3.51 | 0.20 | 0.66 | 2.94 | 0.66 |
| V540C | 0.65 | 2.33 | 0.35 | 1.43 |      |      | 2.02 | 0.60 | 0.85 | 4.00 | 0.15 | 1.86 | 3.67 | 0.82 |
| Q577C | 0.49 | 2.06 | 0.51 | 1.42 |      |      | 1.73 | 0.52 | 0.91 | 4.15 | 0.09 | 1.49 | 3.91 | 0.88 |

[a] The number of fluorescence lifetime components corresponds to the minimum number required to sufficiently describe the experimental data, as judged by a $\chi_r^2$ criterion. [b] $\langle\tau\rangle_x$ is the species weighted average fluorescence lifetime $\langle\tau\rangle_x = \sum_i x_i \tau_i$. [c] $\Phi_F$ is the fluorescence quantum yield of the fluorescent dye species estimated by the species averaged fluorescence lifetime $\langle\tau\rangle_x$, using $\langle\tau\rangle_x$, and $\Phi_F$ of the free dyes as a reference; (Alexa647, $\langle\tau\rangle_x$ =1.0 ns, $\Phi_F = 0.32$) (Alexa488, $\langle\tau\rangle_x$ =4.1 ns, $\Phi_F = 0.92$).

**Supplementary Table 1C |** Complementary inter-dye distance analysis of donor and sensitized acceptor fluorescence decays of sub-ensemble (seTCSPC) obtained from of single-molecule FRET experiments (**Supplementary Figure 3B**) for different FRET labeled (Alexa488, Alexa647) hGBP1 variants. The donor and acceptor fluorescence decays were described by a combination of two normal distributed distances with the central distances of $\bar{R}_{DA}$ of a state. The fractions $x_1$ and $x_2$ correspond to the fraction of the distance $\bar{R}_{DA}(M_1)$ and $\bar{R}_{DA}(M_2)$, respectively. $x_{DOnly}$ is the fraction of molecules with no energy transfer to an acceptor. The distances recovered by eTCSPC (Tab. 1A) and seTCSPC, respectively, agree nicely within the distinct precision of each data set.

| hGBP1 variant | $\bar{R}_{DA}(M_1)$ / Å | $x_1$ | $\bar{R}_{DA}(M_2)$ / Å [a] | $x_2$ [a] | $x_{DOnly}$ |
|---|---|---|---|---|---|
| N18C/Q344C  | 69.3 | 1.00 | -    | -    | 0.18 |
| N18C/V540C  | 60.0 | 0.50 | 34.1 | 0.50 | 0.37 |
| N18C/Q577C  | 63.3 | 0.65 | 45.1 | 0.35 | 0.17 |
| Q254C/Q344C | 76.1 | 1.00 | -    | -    | 0.21 |
| Q254C/V540C | 63.4 | 0.74 | 39.3 | 0.26 | 0.31 |
| Q344C/T481C | 45.0 | 0.44 | 59.3 | 0.56 | 0.17 |
| Q344C/A496C | 47.0 | 0.77 | 36.0 | 0.23 | 0.45 |
| Q344C/Q525C | 51.5 | 0.63 | 36.1 | 0.37 | 0.40 |
| Q344C/V540C | 57.8 | 0.65 | 43.8 | 0.35 | 0.24 |
| T481C/Q525C | 70.6 | 1.00 | -    | -    | 0.20 |
| A496C/V540C | 63.9 | 1.00 | -    | -    | 0.37 |
| Q254C/Q577C | 70.8 | 0.84 | 52.9 | 0.16 | 0.34 |

[a] For cases where a single normal distribution was enough to describe the data no second distance and fraction is reported.

**Supplementary Table 1D |.** Combined and individual uncertainty contributions of the average inter-dye distances determined by eTCSPC measurements.

| Variant | State | Combined uncertainty [a] | | Statistical uncertainty[b] | Reference uncertainty[c] | | Dye simulation [d] | Orientation factor [e] |
|---|---|---|---|---|---|---|---|---|
| | | $\delta_+$ | $\delta_-$ | $\delta_{stat}$ | $\delta_{+,ref}$ | $\delta_{-,ref}$ | $\delta_{AV}$ | $\delta_{\kappa 2}$ |
| N18C/Q344C | (1) | 12.1% | 11.3% | 9.1% | 6.0% | 4.2% | 1.1% | 4.9% |
| N18C/V540C | (1) | 6.3% | 6.3% | 3.1% | 1.2% | 1.1% | 1.4% | 4.5% |
| N18C/Q577C | (1) | 7.4% | 7.3% | 4.6% | 2.3% | 2.0% | 1.2% | 4.9% |
| Q254C/Q344C | (1) | 22.0% | 18.6% | 16.5% | 13.5% | 6.9% | 1.0% | 4.7% |
| Q254C/V540C | (1) | 7.2% | 7.1% | 4.4% | 2.2% | 1.9% | 1.3% | 5.8% |
| Q344C/T481C | (1) | 6.9% | 6.9% | 4.1% | 0.1% | 0.1% | 2.1% | 4.5% |
| Q344C/A496C | (1) | 6.0% | 6.0% | 2.5% | 0.4% | 0.4% | 1.7% | 5.4% |
| Q344C/Q525C | (1) | 6.0% | 6.0% | 2.5% | 0.3% | 0.3% | 1.7% | 6.1% |
| Q344C/V540C | (1) | 6.5% | 6.4% | 3.4% | 1.4% | 1.3% | 1.3% | 4.3% |
| T481C/Q525C | (1) | 9.4% | 9.1% | 6.7% | 4.0% | 3.1% | 1.2% | 6.1% |
| A496C/V540C | (1) | 7.2% | 7.2% | 4.4% | 2.2% | 1.9% | 1.3% | 5.7% |
| Q254C/Q577C | (1) | 13.0% | 12.7% | 11.0% | 4.5% | 3.4% | 1.1% | 5.1% |
| N18C/Q344C | (2) | 8.1% | 7.9% | 5.6% | 3.1% | 2.5% | 1.2% | 4.9% |
| N18C/V540C | (2) | 7.3% | 7.3% | 4.6% | 0.1% | 0.1% | 2.2% | 4.5% |
| N18C/Q577C | (2) | 5.8% | 5.8% | 2.5% | 0.4% | 0.3% | 1.7% | 4.9% |
| Q254C/Q344C | (2) | 10.9% | 10.3% | 8.3% | 5.3% | 3.8% | 1.1% | 4.7% |
| Q254C/V540C | (2) | 7.7% | 7.7% | 4.5% | 0.1% | 0.1% | 2.2% | 5.8% |
| Q344C/T481C | (2) | 5.5% | 5.5% | 2.7% | 0.8% | 0.8% | 1.5% | 4.5% |
| Q344C/A496C | (2) | 34.8% | 34.8% | 34.2% | 0.0% | 0.0% | 3.4% | 5.4% |
| Q344C/Q525C | (2) | 64.6% | 64.6% | 64.2% | 0.0% | 0.0% | 3.9% | 6.1% |
| Q344C/V540C | (2) | 5.4% | 5.4% | 2.6% | 0.3% | 0.3% | 1.8% | 4.3% |
| T481C/Q525C | (2) | 10.0% | 9.7% | 6.7% | 4.0% | 3.1% | 1.2% | 6.1% |
| A496C/V540C | (2) | 7.6% | 7.5% | 4.4% | 2.2% | 1.9% | 1.3% | 5.7% |
| Q254C/Q577C | (2) | 14.0% | 14.0% | 12.9% | 0.7% | 0.7% | 1.5% | 5.1% |

[a] The combined uncertainty used to calculate the relative ($\delta_\pm$) and the absolute ($\Delta_\pm$) uncertainties of the inter-dye distances considers the uncertainty of the dye simulations, $\delta_{AV}$, the uncertainty of the orientation factor, $\delta_{\kappa 2}$, the uncertainty of the reference sample, $\delta_{\pm,ref}$, and the statistical uncertainty determined by the noise of the measurements, $\delta_{stat}$ (see Note S1, section 5). [b] The statistical uncertainties were estimated by sampling the distances that agree with the experimental ensemble TCSPC data (see Tab. S1A). [c] The reference uncertainties were calculated assuming an uncertainty of the reference donor fluorescence lifetime of $\Delta\tau_{D(0)} = 0.15\ ns$. [d] The dye simulation error considers the labeling uncertainty of the dye. In accessible volume simulations (AV) for the dye pair Alexa488/Alexa647 this labeling uncertainty results in an expected error of $\Delta AV = 0.8$ Å (*24*). [e] The uncertainty of the distance due to the orientation factor was estimated using a wobbling in a cone model of the dyes using the experimental anisotropies (see **Tab. S1A**).

**Supplementary Table 2 |** Filtered fluorescence correlation spectroscopy (fFCS) of all FRET-labeled variants of the human guanylate binding protein 1 analyzed by a global model with joint relaxation times and individual amplitudes $A_i$. Selection criteria for the definition of the filters for the high FRET (HF) and low FRET (LF) species.

| hGBP1 variant | Amplitude[a] | | | Burst selection criteria for fFCS filter generation [b] | | | |
|---|---|---|---|---|---|---|---|
| | $A_1$ | $A_2$ | $A_3$ | HF range, $S_g/S_r$ | | LF range, $S_g/S_r$ | |
| **N18C/Q344C** | 0.38±0.09 | 0.50±0.18 | 0.12±0.20 | 0.09 | 2.03 | 5.82 | 13.19 |
| **N18C/V540C** | 0.18±0.03 | 0.37±0.04 | 0.45±0.06 | 0.10 | 1.00 | 4.00 | 10.80 |
| **N18C/Q577C** | 0.24±0.06 | 0.41±0.09 | 0.36±0.11 | 0.40 | 1.50 | 3.00 | 26.00 |
| **Q254C/Q344C** | 0.35±0.09 | 0.00±0.00 | 0.65±0.09 | 0.70 | 1.70 | 2.80 | 5.90 |
| **Q254C/V540C** | 0.34±0.03 | 0.21±0.05 | 0.45±0.06 | 0.18 | 0.56 | 3.50 | 8.90 |
| **Q344C/T481C** | 0.24±0.11 | 0.34±0.14 | 0.41±0.18 | 0.10 | 0.60 | 1.60 | 6.30 |
| **Q344C/A496C** | 0.08±0.03 | 0.30±0.06 | 0.62±0.07 | 0.03 | 0.58 | 1.70 | 22.36 |
| **Q344C/Q525C** | 0.00±0.05 | 0.45±0.09 | 0.55±0.10 | 0.07 | 0.28 | 2.94 | 8.07 |
| **Q344C/V540C** | 0.08±0.02 | 0.33±0.07 | 0.59±0.07 | 0.08 | 0.70 | 1.92 | 8.74 |
| **T481C/Q525C** | 0.58±0.13 | 0.09±0.05 | 0.32±0.14 | 0.60 | 1.52 | 16.27 | 25.40 |
| **A496C/V540C** | 0.21±0.07 | 0.32±0.17 | 0.47±0.18 | 0.07 | 1.77 | 4.61 | 11.40 |
| **Q254C/Q577C** | 0.34±0.07 | 0.45±0.11 | 0.21±0.13 | 0.51 | 1.39 | 5.18 | 11.07 |
| **Correlation time / µs** | 297.6 | 22.6 | 2.0 | | | | |

[a] The correlation times were determined by a joint/global analysis of fFCS curves (**Methods 4**, **eq. 19**, **Fig. S3C**). The amplitudes $A_1$, $A_2$, and $A_3$ are variant specific. The uncertainties were determined by a support plane analysis, which considers the mean and the standard deviation of the individual correlation channels determined by splitting the measurements into smaller sets. [b] Filters defining the high FRET (HF) and the low FRET (LF) species were generated by selecting bursts that are within the given ranges. To select high FRET and low FRET bursts the ratio of the green and red signal intensity ratio, $S_g/S_r$, was used. Sub-ensemble fluorescence decay histograms of the molecules in these ranges were generated and used to calculated filters for fFCS as previously described (*5*).

**Supplementary Table 3A | Experimental restraints for rigid body docking.** Analysis results of DEER-EPR and FRET eTCSPC. The labels Alexa488, Alexa647, and MTSSL are referred to by D, A, and R1, respectively. The names of the labeling sites report on the location of the dyes and the introduced mutation. The measurements recovered average inter-label distances $\langle R_{LL,exp} \rangle$. For FRET mean distances $\bar{R}_{DA,exp}$ and uncertainties of the mean are reported ($\Delta_+$ and $\Delta_-$). The widths of the inter-spin distance distribution $w_+$ and $w_-$ are reported for DEER. The measurements are grouped into three classes. Class informs on $M_1$ and $M_2$ by two distinct distances. Class (b) informed on $M_1$. In class (c) measurements $M_1$ and $M_2$ were not resolved into separate states. The simulated average label-to-label distances correspond to the distances of the pair of structures (M1, M2) best agreeing with SAXS, DEER, and FRET combined.

| | | Labelling site [a] | | Experiment | | | | | | Simulation | | |
|---|---|---|---|---|---|---|---|---|---|---|---|---|
| | | | | $M_1$ | | | $M_2$ | | | $M_1$ | $M_2$ | |
| # | Technique | 1 | 2 | $\bar{R}_{DA,exp}$ / Å | $\Delta_+$ / Å | $\Delta_-$ / Å | $\bar{R}_{DA,exp}$ / Å | $\Delta_+$ / Å | $\Delta_-$ / Å | $\bar{R}_{DA,sim}$ / Å | | Group |
| 1 | FRET | Q344C[D] | N18C[A] | 73.6 | 8.9 | 8.3 | 67.0 | 5.6 | 5.4 | 73.9 | 62.3 | (a) |
| 2 | | V540C[D] | N18C[A] | 57.8 | 3.6 | 3.6 | 36.6 | 2.7 | 2.7 | 63.9 | 40.3 | |
| 3 | | N18C[D] | Q577C[A] | 64.2 | 4.7 | 4.7 | 47.4 | 2.8 | 2.8 | 62.3 | 54.8 | |
| 4 | | Q344C[D] | Q254C[A] | 81.3 | 17.9 | 15.1 | 72.3 | 8.0 | 7.6 | 78.0 | 79.8 | |
| 5 | | V540C[D] | Q254C[A] | 63.6 | 4.6 | 4.5 | 36.8 | 2.7 | 2.7 | 63.2 | 41.3 | |
| 6 | | Q344C[D] | T481C[A] | 37.9 | 2.6 | 2.6 | 54.4 | 3.3 | 3.3 | 49.2 | 57.3 | |
| 7 | | A496C[D] | Q344C[A] | 48.0 | 2.9 | 2.9 | 23.5 | 8.2 | 8.2 | 43.4 | 38.3 | |
| 8 | | Q344C[D] | Q525C[A] | 46.7 | 2.8 | 2.8 | 20.7 | 13.3 | 13.3 | 43.6 | 28.0 | |
| 9 | | Q344C[D] | V540C[A] | 59.3 | 3.8 | 3.8 | 45.5 | 2.8 | 2.8 | 53.1 | 45.5 | |
| 10 | | Q577C[D] | Q254C[A] | 70.8 | 9.2 | 9.0 | 52.9 | 7.4 | 7.4 | 75.2 | 35.4 | |
| # | Technique | 1 | 2 | $\langle R_{SS,exp} \rangle$ / Å | $w_+$ / Å | $w_-$ / Å | | | | $\langle R_{SS,sim} \rangle$ / Å | | (b) |
| 11 | DEER | C225C[R1] | K567C[R1] | 12.0 | 5.5 | 5.5 | -[b] | - | - | 16.4 | 58.1 | |
| 12 | | C225C[R1] | Q577C[R1] | 22.9 | 6.0 | 6.0 | -[b] | - | - | 29.6 | 50.1 | |
| | | | | $M_1$ & $M_2$ | | | | | | $M_1$ | $M_2$ | |
| | | | | $\bar{R}_{DA,exp}$ / Å | $\Delta_+$ / Å | | $\Delta_-$ / Å | | | $\bar{R}_{DA,sim}$ / Å | | |
| 13 | FRET | T481C[D] | Q525C[A] | 69.6 | 6.6 | | 6.3 | | | 68.6 | 72.7 | (c) |
| 14 | | A496C[D] | V540C[A] | 63.6 | 4.6 | | 4.6 | | | 71.7 | 70.7 | |
| # | Technique | 1 | 2 | $\langle R_{SS,exp} \rangle$ / Å | $w_+$ / Å | | $w_-$ / Å | | | $\langle R_{SS,sim} \rangle$ / Å | | |
| 15 | DEER | N18C[R1] | Q344C[R1] | 64.6 | 12.2 | | 12.2 | | | 73.3 | 60.9 | |
| 16 | | N18C[R1] | Q577C[R1] | 53.2 | 7.6 | | 7.6 | | | 63.9 | 51.6 | |
| 17 | | A551C[R1] | Q577C[R1] | 20.5 | 7.7 | | 7.7 | | | 33.0 | 20.5 | |
| 18 | | Q344C[R1] | A496C[R1] | 40.0 | 9.6 | | 9.6 | | | 38.8 | 35.8 | |
| 19 | | Q344C[R1] | Q525C[R1] | 32.0 | 2.7 | | 2.7 | | | 36.7 | 25.0 | |
| 20 | | Q344C[R1] | V540C[R1] | 46.6 | 6.8 | | 6.8 | | | 51.4 | 47.9 | |
| | | | | $\chi_r^2$ | | | | | | 2.07 | 1.89 | |

[a] The names of the labelling sites report on the most likely position of the donor and the acceptor dyes. The distribution among the labelling sites was determined by an analysis of the time-resolved anisotropy decay, anisotropy PDA, and limited proteolysis of the labelled protein. [b] A consistency analysis identifies that $M_2$ must have long distances ($\langle R_{SS,exp} \rangle > 5$ nm) beyond the DEER detection limit for this measurement setting (see **Supplementary Note 1** (section 6) and **Fig. S4E**).

**Supplementary Table 3B | Additional restraints for rigid body docking.**

| Atom 1 | | Atom 2 | | $R_{eq}$ / Å | $k_{ij+}$ / Å | $k_{ij-}$ / Å | Restrain origin |
|---|---|---|---|---|---|---|---|
| Residue | Atom name | Residue | Atom name | | | | |
| 309 | C | 310 | N | 1.5 | 0.5 | 0.5 | Primary sequence |
| 373 | C | 374 | N | 1.5 | 0.5 | 0.5 | |
| 481 | C | 482 | N | 1.5 | 0.5 | 0.5 | |
| 563 | C | 564 | N | 1.5 | 0.5 | 0.5 | |
| 445 | Cα | 348 | Cα | 5.1 | 2 | 2 | X-ray 1DG3 |
| 391 | Cα | 336 | Cα | 8 | 4 | 4 | |
| 381 | Cα | 527 | Cα | 8.2 | 4 | 4 | |
| 323 | Cα | 292 | Cα | 9.3 | 4 | 4 | |

# Supplementary Notes

## Supplementary Note 1. Quality assessment of labeled samples for fluorescence spectroscopy, uncertainty estimation, and consistency analysis

Mutations and labels introduced to different sites of a protein may influence the conformations the protein adopts. Thus, any kind of modification is a putative cause of alterations in protein structure, function, and activity and may, in the worst-case, invalidate conclusions of following experiments. Moreover, for labels which specifically interact with the studied biomolecule, modelling the positional distribution of a labels by their sterically allowed accessible volume (AV) and/or accessible contact volume (ACV) may lead to inaccurate structural models. The ACV explicitly the fraction of fluorescent dyes bound to the molecular surface. The fraction of bound dyes was estimated by the residual anisotropy.

To address these general concerns we: (**1**) select potential labeling sites based on biochemical pre-knowledge, e.g., we avoid active/catalytic sites, (**2**) characterize the effect of the mutations on hGBP1's activity, (**3**) measure the rotational mobility of the fluorescence dyes by their anisotropy, (**4**) use the fluorescence quenching of the donor dyes by their environment in combination with coarse-grained simulations as an indicator for their translational mobility. By (**2**) and (**3**) we probe the effect of the mutations and the labels on the protein. By (**4**) we assure correct references for accurate analysis results of the fluorescence decay. By (**3**) and (**4**) we test the applicability of the coarse-grained AV and ACV model to describe the spatial distribution by which we recover for a given structural model the theoretical spectroscopic properties.

### (1) Selection of labeling sites

To avoid alteration of protein function (nucleotide binding and hydrolysis, oligomerization), neither amino acid positions in direct proximity to the nucleotide binding pocket nor inside the G domain dimerization interface nor charged amino acids on the protein surface were taken into consideration for labeling.(*76*) All chosen positions had an accessible surface area (ASA) value higher than 60 Å$^2$. In the end, eight amino acids distributed over the entire protein were chosen.

### (2) hGBP1's function: Effect of (i) mutations, (ii) labeling, and (iii) temperature

**(i) Mutations**. The used cysteine mutants are based on a cysteine-free hGBP1 construct where all nine native cysteines were mutated to alanine or serine namely: C12A, C82A, C225S, C235A, C270A, C311A, C396A, C407S and C589S. Previously, these mutations were shown to only weakly affect hGBP1's function (*24,55*). Before introduction of new cysteines for site-specific labelling the *GTPase activity* and *nucleotide binding behavior* were characterized. The GTPase activity of the labeled and unlabeled hGBP1 variants was quantified by an assay as previously described.(*77*) Briefly, hGBP1's hydrolytic activity of was controlled by high performance liquid chromatography using a Chromolith Performance RP-18 end-capped column (Merck, Darmstadt) as described earlier.(*77*) 1 µM of protein were incubated with 350 µM GTP at 25°C. The samples were analyzed at different time points. The time dependence of the substrate concentration was used to calculate the specific activities of the different protein mutants (**Fig. S4A**). The assay for measuring the protein activity has an error smaller than 10%. However, besides the relative activity the absolute uncertainty in determining the (active) protein concentration needs to be considered. Hence, the overall uncertainty in determining the absolute protein activities is ~30%. Except of A496C and Q344C/A496C, all mutants produced more GMP than GDP, as known for the wildtype hGBP1.(*78*)

**(ii) Labelling**. To check if the fluorophores bound to cysteines in hGBP1 have an impact on the oligomerization behavior an unlabeled and a labeled construct were analyzed by analytical gel filtration in the presence and the absence of a nucleotide, which induces oligomerization (**Fig. S4A**). For this analysis, the variant N18C/Q577C was chosen, because N18C and Q577C are localized in proximity to dimerization interfaces of the LG and helix α13, respectively. The fluorophores are attached to the sulfhydryl group of the cysteines via a linker of ~20 Å in length. Thus, they potentially interfere with the self-oligomerization of hGBP1. However, the elugrams of the labeled and unlabeled N18C/Q577C did not show any differences (**Fig. S4A**). This indicates that, at least for this mutant, the labels do not influence for hGBP1 assembly. As shown for hGBP1 Cys9, no dimer formation was observed in the presence of 200 µM GppNHp independent of being labelled or not.

In addition to the biochemical activity assays that report on the hydrolytic activity of the GTPase domain, we performed single-molecule FRET measurements of the labeled protein (LP) in the presence of excess unlabeled protein (UP) and GDP-AlFx as a substrate. Under these conditions, hGBP1 forms a dimer and undergoes significant conformational changes as seen by the significant changes of the FRET indicator $F_D/F_A$ in **Fig. S4C**. We found minor differences among three comparable hGBP1 variants, which are affected in the hydrolysis activity to a different degree by the presence of a fluorescent dye. Hence, we conclude that a fluorescent dye, which affects the hydrolysis activity due to its proximity to hGBP1's GTP binding site has only minor influence on the global domain arrangement that is of interest in this study (**Fig. S4C**).

**(iii) Temperature.** Using a steady-state fluorometer, we measured the variants T481C/Q525C, N18C/V540C, and N18C/Q577C. As anticipated, we found a larger change in the FRET efficiency in dependency of the temperature for the variants N18C/V540C and N18C/Q577C as compared to the variant T481C/Q525C (**Fig. S4D (i)**). For T481C/Q525C M1 and M2 are merely indistinguishable (see distances). For these measurements, we found that the largest relative change of the populations happens between 10 °C and 25 °C. From these measurements, no absolute populations can be determined. Hence, we performed after we acquired a temperature-controlled time-resolved fluorescence spectrometer a temperature series. One measurement out of this set of measurements is shown below in **Fig. S4D (ii)**. For this variant, we only found minor changes of the relative population of the states $M_1$ and $M_2$ (**Fig. S4D (iii)**). We compared the different measured variants by normalizing the observed changes (**Fig. S4D (iv)**). We found an average midpoint for all the variant of ~15°C. Hence, the relative population of the states at higher temperatures as found in a living cell resembles the measurements at room temperature.

**(3) Rotational mobility of the fluorescence dyes**

The rotational mobility of the dyes was probed by measuring their time-resolved anisotropy, $r(t)$, using multiparameter fluorescence detection in single-molecule experiments. A formal analysis of $r(t)$ by a multiexponential relaxation model reveals typically "fast" and slow rotational correlation times $\rho_{fast} < 1$ ns and $\rho_{slow} > 20$ ns (**Fig. S4B, upper panel**). The fast component we attribute to the rotation of the dye tethered to the protein. The slow component we attribute to the dye which sticks to the protein surface and thus senses the global rotation of the protein. Hence, the anisotropy decay $r(t)$ reflects local motions of the dye and global rotations of the macromolecule

$$r(t) = \left[(r_0 - r_\infty)e^{-\frac{t}{\rho_{local}}} + r_\infty\right] e^{-\frac{t}{\rho_{global}}} \cong (r_0 - r_\infty)e^{-\frac{t}{\rho_{local}}} + r_\infty e^{-\frac{t}{\rho_{global}}}. \quad (1)$$

Above $r_0$ is the fundamental anisotropy (fixed to $r_0 = 0.38$), $\rho_{global}$ is the global rotation time, $\rho_{local}$ is the local rotation time, and $r_\infty$ is the residual anisotropy. The anisotropy difference $(r_0 - r_\infty)$ relates to the fraction of freely rotating dyes.

To determine $(r_0 - r_\infty)$ for the donor dyes, the two-dimensional single-molecule histograms of the steady-state anisotropy, $r_S$, and the fluorescence lifetime, $\tau$, were analyzed with a Perrin equation derived for dyes with a bi-exponential anisotropy decay (**Fig. S4B**). In this analysis, $r_\infty$ was treated as an unknown parameter, which was determined by optimizing the Perrin equation to the experimental histogram (**Fig. S3A**, blue lines). The Perrin equation for two components is:

$$r_S(\tau) = \frac{r_0 - r_\infty}{1 + \frac{\tau}{\rho_{local}}} + \frac{r_\infty}{1 + \frac{\tau}{\rho_{global}}}. \qquad (2)$$

Using the formalism described in(*79*), we obtain $\kappa^2$ uncertainties ($\Delta R_{DA}(\kappa^2)$) corresponding to each FRET distance for $r_\infty$. Moreover, $r_\infty$ was used as estimate for the fraction of the dyes bound to the surface of the protein, to calibrate the dye's accessible surface volume (ACV) as previously described.(*33*) The labeling-site specific $r_\infty$ are compiled in **Tab. S1A**.

**(4) Translational mobility of the fluorescence dyes**

For all possible labeling sites, we simulated expected fluorescence quantum yields of dynamically quenched donor dyes Alexa488 diffusing within its accessible volume (AV) and accessible surface volume (ACV) using Brownian dynamics simulations with previously published parameters.(*29*) Finding no significant differences to other reference sample, we corroborate that within the model errors AV/ACVs describe the dye behavior (**Fig. S4B, lower panel**).

To conclude, the introduced mutations and the labeling of the dyes has no major influences on the protein function, i.e., the GTP hydrolysis and the GTP induced self-oligomerization. The time-resolved anisotropy measurements and the dynamic quenching simulations agree with a donor dye freely rotating and diffusing within its AV/ACV.

**(5) Uncertainty estimation**

For comparison of an experimentally derived distance to the distances of a structural model different sources of uncertainties of an inter-dye distances need to be combined. Here, the reported estimates of the distance uncertainties consider relative uncertainties, $\delta$, of the accessible volume model (AV), $\delta_{AV}$, the orientation factor, $\delta_{\kappa^2}$, the reference, $\delta_{Reference,\pm}$, and the statistical noise of the data, $\delta_{stat,\pm}$. These uncertainties were combined to $\delta_{R_{DA},\pm}$, a relative uncertainty of the distance:

$$\delta_{R_{DA},\pm} = \sqrt{\delta_{AV}^2 + \delta_{\kappa^2}^2 + \delta_{Reference,\pm}^2 + \delta_{stat,\pm}^2} \qquad (1)$$

$\delta_{AV}$ considers the fact that both dyes were conjugated to the protein by cysteines. Therefore, two FRET species, where the donor is either attached to the first amino acid (DA) or the second (AD), are present in the measured samples. As the donor and acceptor dyes have different geometries, the DA and AD species have distinct distributions of FRET rate constants. We previously demonstrated for the used dyes Alexa488 and Alexa647 well described by AVs, that differences in the FRET rate constant distribution between DA and AD species results in an uncertainty in the distance of $\Delta R_{DA,AV} \sim 1$ Å. This uncertainty was considered by $\delta_{AV}^2 = R_{DA}/\Delta R_{DA,AV}$.(*29*) The uncertainty $\delta_{\kappa^2}^2$ for the orientation factor $\kappa^2$ was determined as previously described using a wobbling in a cone model considering the residual anisotropies of the dyes.(*32*) The asymmetric uncertainty $\delta_{Reference,\pm}$ considered potential reference errors, propagating to systematic errors of an experimentally determined distance $R_{DA,exp}$.

The fluorescence rate constant of the donor in the absence of FRET, $k_D$, serves as reference to recover experimental distances, $R_{DA,exp}$, in the analysis of fluorescence decays. An inaccurate reference for $k_D$ propagates to systematic errors of $R_{DA,exp}$. We estimate the contribution of an inaccurate reference to $\delta_{R_{DA},\pm}$ by $\delta_{Reference,\pm}$

$$\delta_{Reference,\pm} = \left| 1 - \left( 1 \pm \left( \frac{R_{DA,exp}}{R_0} \right)^6 \cdot \delta_{k_D} \right)^{-\frac{1}{6}} \right| \qquad (2)$$

Here, $R_0$ is the Förster radius and $\delta_{k_D}$ is the relative deviation of the experimentally determine $k_D$ from the correct (true) $k_D$. To estimate $\delta_{k_D}$ we use the sample-to-sample variation of the donor fluorescence lifetimes (**Tab. S1B**). The contribution of the statistical error $\delta_{stat,\pm}$ was estimated by support plane analysis and a Monte-Carlo sampling algorithm determining distributions of parameters in agreement with the experimental data.(*63*) Using the relative uncertainty estimates the absolute uncertainties of the distances were calculated.

**(6) Consistency analysis identifies mis-assigned distances**

The fluorescence decays were analyzed by a model function, which assigns distances to the states by their amplitude. The model free analysis of the DEER data (**eq. 7**) recovered inter-spin distance distributions, $p(R_{LL})$, which reflect all conformational heterogeneities with unclear assignment to the corroborated states. The DEER analysis assigns no states to the recovered distributions. Therefore, initially all DEER constraints were assigned to $M_1$ *and* $M_2$ using the width of the distributions as uncertainty. This assignment resulted in structural models inconsistent with the data (**Fig. S4E**). The DEER measurements on C225C/K567C and C225C/Q577C revealed short distances, highlighted by the fast-initial drop of the form factors (**Fig. S2A**, gray traces). Models consistent with $M_2$ predicted long distances (> 5 nm) beyond the DEER detection limit at this measurement settings for these variants, **Fig. S2A**, green traces (for ~ 6-7 nm). Hence, C225C/K567C and C225C/Q577C were considered only to model $M_1$ for highly valuable information on the position of the short helix α13 relative to helix α12. This assignment resulted in a consistent combined set of distances for FRET and DEER used for RBD (**Tab. S3A**). Concluding, by analysis of the self-consistency of the data with the models, we unambiguously assigned recovered distances or average distances to biomolecular states.

**Supplementary Note 2. MD simulations and PC Analysis**

*MD simulations*

We performed molecular dynamics (MD) and accelerated MD (aMD)(*80*) simulations to identify collective degrees of freedom, essential movements, and correlated domain motions of hGBP1 by Principal Component Analysis (PCA).(*80*) The simulations were started from a known crystal structure of the full-length protein (PDB code: 1DG3) protonated with PROPKA(*81*) at a pH of 7.4, neutralized by adding counter ions and solvated in an octahedral box of TIP3P water(*82*) with a water shell of 12 Å around the solute. The obtained system was used to perform unbiased MD simulations and aMD simulations. The Amber14 package of molecular simulation software(*69*) and the ff14SB(*83*) force field were used to perform five unrestrained all-atom MD simulations. Three of the five simulations were conventional MD (2 μs each) and two aMD simulations (200 ns each). The "Particle Mesh Ewald"(*84*) method was utilized to treat long-range electrostatic interactions; the SHAKE algorithm(*85*) was applied to bonds involving hydrogen atoms. For all MD simulations, the mass of solute hydrogen atoms was increased to 3.024 Da and the mass of heavy atoms was decreased respectively according to the hydrogen mass repartitioning method.(*70*) The time step in all MD simulations was 4 fs with a direct-space, non-bonded cutoff of 8 Å. For initial minimization, 17500 steps of steepest descent and conjugate gradient minimization were performed; harmonic restraints with force constants of 25 kcal·mol$^{-1}$ Å$^{-2}$, 5 kcal·mol$^{-1}$·Å$^{-2}$, and zero during 2500, 10000, and 5000 steps, respectively, were applied to the solute atoms. Afterwards, 50 ps of NVT simulations (MD simulations with a constant number of particles, volume, and temperature) were conducted to heat up the system to 100 K, followed by 300 ps of NPT simulations (MD simulations with a constant number of particles, barostat and temperature) to adjust the density of the simulation box to a pressure of 1 atm and to heat the system to 300 K. A harmonic potential with a force constant of 10 kcal·mol$^{-1}$ Å$^{-2}$ was applied to the solute atoms at this initial stage. In the following 100 ps NVT simulations the restraints on the solute atoms were gradually reduced from 10 kcal·mol$^{-1}$ Å$^{-2}$ to zero. As final equilibration step 200 ps of unrestrained NVT simulations were performed. Boost parameters for aMD were chosen by the method as previously suggested.(*38*)

*PC Analysis*

In the MD simulations we found fluctuations of RMSD around the average structure of at most 8 Å RMSD for GTP bound and GTP free hGBP1 (**Fig. S6A**). A correlation analysis of these RMSD trajectories reveals that the dynamics is complex (non-exponential) and predominantly in the 10-100 ns regime (**Fig. 6B**). Structures deviating the most from the X-ray structure kink at the connector of the LG and the middle domain (**Fig. 3G**). A PCA reveals that the first five principal components describe overall more than 60% of the variance of the MD and aMD simulations (**Fig. 3A**). For PCA the GTPase domain (the least mobile domain) was superposed. The mode vectors of the principal components mapped to a crystal structure of hGBP1 (PDB-ID: 1DG3) illustrate the amplitude and the directionality of the principal components (**Fig. 3F**). The first component (1) describes a motion of the middle domain towards the LG domain. In the second component (2) the middle domain and α13 move in opposite directions. The third component (3) is like the first component with a two times smaller eigenvalue. Component (4) is like the second component, except that the middle domain and α12/13 move in the same direction. Component (5) captures a similar directionality of motion for the middle domain and α12/13 as the second component. In component (5) however, the movement of α12/13 describes a breathing motion of the catalytic LG domain. The major motions of the PCA can be described by a rotation of the middle domain relative to the GTPase domain (**Fig. 3F,** cyan sphere).

**Supplementary Note 3. Identification of rigid domain and rigid body decomposition**

The first step for RBD is the segmentation of hGBP1 into rigid domains that can represent the essential motions of the protein (**Fig. 3F**). Moreover, the segmentation should introduce sufficient degrees of freedom to fulfil the experimental constraints. The substrate free (PDB-ID: 2B92) and a substrate bound form of the LG domain (PDB-ID: 2B92) differ by only 1.1 Å RMSD, and the distance network (**Fig. 1A**) for the label based measurements were designed to probe distance changes between the LG, the middle domain, and α12/13. Hence, the LG domain (residue 1 to 309) was modeled as a single RB. In the MD simulations α12/13 moved relative to the middle and the LG domain while the middle domain changes its orientation relative to the LG domain. The DEER measurement on the variant A551C/Q577 informs on the position of α13 relative to helix α12. Consequently, α12 (residue 482 to 563) and α13 (residue 564 to 584) were treated as separate RBs. We note that we treated helix α12 in this work as rigid, because we had no experimental evidence for unfolding or kinking of this helix. Without a rearrangement of the middle domain the motion of α12/13 is highly restricted in a RBD framework. To allow for more flexibility the middle domain was represented by two bodies (residue 310 to 373 and residue 374 to 481). Overall, hGBP1 was decomposed into the LG domain, two RBs for the middle domain, helix α12, and helix α13. Therefore, to capture hGBP1's motions and fulfil the experimentally probed degrees of freedom hGBP1 was described by five RBs. Experimental evidence for such a decomposition are the FRET measurements on the variants A496C/V540C, T481C/Q525C, and Q344C/T481C which probe the conformation of α12 and the middle domain. An analysis of the protein mechanics by the $S^2$ order parameters of the NH bond determined by analysis of the MD simulations and the B-factors of the full-length protein revealed sets of characteristic spikes (**Fig. S6C**). These spikes rationalize the rigid body decomposition and identify flexible regions of the protein, which mainly correspond to flexible loops connecting individual helices. They indicate, that the middle domain, helix α12, and helix α13 are flexibly linked.

**Supplementary Note 4. Rigid body docking**

*Definition of constraints*

As next step of RBD, a set of constraints needs to be defined. The RBD procedure uses for model generation experimental constraints, a repulsion potential as a penalty function for atomic overlaps, and bonds between the RBs. The experimental constraints, bonds connecting the bodies, and the repulsive clash are considered by the terms $\chi^2_{DEER,FRET}$ (**Methods 7**, eq. 23), $\chi^2_{bonds}$, and, $\chi^2_{clash}$, respectively. The overall RBD potential $\chi^2_{RBD}$ is

$$\chi^2_{RBD} = \chi^2_{DEER/FRET} + \chi^2_{bonds} + \chi^2_{clash}. \tag{1}$$

The bond term was a combination of quadratic potentials

$$\chi^2_{bonds} = \sum_{i,j} \left(\frac{R_{ij}-R_{eq,ij}}{k_{ij,\pm}}\right)^2, \tag{2}$$

where $R_{ij} = |\vec{r}_j - \vec{r}_i|$ is the distance between the vectors $\vec{r}_j$ and $\vec{r}_i$, defined by the arrangement of the RBs, $R_{eq,ij}$ is the equilibrium distance of the distance pair, and $k_{ij,\pm}$ is a constant which depends on the sign of $R_{ij} - R_{eq,ij}$.

Overlaps of rigid bodies we penalized by the atomic overlaps in the repulsion potential $\chi^2_{clash}$:

$$\chi^2_{clash} = \sum_{i,j} \begin{cases} 0 & , R_{ij} \geq R_{wi} + R_{wj} \\ \left(\frac{R_{wi}+R_{wj}-R_{ij}}{R_{ctol}}\right)^2 & , R_{ij} < R_{wi} + R_{wj} \end{cases}. \tag{3}$$

Here $R_{ij}$ is the distance between the atoms with the index $i$ and $j$ belonging to different subunits, $R_{wi}$ and $R_{wj}$ are their van der Waals radii, and $R_{ctol}$ is a constant defining the "clash tolerance".

The restraints defining $\chi^2_{DEER,FRET}$ for $M_1$ and $M_2$ are listed in the **Tab. S3A**. Parameters defining contributions to $\chi^2_{bonds}$, namely the connection of the N- and C-term.ni and a set of weak bonds based on a crystal structure (PDB-ID: 1DG3) to stabilize the middle domain, are listed in **Tab. S3B**. The rigid body model does not allow for bending. Therefore, a very weak repulsion potential was used for $\chi^2_{clash}$ ($R_{ctol}$ = 6 Å).

*Rigid body docking*

The final docking step generates structural models fulfilling the constraints summarized by the energy function eq. 1 (above). Starting from a random initial arrangement of the RBs, forces drive the RB assembly towards a configuration with a minimum energy.(*32*) For fast calculations, the forces are applied between the average label position and optimizes the distance between the average label positions $R_{mp}$.

$$R_{mp} = \left|\langle \vec{r}_{L1}^{(i)} \rangle - \langle \vec{r}_{L2}^{(j)} \rangle\right| = \left|\frac{1}{n}\sum_{i=1}^{n}\vec{r}_{L1}^{(i)} - \frac{1}{m}\sum_{j=1}^{m}\vec{r}_{L2}^{(j)}\right|. \tag{4}$$

Here $\vec{r}_{L1}^{(i)}$ and $\vec{r}_{L2}^{(j)}$ are the coordinates of the two labels in the conformation (*i*) and (*j*). Using the mean position of the dyes instead of the full spatial distribution of the dyes reduces the complexity of the RBD and increases its speed (*32*). DEER and fluorescence decays measurements recover inter-label distance distributions, *p*($R_{LL}$), and not the distance between the label positions. For a uniformly populated AV

$$\langle R_{LL} \rangle = \left|\langle \vec{r}_{L1}^{(i)} - \vec{r}_{L2}^{(i)} \rangle\right| = \frac{1}{nm}\sum_{j=1}^{n}\sum_{j=1}^{m}\left|\vec{r}_{L1}^{(i)} - \vec{r}_{L2}^{(i)}\right| \tag{5}$$

To use average distances $\langle R_{LL} \rangle$ during RBD a transfer function converted the experimental average inter label distance $\langle R_{LL} \rangle$ to $R_{mp}$.(*32*) After docking, the spatial distribution of the labels were simulated for the generated structural models to calculate average distances $\langle R_{LL} \rangle$ as a set of model distances, {$R_{\text{model}}$}.